\documentclass[useAMS,usenatbib]{mn2e}

\usepackage[dvips]{graphicx}
\usepackage{amssymb}
\usepackage{amsmath}
\usepackage{amstext}

\newcommand{\HI}{H\,{\sc i}}
\newcommand{\HII}{H\,{\sc ii}}

\newcommand{\Ha}{H$\alpha$}
\newcommand{\Hb}{H$\beta$}

\newcommand{\abox}{12+log(O/H)}

\newcommand{\nodata}{...}

\newcommand{\Te}{$T_{\rm e}$}

\newcommand{\TeOiii}{$T_e$[\ion{O}{iii}]}
\newcommand{\TeOii}{$T_e$[\ion{O}{ii}]}

\newcommand{\CHb}{$c$(H$\beta$)}
\newcommand{\Wabs}{$W_{abs}$}

\DeclareRobustCommand{\ion}[2]{%
\relax\ifmmode
\ifx\testbx\f@series
{\mathbf{#1\,\mathsc{#2}}}\else
{\mathrm{#1\,\mathsc{#2}}}\fi
\else\textup{#1\,{\mdseries\textsc{#2}}}%
\fi}

\title[Eliminating Error in Chemical Abundances of Extragalactic \HII\ Regions]
      {Eliminating Error in the Chemical Abundance Scale for Extragalactic H\, II Regions}

\author[\'A.R. L\'opez-S\'anchez et al.]
       {\'A.R. L\'opez-S\'anchez$^{1,2}$\thanks{E-mail: alopez@aao.gov.au}
      M. A. Dopita$^{3,4.5}$,
      L. J. Kewley$^{5}$,
 \newauthor 
 	H. J. Zahid$^{5}$,
 	D. C. Nicholls$^{3}$ 
	\& J. Scharw\"achter$^{3}$\\
	$^1$Australian Astronomical Observatory, PO BOX 296, Epping, NSW 1710, Australia\\
         $^2$Department of Physics and Astronomy, Macquarie University, NSW 2109, Australia\\
	$^3$Research School of Astronomy and Astrophysics, Australian National University, Cotter Rd. Weston, ACT 2611, Australia\\
	$^4$Astronomy Department, Faculty of Science, King Abdulaziz University, P.O. Box 80203, Jeddah 21589, Saudi Arabia \\
	$^5$Institute for Astronomy, University of Hawaii, 2680 Woodlawn Drive, Honolulu, HI 96822 \\
		}

\date{Received date: 26 Jan 2012; accepted date: 22 Mar 2012}

\begin{document}

\maketitle


\begin{abstract}
In an attempt to remove the systematic errors which have plagued the calibration of the \HII\ region abundance sequence, we have theoretically modeled the extragalactic \HII\ region sequence. We then used the theoretical spectra so generated  in a double blind experiment to recover the chemical abundances using both the classical electron temperature + ionization correction factor technique, and the technique which depends on the use of strong emission lines (SELs) in the nebular spectrum to estimate the abundance of oxygen. We find a number of systematic trends, and we provide correction formulae which should remove systematic errors in the electron temperature + ionization correction factor technique. We also provide a critical evaluation of the various semi-empirical SEL techniques. Finally, we offer a scheme which should help to eliminate systematic errors in the SEL-derived chemical abundance scale for extragalactic \HII\ regions.
\end{abstract}



\section{Introduction} \label{Section1}

To understand how galaxies in the early universe evolved into those that we see locally requires an understanding of both chemical and star formation history of galaxies over cosmic time.   
Indeed, observational data compiling hundreds of thousands galaxies at different redshifts are suggesting that there actually is a deep physical connection between metallicities, masses and star-formation rate \citep[e.g.][]{Tremonti04,Brinchmann04,KE08,Lara10a,Lara10b}, although these results are mainly based on trends of the chemical abundances and not on their absolute metallicities.

Theoretical simulations can now predict the evolution of chemical abundances in galaxies \citep[e.g.][]{MollaDiaz05,DaveOppenheimer07}. However, it is difficult to put these predictions to the test because the absolute calibration of the chemical abundance scale from \HII\ regions is currently discrepant by more than a factor of two. See, for example the work of \citet{KE08} where abundances were calculated for 45,000 galaxies from the Sloan Digital Sky Survey (SDSS) \citep{York00,Abazajian03}  using 10 different calibrations given in the literature. Although all methods show some correlation with abundance, there is typically a scatter ranging over a factor of 2-3 in the absolute oxygen abundance derived from the various techniques.  The same result was found by \citet{LSE10b}, who also considered in their analysis the metallicities derived using a direct determination of the electron temperature of the ionized gas. Without a proper estimation of the absolute chemical abundances in galaxies, the comparison between observational data and theoretical models is not entirely satisfactory.

Thanks to its strong emission lines, the easiest element to measure in \HII\ region emission spectra is oxygen. This is fortunate, since O is an $\alpha$-process element made directly in short-lived massive stars. It comprises about 50\% of all the heavy elements by mass throughout the universe, and is therefore representative of all the heavy elements. By contrast, Fe is made in lower mass stars, and is not released promptly into the interstellar medium of galaxies. The O/H abundance in \HII\ regions --typically expressed in terms of \abox , where O/H is the ratio of the number of oxygen to hydrogen atomic density . This has been determined in three semi-empirical ways, as well as through direct theoretical photoionization modeling:
\begin{enumerate}
\item[1.] By direct measurement of the electron temperature, from line ratios such as the [\ion{O}{iii}] $\lambda$4363/$\lambda$5007\ \citep[see][]{OsterbrockFerland06},
followed by an analysis of the various ionization fractions in the zones of the \HII\ region which produce optical emission lines. We dub this the {\bf \Te\ method}. This has been very extensively applied by observers over the past 30 years \citep[e.g.][]{PC69,S78,Aller84,Diaz87,ITL94,EP95,Vilchez96,Sta05,Bresolin05,Izotov06,Esteban04, LSE09}. 

\item[2.] By direct measurement of recombination lines of ions of heavy elements such as  \ion{O}{ii}  and, in some cases, \ion{O}{i} and comparing these with the recombination lines of hydrogen
 \citep[e.g.][]{Esteban98,Esteban04,Esteban09,GRE04,GRE07, LSEGRPR07, Peimbert07}.
We refer to this technique as the {\bf RL method}, although it  is generally not useful in the determination of the cosmic abundance scale, since the recombination lines of heavy elements are generally too weak ($\sim10^{4-6}$ times fainter than \Hb) to measure. 

\item[3.] By measurement of the strong emission lines in the \HII\ region spectrum calibrated using 
\begin{enumerate}
\item[a.] photoionization modeling \citep[e.g.][]{Pagel79,McGaugh91,KD02,KK04,Dopita06a,DorsCopetti06}, or 
\item[b.] \HII\ regions for which the oxygen abundance is known from the \Te\ method \citep[e.g.][]{P01a,P01b,D02,PP04,PT05,Perez-MonteroDiaz05,PVT10,PM11}.
\end{enumerate}
We will call this the {\bf Strong Emission Line (SEL) method}, and it is nowadays extensively used to determine metallicities in large galaxy surveys.
\end{enumerate}

Each of these methods has its particular problems, and range of application. More importantly, where all methods can be used, the derived O/H abundances often show systematic disagreements between the various methods. These may amount to factors of two or more. In this paper, we seek to resolve some of the causes of these discrepancies by re-deriving the chemical abundances of model \HII\ regions using the $T_e$ method and the standard SEL techniques. We use a double-blind approach, so that the analysis of the photoionizaton models was conducted without any \emph{a priori} knowledge of the chemical abundances and physical parameters used in the photoionization models.

We will now briefly indicate how these alarming abundance discrepancies between the various methods may arise.

\subsection{The $T_e$ and RL methods}  \label{Section1.1}
At normal nebular temperatures ($T_e \sim 10^4$\,K), the auroral [\ion{O}{iii}]~$\lambda$4363 line is only a few percent as strong as the nebular $\lambda$5007 line, so that high quality spectra are required to give an adequate signal-to-noise ratio to measure the fainter line. When the S/N ratio is inadequate, the strength of the fainter line tends to be overestimated, leading to an underestimate in the derived abundance. This is also true for other commonly-used temperature sensitive line ratios.

A more insidious problem was first pointed out by \citet{PC69}. This still has not been resolved to everyone's satisfaction. Normal \HII\  regions are not homogeneous but contain temperature gradients, dense inclusions in which collisional de-excitation of cooling lines leads to higher temperatures, as well as colliding supersonic flows in which both density and electron temperatures may be raised as a result of shocks. In all such regions of enhanced temperature, the [\ion{O}{iii}]~$\lambda$4363/$\lambda$5007 emission line flux ratio is raised by a factor which depends only on the temperature. This would not be a problem except for the fact that the line emissivity is also raised in such regions by a factor proportional to the square of the electron density, $n_e^2$. Thus, the temperature estimate provided by the forbidden line ratio is dominated by the line ratio characterizing these over-dense inclusions, rather than providing a measure of the electron temperature of the nebula as a whole. The electron temperature is therefore systematically overestimated. 
Hence, the presence of temperature gradients or fluctuations in the ionized gas leads to the abundances based on the  \Te\ method being systematically underestimated \citep{P67,Peimbert07,Sta02,Sta05}.

 Indeed, detailed studies that compared heavy elements abundances derived using both the  \Te\ method and the RL method in Galactic \citep{Esteban04,GRE05,GRE06,GRE07,Mesa-Delgado09b,Mesa-Delgado10,Mesa-Delgado11} and extragalactic \citep{Tsamis03,Peimbert03,LSEGRPR07,Esteban09} \HII\ regions find a  very good agreement between both results when temperature fluctuations are considered. However, their origin and existence is still controversial because they are not well reproduced by standard photoionization models \citep{KingdonFerland95,RGR10}, and hence additional mechanisms are proposed to explain the presence of temperature fluctuations in the ionized gas \citep[see reviews by][]{Esteban02RMx,PP06}.

\begin{figure*}
\centering
\includegraphics[angle=0,width=0.85\linewidth]{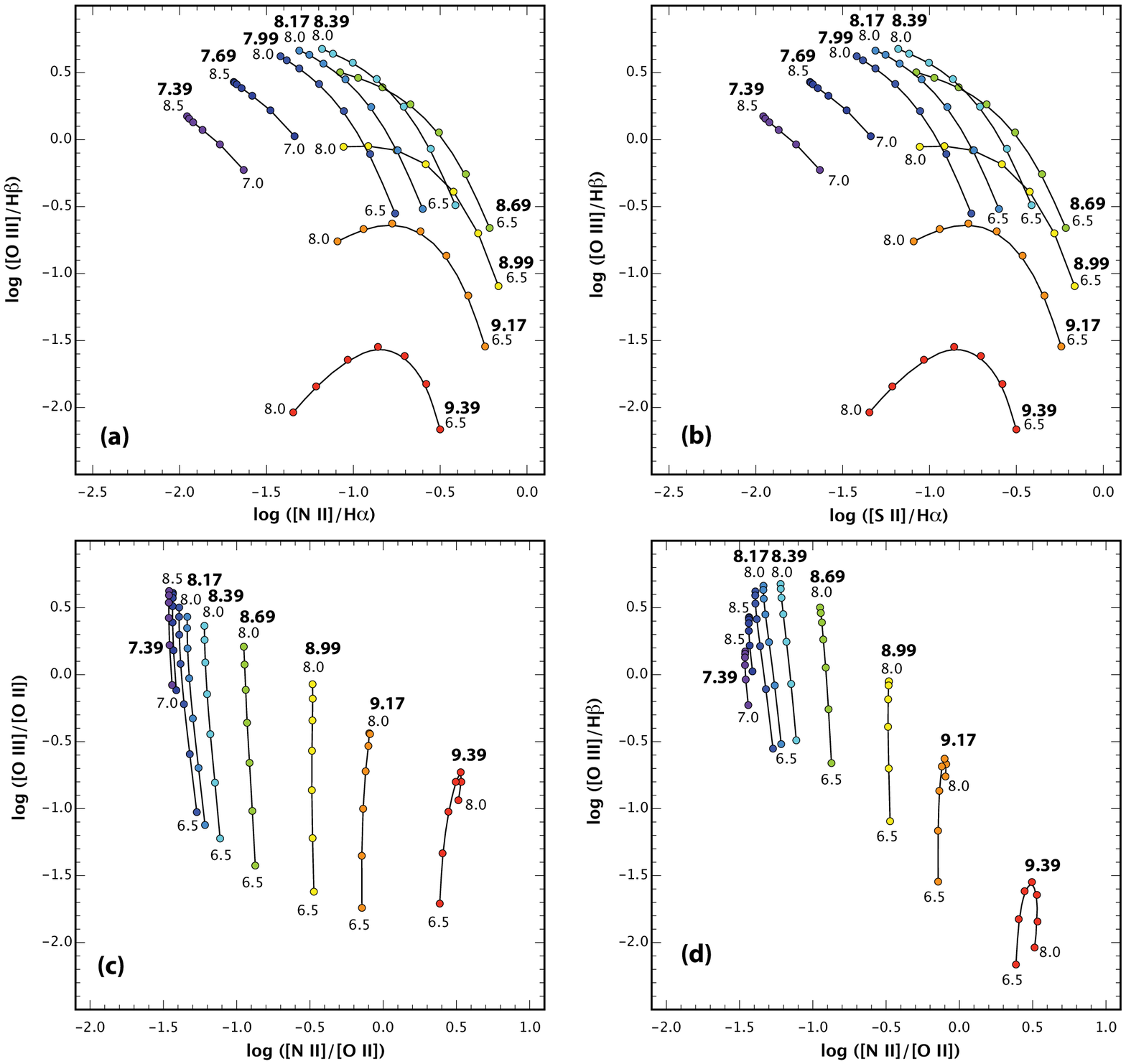}   
\caption{The complete model grid plotted on the \citet{VO87} diagnostics,  [\ion{O}{iii}]/H$\beta$ \emph{vs.} [\ion{N}{ii}]/H$\alpha$; (panel~a) and [\ion{O}{iii}]/H$\beta$ \emph{vs.} [\ion{S}{ii}]/H$\alpha$ (panel~b). In the lower panels we show the  \citet{Do00} diagnostics  [\ion{O}{iii}]/ [\ion{O}{ii}] \emph{vs.} [\ion{N}{ii}]/[\ion{O}{ii}] (panel~c) and  [\ion{O}{iii}]/H$\beta$ \emph{vs.} [\ion{N}{ii}]/[\ion{O}{ii}] (panel~d). The models are identified by their input (gas + dust) $12 + \log({\rm O/H})$ (written in bold face and color coded) and their input ionization parameter, $\log q$. Models are stepped by 0.25 in $\log q$.}
\label{Diagnostic_Diags}
\end{figure*}

\subsection{The SEL method}  \label{Section1.2}

The strong emission line (SEL) method was first proposed by \citet{Pagel79}. It relies upon the ratio of the sum of the strong forbidden oxygen emission lines with respect to \Hb, the so-called $R_{23}$ ratio; $R_{23}$ =  $I$\,(\, [\ion{O}{ii}]\, $\lambda\lambda$3726,3729 + \mbox{[\ion{O}{iii}]\, $\lambda\lambda$4959,5007\,)\, /\, $I$(\Hb).} 
Since then many other such ratios have been proposed. The most widely used of these are the calibrations of \citet{McGaugh91}  and \citet{KD02}, based on detailed photoionization modelling, and the empirical relations
provided by \citet{P01a,P01b,PT05} and \citet*{PVT10}. Both kinds of calibrations strive to improve the accuracy by making use of the [\ion{O}{iii}]/[\ion{O}{ii}] ratio as ionization parameter, which accounts for the large scatter found in the $R_{23}$ versus oxygen abundance calibration, which is larger than observational errors \citep{Kobulnicky99}

Hence, the SEL method is ideally suited to global abundance determinations in distant galaxies, since only the strong lines are  visible against the stellar continuum. The abundance scale for ensembles of \HII\ regions in galaxies has recently been calibrated by \citet{Dopita06a}. The basic problem with all ratios of optical forbidden lines to hydrogen recombination lines is that they are two-valued as a function of chemical abundance. This abundance ambiguity can only be raised by the simultaneous use of several such ratios.  An additional problem is that not all SEL method give similar oxygen abundances: 
the SEL methods based on calibrations using photoionization models generally
tend to overpredict \citep[but not always, see][]{Perez-Montero10,Dors+11}
the observed oxygen abundances derived using the \Te\ method by 0.2--0.4~dex \citep{Peimbert07,Bresolin09,LSE10b,Moustakas+10}. As we should expect, that is not the case of the SEL techniques that are based on calibrations with objects for which the oxygen abundances are well known from the \Te\ method. 
Generally speaking, the SEL method generally returns abundances which are higher than the $T_e$ method, 
but in better agreement with those derived from recombination lines of heavy elements \citep{LSE10b}.

\subsection{A double-blind test}  \label{Section1.3}

In this paper, we seek to eliminate some of the systematic errors in abundance determinations by applying the  \Te\ method and the SEL method to theoretically-generated \HII\ region spectra, rather than using observed spectra. This has the advantage of eliminating observational errors but, more importantly, it also provides a direct test of whether the various methods can recover the chemical abundances and physical conditions which were fed into the theoretical models. In order to ensure objectivity, L\'opez-S\'anchez (who applied the  $T_e$ and the SEL method) was provided a randomized selection of theoretical \HII\ region models generated by Dopita. The data provided consisted of a set of line fluxes (relative to H$\beta$) as delivered by the model. The models were identified only by letters, and no information was given about the input physical parameters which were input to generate the models. This is the classical double-blind technique often applied in medicine, but rarely in astrophysics.

In Section 2, we provide details of the techniques used to generate the theoretical spectra, in Section 3 we apply the $T_e$ to the theoretical spectra and examine how well the input parameters have been recovered. Where possible, we  provide correction formulae to apply. Section~4 presents the results provided by the most common SEL techniques. In this section, we also compare the oxygen abundances provided by the SEL methods with the derived \Te\ and RL abundances in high-quality Galactic and extragalactic \HII\ spectra. Finally, Section~5 gives the conclusions and provides our recommended procedure needed to accurately derive the oxygen abundance in \HII\ regions.


\begin{table*} 
\caption{\label{Table1}
The physical and chemical properties of the photoionization models  used in the double-blind abundance fitting, organized according to chemical abundance \abox, the nominal ionization parameter input into the model, the ionization parameter at the inner surface of the ionized gas, $ q_{\rm in}$, and mean ionization parameter in the ionized region of the model, $<q(\rm H)>$. For each model we give the emission line weighted mean temperatures (in K) for the [\ion{N}{ii}], [\ion{O}{iii}], [\ion{S}{ii}] and [\ion{O}{ii}] lines, and the mean recombination temperature for ionized hydrogen, $T(rec)$.}
\label{empirical} 
\begin{tabular}{ ccc cc ccccc }   %
\hline

Name &12+log(O/H) & $ \log q$& $ q_{\rm in}$&$<q(\rm H)>$&$<T_{\rm [N~\textsc{ii}]}>$&$<T_{\rm [O~\textsc{iii}]}>$&$<T_{\rm [S~\textsc{ii}]}>$&$<T_{\rm [O~\textsc{ii}]}>$& $T(rec)$ \\
\hline
 \\
X&9.39&8.00&5.417E+07&2.598E+07&3363&2987&4277&4284&3439 \\
J&9.39&7.50&2.510E+07&1.171E+07&3041&3620&4201&4194&3908 \\
H&9.39&7.00&1.005E+07&4.426E+06&2881&4282&4352&4357&4307 \\
C&9.39&6.50&3.393E+06&1.376E+06&2617&4568&4494&4558&4403 \\
 \\
G&9.17&7.75&5.358E+07&2.432E+07&4478&4619&5544&5487&5004 \\
AA&9.17&7.25&2.140E+07&9.721E+06&4111&5347&5633&5586&5511 \\
O&9.17&6.75&7.393E+07&3.252E+06&3766&5765&5745&5782&5745 \\
 \\
I&8.99&8.00&1.212E+08&4.573E+07&5988&5585&6780&6688&5928 \\
R&8.99&7.50&4.210E+07&1.895E+07&6014&6162&6801&6688&6433 \\
AC&8.99&7.00&1.494E+07&6.827E+06&5826&6736&6898&6872&6836 \\
Z&8.99&6.50&4.883E+07&2.119E+06&5351&6936&6865&6935&6902 \\
 \\ 
N&8.69&7.75&1.025E+08&3.850E+07&8604&8069&8782&8713&8294 \\
A&8.69&7.25&3.265E+07&1.460E+07&8680&8448&8751&8729&8618 \\
AG&8.69&6.75&1.076E+07&4.837E+06&8738&8764&8693&8791&8780 \\
 \\
U&8.39&8.00&2.720E+08&6.599E+07&10490&10492&10173&10610&10520 \\
F&8.39&7.50&7.609E+07&2.951E+07&10414&10492&10124&10517&10490 \\
W&8.39&7.00&2.314E+07&1.030E+07&10290&10538&10000&10416&10420 \\
AF&8.39&6.50&7.096E+06&3.055E+06&9960&10282&9663&10088&10040 \\
  \\
T&8.17&7.75&1.720E+08&4.959E+07&11335&12146&10784&11676&11960 \\
S&8.17&7.25&4.807E+07&1.997E+07&11207&11875&10784&11501&11630 \\
L&8.17&6.75&1.410E+07&6.185E+06&10815&11379&10353&11060&11020 \\
  \\
Q&7.99&8.00&3.853E+08&6.985E+07&11843&13584&11123&12474&13250  \\
AE&7.99&7.50&1.005E+08&3.513E+07&11792&13180&11140&12366&12820 \\
AB&7.99&7.00&2.787E+07&1.207E+07&11483&12420&10897&11912&11990 \\
E&7.99&6.50&7.929E+06&3.320E+06&10793&11434&10284&11077&10980 \\
  \\
B&7.69&8.25&7.843E+08&8.498E+07&12091&14699&11294&12997&14250 \\
V&7.69&7.75&2.068E+08&5.365E+07&12094&14394&11320&12969&13930 \\
M&7.69&7.25&5.561E+07&2.248E+07&11973&13558&11281&12676&13070 \\
K&7.69&6.75&1.542E+07&6.641E+06&11428&12374&10783&11874&11840 \\
 \\
Y&7.39&8.25&8.636E+08&8.379E+07&12220&15354&11398&13299&14830 \\
AD&7.39&7.75&2.191E+08&5.455E+07&12227&14987&11421&13269&14450 \\
P&7.39&7.25&5.769E+08&2.316E+07&12120&13985&11392&12952&13430 \\
D&7.39&6.75&1.575E+08&6.754E+06&11559&12615&10866&12065&12030 \\
\hline
\end{tabular}
\end{table*}

\section{The Theoretical Models}  \label{Section2}

We have generated a sequence of photoionization models using the \textsc{Mappings~iii}s code, an updated version of the code originally described in \citet{SD93}. These models are very similar to those produced by \citet{Do00}, and subsequently used by \citet{KD02}, but the chemical abundances have been specifically modified so as to provide a better fit to the Sloan Digital Sky Survey (SDSS) observed sequence of extragalactic  \HII\ regions \citep{Kewley06}.

Our grid of photoionization models uses as stellar input the fluxes computed by the \textsc{Starburst~99}~code  \citep{Leitherer99} appropriate for continuous star formation. The input spectrum and the assumed stellar IMF input into the \textsc{Starburst~99}~code is the same as used by \citet{Do00}. The later \textsc{Starburst~99} models \citep{VL05}, which incorporate a fully self-consistent mass-loss formulation were not used as these generate a stellar radiation field that is somewhat too soft relative to these earlier models \citep[see][for a detailed discussion of the issue]{Dopita06a}. To summarise, the most likely cause of this discrepancy is probably that the stellar wind is clumpy as a result of radiation pressure instabilities, so a multi-dimensional model atmosphere would need to be constructed. Effects of stellar rotation can also modify the emergent ionizing spectrum \citep{Levesque10}. For consistency, we have interpolated the stellar fluxes to match the chemical abundances actually used in the model, allowing for the abundance scale shift caused by the re-calibration of the solar metallicity \citep{Grevesse10}.

The extragalactic \HII\ region sequence is clearly defined in the observational plane by the standard diagnostic line ratio plots of \citet{BPT81} and \citet{VO87}. These plot the ratios [\ion{N}{ii}]\,$\lambda 6584$/\Ha, [\ion{S}{ii}]\,$\lambda\lambda 6717,31$/\Ha\ and [\ion{O}{i}]\,$\lambda 6300$/\Ha\ against the [\ion{O}{iii}]\,$\lambda 5007$/\Hb\ emission line ratio.  \citet{Kewley06} gave these diagnostics for some 45,000 galaxies drawn from the SDSS. Our photoionisation models closely reproduce the observational sequence of extragalactic \HII\ regions, bearing in mind that the ionization parameter tends to become higher as the abundance decreases, thanks to the decreasing importance of the ram pressure of the stellar winds \citep{Dopita06a}.

The solar metallicity for our photoionisation models was defined by the solar oxygen abundance given by \citet{Grevesse10}, \abox=8.69$\pm$0.05. However, the abundance of helium and carbon were assumed to vary according the formula given by \citet{Dopita06a}. For nitrogen, we use a form similar to that given in \citet{Dopita06b}:
\begin{equation}
12+ \log({\rm N/H})=7.6 + \log[Z/Z_{\odot} +(Z/Z_\odot)^2].   \label{calN_Z}
\end{equation}
Here the zero point of the nitrogen abundance has been adjusted to prove a better fit in the corresponding \citet{VO87} diagnostic diagram. Because nitrogen is an important coolant at high metallicity, this has the incidental effect of improving the fit of the models to the observed extragalactic \HII\ region abundance sequence in the other two diagnostic diagrams. For all models we include dust, and dust physics according to the description in \citet{Dopita06a}, where the depletion factors of the various elements from the gaseous phase are also given.

Our models form a grid in chemical abundance or metallicity, defined by the quantity \abox, and the ionization parameter, $\log q$,  the ratio of the number of photons passing per unit area and time to the number density of hydrogen atoms. The $q$ parameter is related to the dimensionless ionization parameter $U$ by $U = q/c$, where $c$ is the speed of light. 

Our photoionization models are spherical, with a target ionization parameter for the inner surface. This is unlike the  \citet{KD02} models, which were plane parallel. Our models are also isobaric with $\log (P/k) = 5.5$ cm$^{-3}$\,K, where $k$ is the Boltzmann's Constant, so that the density varies throughout the model, but is typically $10-30$ cm$^{-3}$. Unlike the  \citet{KD02} models, our models allow for photoelectric heating of the gas by dust grains, and include the effects of radiation pressure, which is becoming appreciable at the higher values of the ionization parameter.

\begin{figure*}
\centering
\includegraphics[angle=0,width=0.84\linewidth]{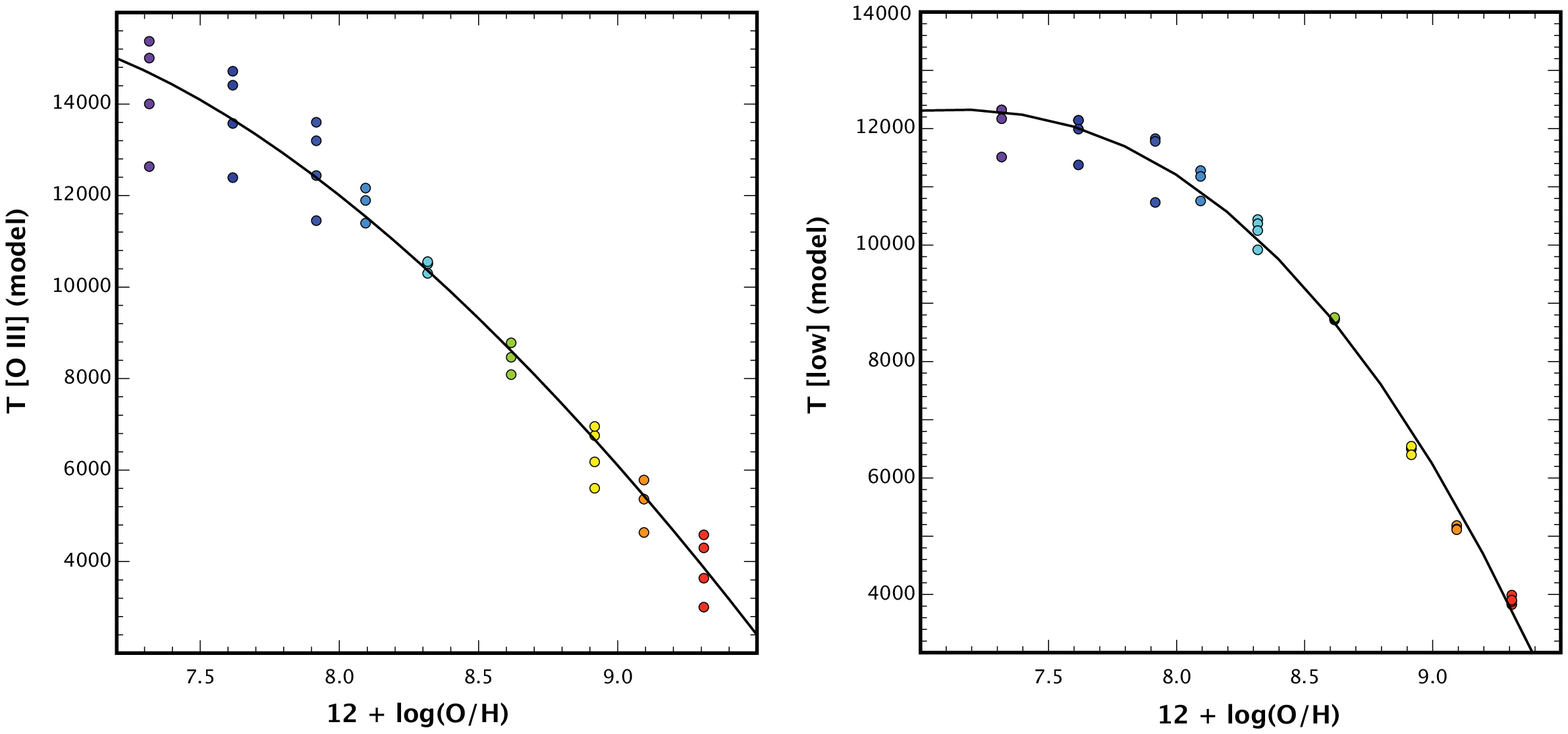}   
\caption{The relation between the oxygen abundance, 12 + log(O/H), and the electron temperature, \Te\, given by the models in the high-ionization zone containing [\ion{O}{iii}] (left) and in the low-ionization zone containing the ions [\ion{S}{ii}],  [\ion{O}{ii}], and [\ion{N}{ii}] (right). The empirical fits we have derived for both high- and low-ionization zones as given by Eqs.~\ref{ZT_OIII} and \ref{ZT_OII} are shown as solid lines on these plots.}
\label{Z_Te}
\end{figure*}

\subsection{The Double-Blind Test Models}  
\label{Section2.1}

From the theoretical model \HII\ region grid we have taken, for a fixed metallicity, every second model in $\log q$, and applied to these the standard observational techniques for deriving abundances. In order that no bias was incurred in the analysis, the spectra were randomly selected from the input grid of models, and the raw spectra labelled simply A,B,C....Z, AA..AG were passed by Dopita to L\'opez-S\'anchez for analysis. This analysis returned physical conditions, ionization parameters, and chemical abundances given by each of the semi-empirical techniques commonly used by observers, which included both the \Te\ and the SEL methods. We can then compare the results of this analysis to the known inputs into the original models to discover where there exist systematical errors, and to evaluate the magnitude of the random errors that can be generated by these techniques. The advantage of this approach is that it totally eliminates errors generated by the observations, and provides spectra of essentially infinite signal to noise for analysis.

In an ideal world, the parameters returned by the abundance analysis would be the same as those input into the photoionization models. However, in practice the different abundance techniques have different implicit assumptions about both the atomic physics relating to particular ionic species, and more generally, about the physics of the \HII\ region itself. Therefore, we should not expect the results to agree exactly.

The physical and abundance parameters of the input models are given in Table~\ref{Table1}. Note that we give three values of the ionization parameter, the nominal ionization parameter input into the model, $q$, the ionization parameter at the inner surface of the ionized gas, $ q_{\rm in}$, and mean ionization parameter in the model, $<q(\rm H)>$. These are related to each other, but not in any obvious way. For example, $q$ is calculated on the basis of an assumption that the temperature in the inner zone of the \HII\ region is 10,000\,K. Because of our isobaric assumption, $ q_{\rm in}$ is lower than $q$ for the high-abundance models, and higher than $ q_{\rm in}$ for the low-abundance models. The ratio of $ q_{\rm in}$ to  $<q(\rm H)>$ is always greater than two due to the spherical divergence of the radiation field, and can reach values as high as ten where radiation pressure gradients start to play a role in determining the gas pressure.

As stated above, the input stellar spectrum, the physical conditions and the chemical abundance set were chosen so as to replicate, as far as possible, the observed sequence for extragalactic \HII\ regions \citep{Do00,Kewley06}. In Figure~\ref{Diagnostic_Diags} we show the two main diagnostic diagrams from \citet{VO87} and the diagrams which \citet{Do00} used to separate the chemical abundance from the ionization parameter at the high-abundance end of the scale.

These models display quite a tight relationship between the oxygen abundance, \abox, and the electron temperature, \Te, given by the high-ionization zone from the  [\ion{O}{iii}]~($\lambda$4959+$\lambda$5007)/$\lambda$4363 ratio, and in the low-ionization (H$^+$ + He$^0$) zone emitting the  [\ion{O}{ii}]~($\lambda$3726+$\lambda$3729), [\ion{N}{ii}]~$\lambda$6584 and   [\ion{S}{ii}]~($\lambda$6717+$\lambda$6731) lines. This is shown in Fig~\ref{Z_Te}.  Note that there is larger scatter in the plot for the [\ion{O}{iii}] zone, because the temperature in this zone is more sensitive to the input ionization parameter. On the left hand plot, the left-hand upper points correspond to high ionization parameter, while the reverse is true for the points in the lower right-hand side of the diagram. There is much less variation in the local ionization parameter in the low-ionization zone of the model \HII\ regions, so that the scatter is less for the right-hand panel of Fig.~\ref{Z_Te}.

For each zone of the model \HII\ region we have generated empirical fits to the oxygen abundance -- electron temperature relationship. For  [\ion{O}{iii}] we find
\begin{equation}
T_e {\rm [O\,\textsc{iii}]}  = 25200 - 1400x -2000|x-7.0|^{1.7}, \label{ZT_OIII}
\end{equation}
while for the low-ionization zone we have
\begin{equation}
T_e {\rm (low)} = 9500 + 400x -2000|x-6.9|^{2.35,}  \label{ZT_OII}
\end{equation}
where $x = 12+\log(\rm O/H)$ in both cases.
These curves emphasize once again that the electron temperature is a key diagnostic for the chemical abundance in  \HII\ regions.

\begin{figure*} 
\centering
\includegraphics[angle=0,width=0.84\linewidth]{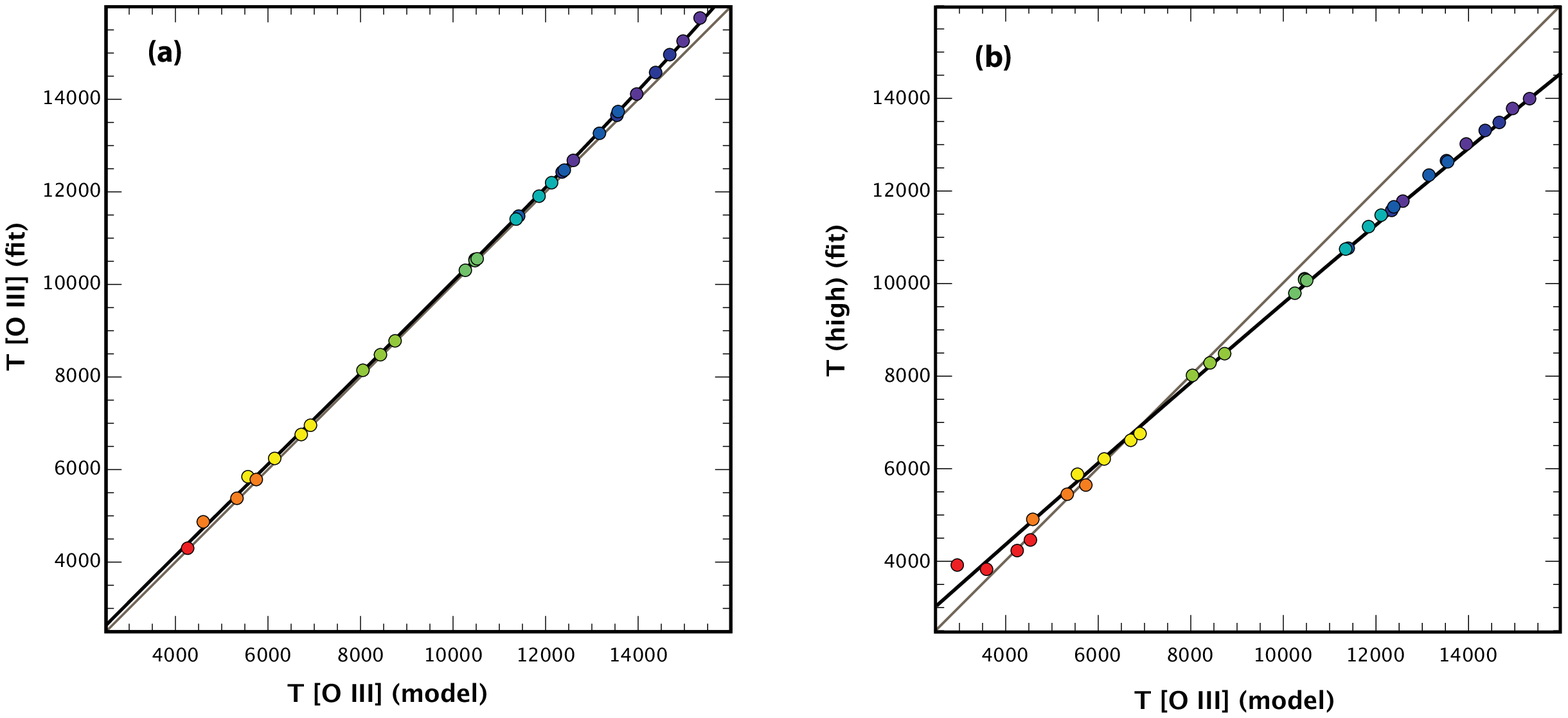}   
\caption{Comparison between the  \Te[\ion{O}{iii}] given by the model (x-axis) and the temperature derived by analysis of the model spectra. The left hand panel directly compares the  [\ion{O}{iii}]  temperature from the model and the derived  [\ion{O}{iii}]  temperature, while the right -hand panel compares the fitted temperature drawn from the mean of the [\ion{O}{iii}] ,  [\ion{S}{iii}]  and  [\ion{Ar}{iii}]  temperatures with the  [\ion{O}{iii}]  temperature from the model. The empirical fits we have derived are shown as a solid line for both panels. As in figure \ref{Z_Te}, the points are color coded according to  the input abundance set. }
\label{te_high}
\end{figure*}

\section{Application of the \Te\ method}  \label{Section3}

\subsection{Physical conditions}  
\label{Section3.1}

We have taken care to analyze the model spectra in exactly the same way as an observer would do when presented with the spectrum of real \HII\ regions. If there are no systematic effects in the analysis technique, this exercise should yield the same electron temperatures and chemical abundances as was input into the models. One potential source of difference is the fact that the atomic data used in the nebular models is not necessarily the same as are used in the analysis of the spectra, although in many cases they are. 
The atomic data used for our models are detailed in \citet{SD93} (see their Section 3.5.2), which are similar to the atomic data included in IRAF task {\it temden} in the {NEBULAR} package by \citet{SD95} (see their Table~1). 
%
The output of the various nebular modeling codes have been compared against each other for a number of standard models on more than one occasion through the so-called \emph{Lexington Benchmarks} code-off exercise \citep{Ferland95}. Any egregious errors made by any one of them have subsequently been eliminated from the appropriate codes. We can therefore be fairly certain that other photoionization codes such as \textsc{Cloudy} \citep[e.g.][]{Ferland94} will produce very similar output to the \textsc{Mappings~III}s code used here.

The model spectra were created without considering any extinction or reddening, and not stellar absorption in the \HI\ Balmer lines, to decrease the uncertainties when analyzing the data. Thus, for the analysis of the model spectra we have used a reddening coefficient, \CHb, and a equivalent width of the stellar absorption underlying the \HI\ Balmer lines, \Wabs, equal to 0, following \citet{LSE09}.  

We derived the electron temperature, $T_e$, and density, $n_e$, of the ionized gas of the models using several emission line ratios. The values obtained for each region are compiled in Table~\ref{physical}. All determinations were computed with the IRAF task {\it temden} in the {NEBULAR} package \citep{SD95}. We used the updated atomic dataset for the O$^+$, S$^+$, and S$^{++}$ ions as input to the NEBULAR routine. The references are indicated in Table~4 of \citet{GRE05}.

In accord with Fig.~\ref{Z_Te}, and with equations~\ref{ZT_OIII} and \ref{ZT_OII}, we assumed a two-zone approximation for the \HII\ region to describe the temperature structure of the nebulae. In the high-ionization H$^+$ + He$^+$ zone, we used both the [\ion{O}{iii}] temperature, $T_e$[\ion{O}{iii}], and the  mean of $T_e$[\ion{O}{iii}], $T_e$[\ion{S}{iii}] and $T_e$[\ion{Ar}{iii}],  $T_e {\rm (high)}$,  as the representative temperature for high ionization potential ions. In the low-ionization zone  containing H$^+$ + He$^0$  we adopted the mean of $T_e$[\ion{O}{ii}], $T_e$[\ion{S}{ii}] and $T_e$[\ion{N}{ii}], $T_e {\rm (low)}$,  for the low ionization potential ion temperatures. 

The line ratios used to measure the temperatures are: 

\smallskip

[\ion{O}{iii}] ($\lambda$4959+$\lambda$5007)/$\lambda$4363, 
\smallskip
 
[\ion{S}{iii}] ($\lambda$9069+$\lambda$9530)/$\lambda$6311, 
\smallskip

[\ion{Ar}{iii}] ($\lambda$7135+$\lambda$7751)/$\lambda$5192, 
\smallskip

[\ion{N}{ii}] ($\lambda$6548+$\lambda$6583)/$\lambda$5755,
\smallskip

[\ion{O}{ii}] ($\lambda$3726+$\lambda$3729)/($\lambda$7319+$\lambda$7330), and
\smallskip

[\ion{S}{ii}] ($\lambda$6717+$\lambda$6731)/($\lambda$4069+$\lambda$4076). 
\smallskip
 
We note that some analyses of high-quality, deep spectrophotometric data of bright \HII\ galaxies \citep[e.g.][]{Hagele08,Hagele11,Perez-Montero10} consider the electron temperature derived for a particular ion only to obtain the ionic abundance of that particular ion. However, the two-zone approximation for the \HII\ region is most widely used in extragalactic analyses.

Since, in the analysis of the spectra, no information about the pressures or density adopted in the models was given,  we used  the standard density-sensitive line ratios  [\ion{O}{ii}]~$\lambda$3726/$\lambda$3729, [\ion{N}{i}]~$\lambda$5198/$\lambda$5200,  [\ion{Cl}{iii}]~$\lambda$5517/$\lambda$5537 and [\ion{S}{ii}]~$\lambda$6717/$\lambda$6731 to derive $n_e$. In practice, all models have a common pressure $P/k = 3\times 10^5$cm$^{-3}$K, all density-sensitive ion line ratios should be at their low-density limit. Table~\ref{physical} compiles the value derived in all cases, and the average value adopted for each model. Once $n_e$ was obtained we used it to derive $T_e$ using the [\ion{O}{iii}], [\ion{S}{iii}], [\ion{Ar}{iii}],  [\ion{O}{ii}], [\ion{S}{ii}], and [\ion{N}{ii}] line ratios, and we iterated until convergence was attained. 

\begin{figure}
\centering
\includegraphics[angle=0,width=0.8\linewidth]{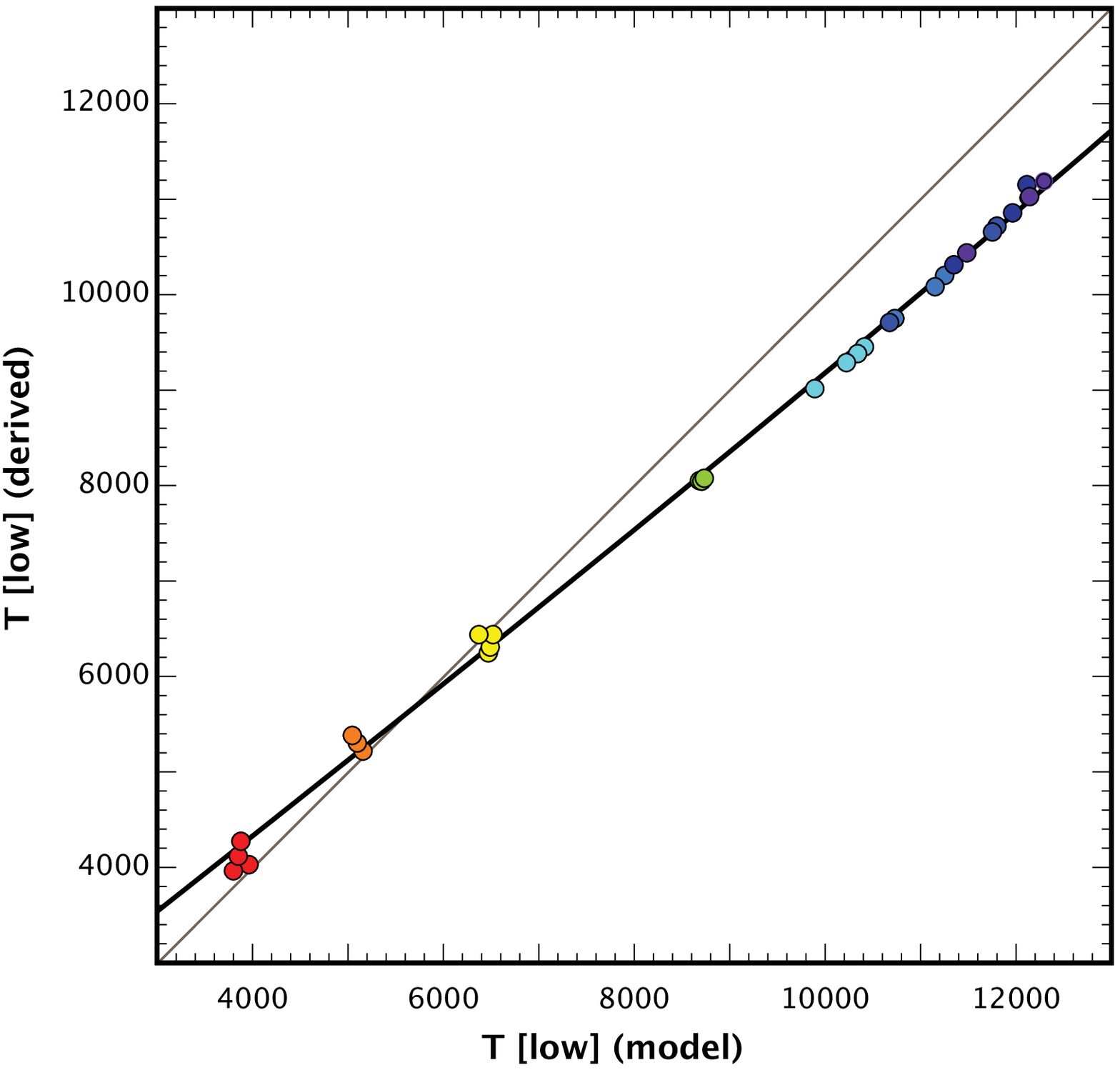}   
\caption{Comparison between the (emission weighted)  \Te\ for the low-ionization species [\ion{N}{ii}], [\ion{O}{ii}] and [\ion{S}{ii}], as returned by the model (x-axis) and the \Te\ derived for these same species from the temperature-sensitive line ratios (y-axis). The fit to the data (Eq.~\ref{T_low}) is also plotted.} \label{te_low}
\end{figure}

\begin{table*} 
\caption{Total abundances derived from the analysis of line intensity ratios of the model spectra using IRAF task {\it ionic} considering the electron temperature derived from all the high-ionization species, \Te(high), and the low ionization species, \Te(low), separately. All abundances are expressed in the form of 12 + log (X/H) or log (X/O).}
\label{abundances}
\scriptsize{ 
\begin{tabular}{ cccc cc cc cc cc cc }   %
\hline

 Model       &       & O$^{++}$/O$^+$        &     O/H                         &     N/H                         &     N/O                         &     S/H                         &     S/O                                 
                         &     Ne/H                       &     Ne/O                       &     Ar/H                         &     Ar/O                         &     Cl/H                         &     Cl/O                            \\

\hline
\noalign{\smallskip}

A  &   &   -0.592   &  {\bf  8.816 }    &    7.432   &  -1.384     &    7.147   &  -1.669     &    8.374   &  -0.443     &    6.883   &  -1.933     &    4.424   &  -4.393   \\  
B  &   &    0.326   &   {\bf 7.777  }   &    6.129   &  -1.648     &    5.974   &  -1.803     &    7.109   &  -0.668     &    5.512   &  -2.264     &    3.426   &  -4.351  \\  
C$^a$  &   &   -1.771   &   {\bf 8.880 }    &    8.264   &  -0.616     &    7.601   &  -1.279     &    9.125   &   0.245     &    \nodata   &   \nodata   &    4.814   &  -4.066   \\
D  &  &   -0.680   &   {\bf 7.571}     &    5.923   &  -1.649     &    5.843   &  -1.728     &    7.230   &  -0.342     &    5.618   &  -1.954     &    3.240   &  -4.332   \\
E  & &   -1.258   &  {\bf  8.197 }    &    6.629   &  -1.567     &    6.466   &  -1.731     &    8.165   &  -0.032     &    6.632   &  -1.565     &    3.761   &  -4.436  \\
F &   &   -0.126   &  {\bf  8.495 }    &    6.959   &  -1.536     &    6.810   &  -1.686     &    7.912   &  -0.583     &    6.340   &  -2.155     &    4.124   &  -4.371   \\   
G  &  &   -0.774   &  {\bf  9.157 }    &    8.105   &  -1.051     &    7.595   &  -1.562     &    8.644   &  -0.512     &    7.301   &  -1.855     &    4.822   &  -4.335   \\ 
H$^a$  &  &   -0.889   &  {\bf  8.763 }    &    8.205   &  -0.558     &    7.669   &  -1.094     &    8.600   &  -0.163     &    7.441   &  -1.323     &    4.884   &  -3.879   \\
I  &   &   -0.654   &  {\bf  9.050 }    &    7.796   &  -1.254     &    7.306   &  -1.744     &    8.495   &  -0.555     &    7.033   &  -2.017     &    4.568   &  -4.482   \\ 
J$^a$    &   &   -0.267   &  {\bf  8.560 }    &    8.080   &  -0.480     &    7.723   &  -0.837     &    8.164   &  -0.396     &    7.011   &  -1.548     &    4.905   &  -3.654   \\
K  &  &   -0.697   &  {\bf  7.857 }    &    6.235   &  -1.623     &    6.142   &  -1.715     &    7.530   &  -0.327     &    5.931   &  -1.927     &    3.444   &  -4.413   \\ 
 L  & &  -0.925   &  {\bf  8.362 }    &    6.807   &  -1.555     &    6.627   &  -1.735     &    8.103   &  -0.259     &    6.573   &  -1.789     &    3.914   &  -4.448   \\
 M &  &   -0.067   &  {\bf  7.805 }    &    6.153   &  -1.653     &    6.094   &  -1.711     &    7.212   &  -0.594     &    5.597   &  -2.208     &    3.423   &  -4.383   \\
 N &  &   -0.103   &  {\bf  8.733 }    &    7.332   &  -1.401     &    7.116   &  -1.617     &    8.160   &  -0.573     &    6.630   &  -2.103     &    4.411   &  -4.323   \\
 O & &   -1.863   &  {\bf  9.358 }    &    8.320   &  -1.038     &    7.611   &  -1.747     &    9.332   &  -0.026     &    7.996   &  -1.362     &    4.835   &  -4.523  \\
 P   &  &   -0.036   &  {\bf  7.502 }    &    5.831   &  -1.670     &    5.786   &  -1.716     &    6.900   &  -0.602     &    5.283   &  -2.219     &    3.121   &  -4.381  \\
Q &  &    0.236   &  {\bf  8.079 }    &    6.456   &  -1.622     &    6.308   &  -1.770     &    7.422   &  -0.656     &    5.829   &  -2.250     &    3.721   &  -4.358   \\
R &  &   -0.647   &  {\bf  9.073 }    &    7.877   &  -1.196     &    7.439   &  -1.633     &    8.597   &  -0.476     &    7.174   &  -1.899     &    4.698   &  -4.374   \\
S &   &   -0.255   &  {\bf  8.302 }    &    6.708   &  -1.594     &    6.598   &  -1.704     &    7.747   &  -0.555     &    6.163   &  -2.139     &    3.900   &  -4.402   \\
T &  &    0.112   &  {\bf  8.256 }    &    6.657   &  -1.599     &    6.539   &  -1.717     &    7.621   &  -0.635     &    6.036   &  -2.221     &    3.896   &  -4.360   \\
U &  &    0.160   &  {\bf  8.456 }    &    6.920   &  -1.536     &    6.760   &  -1.696     &    7.824   &  -0.632     &    6.252   &  -2.204     &    4.117   &  -4.338   \\
V &  &    0.230   &  {\bf  7.780 }    &    6.129   &  -1.651     &    6.010   &  -1.770     &    7.125   &  -0.655     &    5.524   &  -2.256     &    3.423   &  -4.357   \\
W & &   -0.677   &  {\bf  8.564 }    &    7.055   &  -1.510     &    6.847   &  -1.717     &    8.167   &  -0.397     &    6.634   &  -1.930     &    4.132   &  -4.432   \\ 
X$^a$   &  &   -0.567   &  {\bf  8.303 }    &    7.803   &  -0.501     &    7.492   &  -0.811     &    7.765   &  -0.538     &    6.544   &  -1.759     &    4.640   &  -3.663   \\ 
Y  &  &    0.309   &  {\bf  7.486 }    &    5.819   &  -1.667     &    5.664   &  -1.822     &    6.814   &  -0.672     &    5.211   &  -2.275     &    3.127   &  -4.359   \\ 
 Z &  &   -2.018   &  {\bf  9.183 }    &    8.028   &  -1.155     &    7.455   &  -1.728     &    9.371   &   0.188     &    7.939   &  -1.244     &    4.695   &  -4.488  \\
AA  &  &   -1.189   &  {\bf  9.320 }    &    8.267   &  -1.053     &    7.622   &  -1.698     &    8.962   &  -0.357     &    7.644   &  -1.676     &    4.854   &  -4.466   \\  
AB  &  &   -0.447   &  {\bf  8.149 }    &    6.537   &  -1.612     &    6.429   &  -1.720     &    7.672   &  -0.477     &    6.077   &  -2.072     &    3.728   &  -4.422   \\
AC  &   &   -1.232   &  {\bf  9.163 }    &    7.988   &  -1.174     &    7.449   &  -1.714     &    8.923   &  -0.239     &    7.563   &  -1.600     &    4.699   &  -4.464   \\  
AD &  &    0.243   &  {\bf  7.481 }    &    5.813   &  -1.668     &    5.697   &  -1.784     &    6.822   &  -0.660     &    5.218   &  -2.264     &    3.123   &  -4.358   \\
AE &  &    0.048   &  {\bf  8.092 }    &    6.467   &  -1.625     &    6.373   &  -1.719     &    7.466   &  -0.625     &    5.868   &  -2.224     &    3.719   &  -4.373   \\
AF  &  &   -1.473   &  {\bf  8.599 }    &    7.132   &  -1.467     &    6.870   &  -1.729     &    8.644   &   0.045     &    7.177   &  -1.423     &    4.151   &  -4.448  \\
 AG  &  &   -1.281   &  {\bf  8.873 }    &    7.525   &  -1.348     &    7.161   &  -1.713     &    8.759   &  -0.114     &    7.347   &  -1.526     &    4.429   &  -4.444   \\

\hline
\end{tabular}
}
\flushleft
$^a$ For these models, there is not estimation of the high and low electron temperatures using the [\ion{O}{iii}] and [\ion{O}{ii}] ions because their auroral lines are not tabulated (see Table~\ref{physical}). Furthermore, {\it ionic} task does not allow electron temperatures lower than 5,000\,K for the low ionization ions, so we adopted  \mbox{\Te(low) = 5,000\,K} in these four cases.

\end{table*}

\begin{table*} 
\caption{Total abundances derived from the analysis of line intensity ratios of the model spectra using IRAF task {\it ionic} considering only the electron temperature derived from the [\ion{O}{iii}] lines in the high-ionization zone, \Te[\ion{O}{iii}],  and that derived from the [\ion{O}{ii}] lines, \Te[\ion{O}{ii}], in the low-ionization zone. All abundances are expressed in the form of 12 + log (X/H) or log (X/O).}
\label{abundances_ox} 
\scriptsize{
\begin{tabular}{ cccc cc cc cc cc cc }   %
\hline

 Model       &       & O$^{++}$/O$^+$        &     O/H                         &     N/H                         &     N/O                         &     S/H                         &     S/O                                 
                         &     Ne/H                       &     Ne/O                       &     Ar/H                         &     Ar/O                         &     Cl/H                         &     Cl/O                \\

\hline
\noalign{\smallskip}

A &  &   -0.511   &  {\bf  8.707 }    &    7.362   &  -1.344     &    7.114   &  -1.593     &    8.269   &  -0.438     &    6.797   &  -1.910     &    4.388   &  -4.319  \\
B &  &    0.344   &  {\bf  7.616 }    &    6.021   &  -1.595     &    5.884   &  -1.732     &    6.968   &  -0.648     &    5.406   &  -2.210     &    3.337   &  -4.279 \\
C$^a$ &   &   -1.771   &   {\bf 8.880 }    &    8.264   &  -0.616     &    7.601   &  -1.279     &    9.125   &   0.245     &    \nodata   &   \nodata   &    4.814   &  -4.066   \\
D &   &   -0.662   &  {\bf  7.435 }    &    5.828   &  -1.607     &    5.770   &  -1.666     &    7.109   &  -0.327     &    5.521   &  -1.915     &    3.071   &  -4.365  \\
E  &   &   -1.239   &  {\bf  8.065 }    &    6.538   &  -1.527     &    6.397   &  -1.669     &    8.047   &  -0.018     &    6.538   &  -1.527     &    3.686   &  -4.380   \\
F & &   -0.087   &  {\bf  8.397 }    &    6.896   &  -1.501     &    6.759   &  -1.638     &    7.823   &  -0.574     &    6.271   &  -2.126     &    4.076   &  -4.321  \\
G & &   -0.559   &  {\bf  9.027 }    &    8.031   &  -0.996     &    7.609   &  -1.418     &    8.511   &  -0.515     &    7.185   &  -1.842     &    4.837   &  -4.190 \\
H$^b$ &  & -0.934   &  {\bf  8.758 }    &    8.200   &  -0.558     &    7.649   &  -1.109     &    8.598   &  -0.160     &    7.443   &  -1.315     &    4.860   &  -3.898 \\
I &  & -0.140   &  {\bf  8.888 }    &    7.730   &  -1.158     &    7.408   &  -1.480     &    8.322   &  -0.566     &    6.891   &  -1.997     &    4.671   &  -4.217 \\
J$^a$     &   &   -0.267   &  {\bf  8.560 }    &    8.080   &  -0.480     &    7.723   &  -0.837     &    8.164   &  -0.396     &    7.011   &  -1.548     &    4.905   &  -3.654   \\
K & &   -0.698   &  {\bf  7.740 }    &    6.153   &  -1.587     &    6.073   &  -1.667     &    7.427   &  -0.313     &    5.849   &  -1.891     &    3.371   &  -4.369  \\
 L &  &   -0.903   &  {\bf  8.236 }    &    6.720   &  -1.516     &    6.562   &  -1.674     &    7.990   &  -0.247     &    6.482   &  -1.755     &    3.844   &  -4.392   \\
M & &  -0.043   &  {\bf  7.675 }    &    6.064   &  -1.610     &    6.024   &  -1.651     &    7.095   &  -0.579     &    5.506   &  -2.168     &    3.351   &  -4.323 \\
N & &   -0.010   &  {\bf  8.654 }    &    7.291   &  -1.363     &    7.100   &  -1.554     &    8.084   &  -0.570     &    6.575   &  -2.079     &    4.388   &  -4.266  \\
 O &   &   -1.759   &  {\bf  9.192 }    &    8.208   &  -0.984     &    7.549   &  -1.643     &    9.172   &  -0.020     &    7.887   &  -1.305     &    4.768   &  -4.424   \\ 
P &  &  -0.013   &  {\bf  7.367 }    &    5.740   &  -1.627     &    5.713   &  -1.654     &    6.780   &  -0.586     &    5.189   &  -2.177     &    3.047   &  -4.320  \\
Q & &    0.248   &  {\bf  7.943 }    &    6.364   &  -1.579     &    6.232   &  -1.711     &    7.303   &  -0.640     &    5.737   &  -2.206     &    3.645   &  -4.298  \\
R & &   -0.508   &  {\bf  8.962 }    &    7.811   &  -1.151     &    7.429   &  -1.533     &    8.486   &  -0.476     &    7.080   &  -1.882     &    4.686   &  -4.276 \\
S & &   -0.228   &  {\bf  8.185 }    &    6.629   &  -1.556     &    6.538   &  -1.647     &    7.642   &  -0.543     &    6.080   &  -2.105     &    3.837   &  -4.348 \\
T & &    0.139   &  {\bf  8.143 }    &    6.583   &  -1.560     &    6.480   &  -1.663     &    7.520   &  -0.623     &    5.958   &  -2.185     &    3.835   &  -4.308 \\
U &  &  0.198   &  {\bf  8.363 }    &    6.862   &  -1.501     &    6.717   &  -1.646     &    7.740   &  -0.623     &    6.189   &  -2.174     &    4.071   &  -4.292  \\
V & &   0.243   &  {\bf  7.639 }    &    6.033   &  -1.606     &    5.930   &  -1.709     &    7.001   &  -0.638     &    5.428   &  -2.210     &    3.344   &  -4.294 \\
W & &   -0.645   &  {\bf  8.451 }    &    6.977   &  -1.473     &    6.793   &  -1.658     &    8.063   &  -0.388     &    6.550   &  -1.901     &    4.074   &  -4.377 \\
X$^a$   &  &   -0.567   &  {\bf  8.303 }    &    7.803   &  -0.501     &    7.492   &  -0.811     &    7.765   &  -0.538     &    6.544   &  -1.759     &    4.640   &  -3.663   \\ 
Y & &    0.297   &  {\bf  7.319 }    &    5.701   &  -1.619     &    5.562   &  -1.757     &    6.670   &  -0.650     &    5.098   &  -2.221     &    3.031   &  -4.289  \\
Z & &   -1.929   &  {\bf  9.027 }    &    7.922   &  -1.105     &    7.393   &  -1.634     &    9.222   &   0.195     &    7.844   &  -1.183     &    4.626   &  -4.400  \\
AA & &   -1.054   &  {\bf  9.173 }    &    8.172   &  -1.002     &    7.591   &  -1.583     &    8.818   &  -0.356     &    7.522   &  -1.651     &    4.821   &  -4.352  \\
AB & &  -0.425   &  {\bf  8.022 }    &    6.450   &  -1.572     &    6.362   &  -1.660     &    7.558   &  -0.464     &    5.986   &  -2.036     &    3.657   &  -4.365 \\
AC & &   -1.134   &  {\bf  9.022 }    &    7.895   &  -1.127     &    7.406   &  -1.616     &    8.787   &  -0.235     &    7.450   &  -1.571     &    4.652   &  -4.370  \\
AD & &    0.251   &  {\bf  7.330 }    &    5.709   &  -1.621     &    5.610   &  -1.720     &    6.690   &  -0.641     &    5.116   &  -2.215     &    3.037   &  -4.293 \\
AE & &    0.071   &  {\bf  7.967 }    &    6.383   &  -1.584     &    6.306   &  -1.661     &    7.355   &  -0.611     &    5.781   &  -2.185     &    3.650   &  -4.316 \\
AF & &  -1.445   &  {\bf  8.476 }    &    7.046   &  -1.429     &    6.809   &  -1.667     &    8.531   &   0.056     &    7.088   &  -1.388     &    4.083   &  -4.392  \\
AG & &   -1.214   &  {\bf  8.740 }    &    7.434   &  -1.305     &    7.109   &  -1.630     &    8.633   &  -0.106     &    7.244   &  -1.495     &    4.373   &  -4.367 \\

\hline
\end{tabular}
}
\flushleft

$^a$ For these models there is not estimation of the high and low electron temperatures using the [\ion{O}{iii}] and [\ion{O}{ii}] ions because their auroral lines are not tabulated (see Table~\ref{physical}). The results shown here are the same that in Table~\ref{abundances} because we used the \Te(high) and \Te(low).
Furthermore, {\it ionic} task does not allow electron temperatures lower than 5,000\,K for the low ionization ions, so we adopted   \Te(low) = 5,000\,K in these four cases.

\end{table*}

\subsection{Results for $T_e$}  \label{Section3.2}

The temperatures derived from the [\ion{O}{iii}]~($\lambda$4959+$\lambda$5007)/$\lambda$4363 ratio, and the mean temperature derived for the high-ionization species  [\ion{O}{iii}],  [\ion{S}{iii}]  and  [\ion{Ar}{iii}], are compared with the  [\ion{O}{iii}] emission-line weighted temperature given by the model in Fig~\ref{te_high}. The agreement for  [\ion{O}{iii}] is very good, while the mean of the high-ionization species tends to underestimate the true temperature. 

We empirically derive the following third-order correction formula to transform the measured  [\ion{O}{iii}]  line temperatures into the model  [\ion{O}{iii}]  line temperatures:
\begin{equation}
t_4^c = t_4(1-0.05(t_4-1)^2) \label{OIII_cor},
\end{equation}
where $t_4 = T_e/10,000$\,K is the temperature measured from the spectra, and $t_4^c$ is the temperature corrected to that  delivered by the corresponding model in units of $10^4$\,K.  The corrections to  \Te[\ion{O}{iii}]  are almost always smaller than the intrinsic uncertainty in the derived temperature -- typically of order 500--1000\,K.

For the mean of the high excitation species [\ion{O}{iii}] ,  [\ion{S}{iii}]  and  [\ion{Ar}{iii}] the corresponding correction formula is:
\begin{equation}
t_4^c = t_4(1.12+0.02t_4^2) -0.087, \label{T_high}
\end{equation}
with $t_4$ as the average value of the electron temperature derived using the [\ion{O}{iii}],  [\ion{S}{iii}]  and  [\ion{Ar}{iii}] ratio.

For the low-ionization zones of the model \ion{H}{ii} regions we use the mean of the  [\ion{O}{ii}] ,  [\ion{S}{ii}]  and  [\ion{N}{ii}]  temperatures given by the models, as was also done in the fitting exercise. The result of the comparison is shown in Fig~\ref{te_low}. 
As we see, we find a clear linear correlation between model and fit, although the slope is not exactly unity. The empirical fit is given by the formula:
\begin{equation}
t_4^c = t_4(1.28-0.03t_4^2) -0.15, \label{T_low}
\end{equation}
where $t_4$ and $t_4^c$ are the temperature measured from the spectrum and the corrected to that  delivered by the corresponding model in units of $10^4$\,K, respectively. The correlation coefficients of the fits in Eqn. \ref{OIII_cor}, \ref{T_high} and \ref{T_low} are all in excess of 0.98.

These differences between the temperatures delivered by the model are the result of the temperature structure within the ionized region. The temperatures delivered by the \textsc{Mappings~IIIs} code are weighted by the emissivity of, for example, the [\ion{O}{iii}]~$\lambda$5007 line for [\ion{O}{iii}]. The [\ion{O}{iii}]~$\lambda$4363 line is more strongly affected by temperature stratification as pointed out by \citet{PC69}, so that the temperature delivered by measurement of the  [\ion{O}{iii}]~($\lambda$4959+$\lambda$5007)/$\lambda$4363 ratio is different to the strong-line weighted mean temperature delivered by the code.

For observers who are only able to measure the  [\ion{O}{iii}]  temperature in the high-ionization zone (H$^+$ + He$^+$), and need a correction formula to estimate the temperature in the low-ionization (H$^+$ + He$^0$) zone, the models provide a convenient relationship that fits to an accuracy of $\pm300$\,K for temperatures above 10,000\,K:
\begin{equation}
T_e ({\rm low}) =T_e {\rm [O~\textsc{iii}]} + 450-70\exp[(T_e [{\rm O~\textsc{iii}}]/5,000)^{1.22}]. \label{T_hi-low}
\end{equation}
Note that the temperature  sensitivity to the abundance is strong, and that in high-abundance \HII\ regions, the low ionization species deliver a lower electron temperature. Note however, that the form and shape of Eq.~\ref{T_hi-low} is rather different from the one proposed by \citet{Garnett92}, \mbox{$T_e ({\rm low}) =0.7 \times T_e {\rm [O~\textsc{iii}]} +3,000$.} However, both the temperature offset and the slope are similar in the range \mbox{$8,000 \leq T_e {\rm [O~\textsc{iii}]} \leq 12,000$} where most of the high-quality nebula spectra have been obtained.

\begin{figure*}
\centering
\includegraphics[angle=0,width=0.84\linewidth]{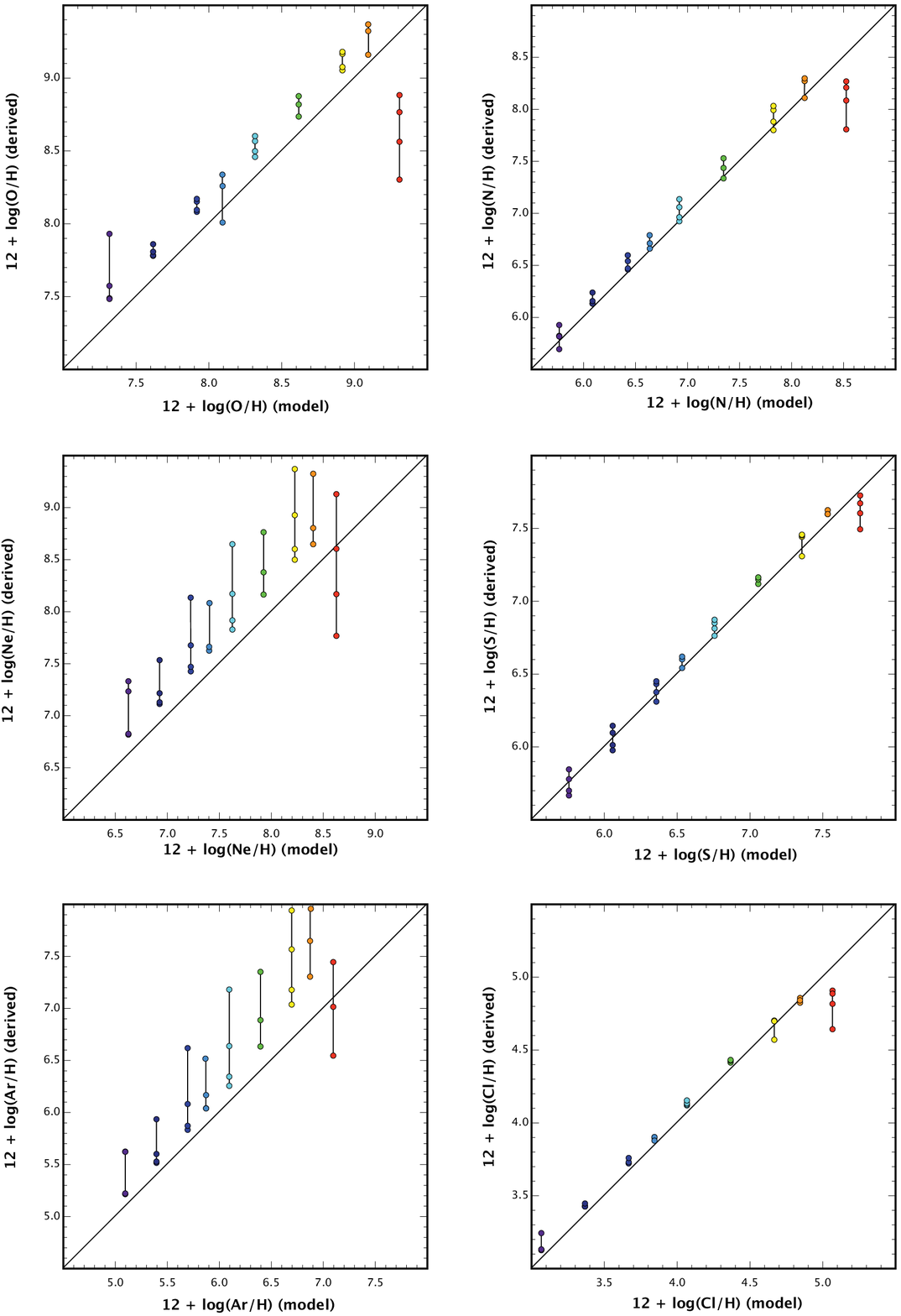}   
\caption{A comparison of the total abundances derived  from the analysis of line intensity ratios of the model spectra and using \Te(high) and \Te(low) as representative for the electron temperature of the high- and low-ionization species, respectively  (see Table~\ref{abundances}), with the total abundances provided by the models.
The fit is fairly good for the species for which the optical line emission arises principally in the low-ionization zone, N, S and Cl, but there is clearly a systematic error for the other species, O, Ne and Ar. For 12+log(O/H) $= 9.39$, the \Te\ method fails because IRAF task {\it ionic}  does not allow electron temperatures lower than 5,000\,K for the low ionization ions, and hence  \Te(low) = 5,000\,K  was adopted.}
\label{Z_Te_high}
\end{figure*}

\begin{figure*}
\centering
\includegraphics[angle=0,width=0.82\linewidth]{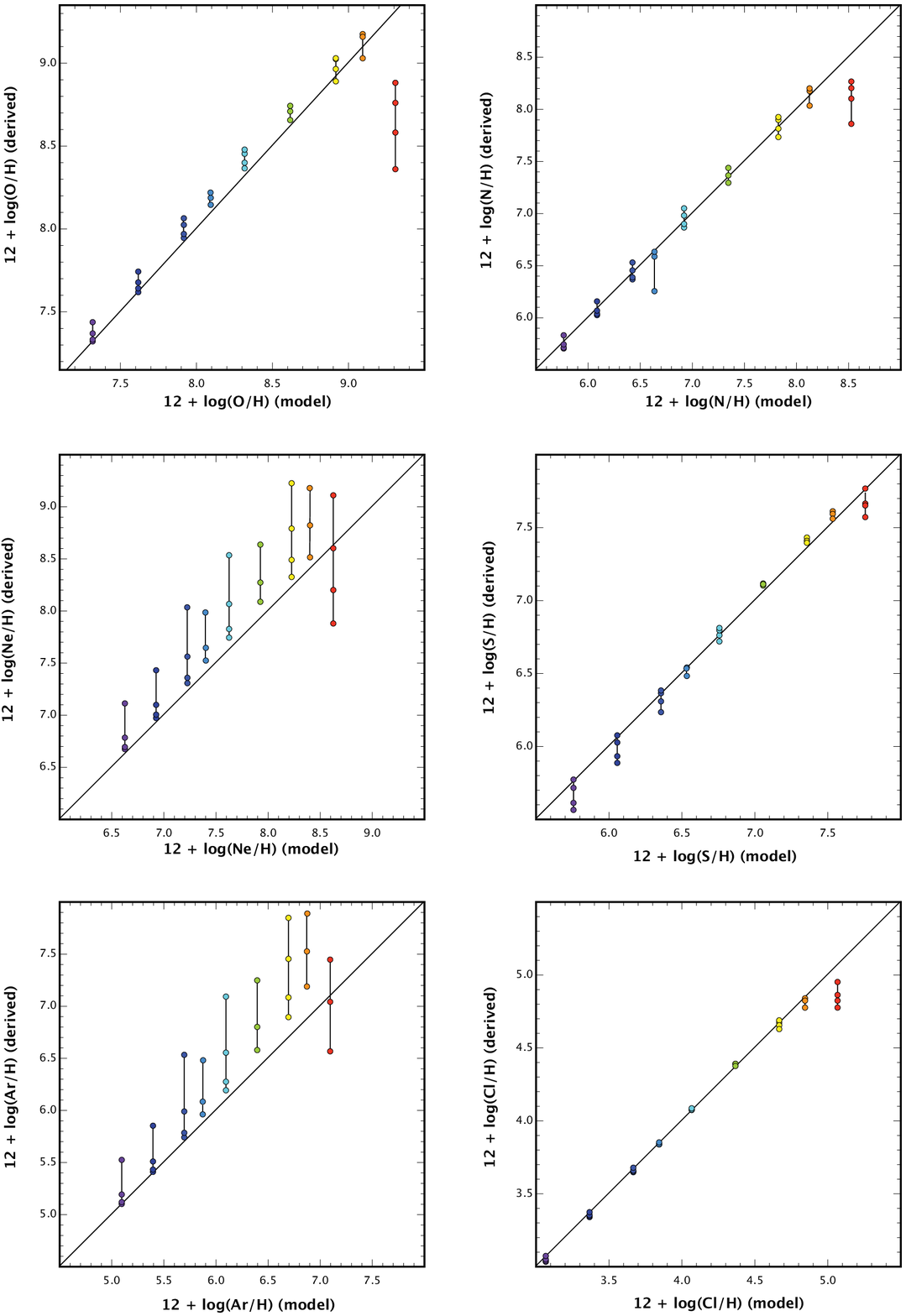}   
\caption{A comparison of the total abundances derived by using \TeOiii\ using the [\ion{O}{iii}] ($\lambda$4959+$\lambda$5007)/$\lambda$4363 ratio in the high-ionization zone and the \TeOii\ using the [\ion{O}{ii}]  ($\lambda$3726+$\lambda$3729)/($\lambda$7319+$\lambda$7330) ratio in the low-ionization zone (see Table~\ref{abundances_ox}) with the total abundances provided by the models. As in Fig.~\ref{Z_Te_high}, the fit is fairly good for the species for which the optical line emission arises principally in the low-ionization zone, N, S and Cl. Although it is slightly overpredicting the model abundances, the fit is now acceptable for O, except for 12+log(O/H) $= 9.39$ because of the aforementioned problem when calculating \TeOii\ for the low ionization ions. However, the fit still has a systematic error in the case of Ne (which is seen only in its high-ionization stage Ne$^{++}$ and Ar (for which the emission is also dominated by its high ionization lines, and which is strongly affected by charge-exchange reactions). Random errors are also larger for these species.}
\label{Z_Te_OIII}
\end{figure*}

\subsection{Ionic abundances}  \label{Section3.3}

The IRAF package NEBULAR  \citep{SD95} has been used to analyze the intensities of the forbidden lines using the temperatures and densities derived for the high- and low-ionization zones to derive ionic abundances of O$^+$,  O$^{++}$, N$^+$, S$^+$, S$^{++}$, Ne$^{++}$, Ar$^{++}$, Ar$^{+3}$, Cl$^{++}$, and Cl$^{+3}$. 
The ionic abundances 
so derived 
are listed in the Appendix in Table~\ref{ionic}.

\subsection{Total abundances}  \label{Section3.4}

In all cases we adopted O/H = O$^+$/H$^+$ +  O$^{++}$/H$^+$ to determine the total oxygen abundance. Although a very weak nebular \ion{He}{ii} $\lambda$4686 line is produced in several models, the relative contribution of He$^{++}$ to the  total amount of helium is negligible, implying that O$^{3+}$ has also a very low abundance in the nebula, thus we did not consider its contribution to the total O/H ratio.  For the rest of the elements, we have  to adopt a set of ionization correction factors (ICFs) to correct for the unseen ionization stages. The ICFs adopted here are basically the same used by \citet{LSE09}.

To derive the nitrogen abundance we assumed the standard ICF by \citet{PC69}: N/O = N$^+$/O$^+$, 
which is the typical assumption considered in the analysis of the ionized gas in extragalactic \HII\ regions
\citep[e.g.][]{LSE10b}. 

We measured two ionization stages of sulphur, S$^+$ and S$^{++}$, in all spectra. However,  a significant contribution of S$^{3+}$ is expected. We  adopted the ICF given by \citet{S78}, which is based on photoionization models of  \HII\ regions and is expressed as a function of the O$^+$/O ratio.

In the case of neon we also applied the classical ICF proposed by \citet{PC69}, that assumes that the ionization structure of Ne is similar to that of O. This is a  good approximation for high ionization objects, where a small fraction of Ne$^+$ is expected. 

For argon we have determinations of the Ar$^{++}$ and Ar$^{3+}$ abundances. However, some contribution of Ar$^{+}$ is expected.  Hence, the total argon abundance was calculated by considering the ICFs proposed by \citet*{ITL94}. 

We  measured lines Cl$^{++}$ and Cl$^{3+}$. As can be seen in Table~\ref{ionic}, the dominant ionization stage is Cl$^{++}$ and the contribution of Cl$^+$ to the total abundance is rather small. To take into account the Cl$^+$ fraction we have adopted the relation by \citet{PTP77}.

The results we obtain depend on whether we use all of the high-ionization species to determine $T_e$ in the high-ionization zone, $T_e$(high), and all the low-ionization species to compute $T_e$ in the low-ionization zone, $T_e$(low), or whether we use only the temperature derived from [\ion{O}{iii}] line ratio, \Te[\ion{O}{iii}]. The total abundances we obtain by these two approaches are shown in Tables~\ref{abundances} and \ref{abundances_ox}, and are shown in graphical form in Fig~\ref{Z_Te_high} and \ref{Z_Te_OIII}, respectively.

When all of the high ionization species are used to derive \Te(high), the oxygen abundance is systematically overestimated by \mbox{$\sim0.2$~dex} and the other species seen principally in their high-excitation stages, Ne and Ar, are overestimated by somewhat larger amounts. 
For 12+log(O/H)=9.39, the \Te\ method fails because IRAF task {\it ionic}  does not allow electron temperatures lower than 5,000\,K for the low ionization ions.
If we use only the temperature derived from the [\ion{O}{iii}]  lines, \TeOiii\ using the [\ion{O}{iii}] ($\lambda$4959+$\lambda$5007)/$\lambda$4363 ratio in the high-ionization zone, and the temperature derived from the [\ion{O}{ii}]  lines, \TeOii\ using the [\ion{O}{ii}] 
($\lambda$3726+$\lambda$3729)/($\lambda$7319+$\lambda$7330) ratio in the low-ionization zone, the systematic error is much less, specially for the total oxygen abundance, although it is still slightly overpredicted.

Although there are small systematic errors in this procedure, all are correctable. It is clear that the \Te\ method returns reliable abundances. Provided that the [\ion{O}{iii}] ($\lambda$4959+$\lambda$5007)/$\lambda$4363  ratio is used to determine the $T_e$ in the high-ionization zone, the scatter in the derived abundances of O, N, S and Cl is typically $\pm 0.1$~dex, provided that the O abundance is below 12+log(O/H) $< 9$.  For high abundances, the \Te\ method returns systematically low abundances, where it can be applied. This is in agreement with the work of \citet{Sta02} and \citet{Sta05}, who predicted that temperature gradients
in these high-metallicity \HII\ regions can cause the abundances to be underestimated by as much as $\sim0.4$~dex.

Both for Ne and Ar, the systemic error and the scatter are distressingly large.
We do not think that this is a consequence of using slightly different atomic data for models and line analysis using IRAF.
 However the cause is likely to be different in the two cases. In the case of Ne, the error is probably caused by the fact that only Ne$^{++}$ is observable, and the [\ion{Ne}{iii}]~$\lambda\lambda$3869, 3968 doublet has sufficiently high excitation energy that the line emissivity is strongly biased towards the hottest part of the \HII\ region. In the case of Ar, charge-exchange processes are very important in determining the ionic balance, and these strongly affect the ionization balance and therefore the ionization correction factors \citep{Dopita97}. 
Furthermore, the assumptions of the ICFs used to derived the total abundances of Ne and Ar from their ionic abundances may be not valid in some cases \citep[e.g.][]{Perez-Montero07}.

\begin{figure}
\centering
\includegraphics[angle=0,width=0.8\linewidth]{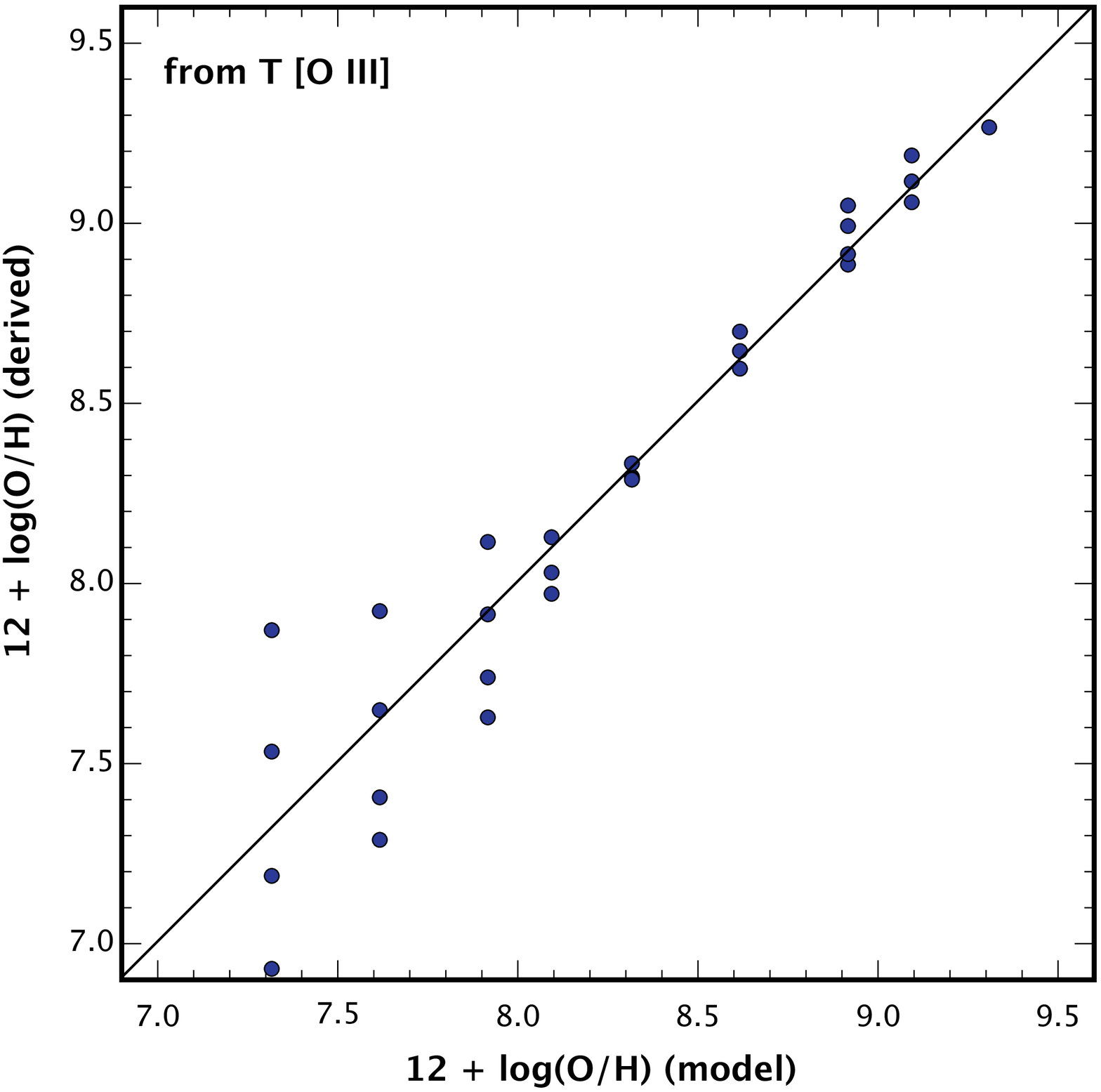}   
\caption{A comparison of the abundances derived by using only the \TeOiii\ measured from the [\ion{O}{iii}] ($\lambda$4959+$\lambda$5007)/$\lambda$4363 ratio.  For abundances 12+log(O/H) $> 8.0$ the abundances are recovered to an accuracy of $\pm0.1$ dex. At lower abundances, the scatter is much increased because of the increased sensitivity of  \TeOiii\   with the ionization parameter in this region.}
\label{Z_Z[OIII]}
\end{figure}


Given that Fig~\ref{Z_Te} shows that the electron temperature and the abundance are closely correlated, it is interesting to see just how well the oxygen abundance can be recovered from the measured [\ion{O}{iii}] temperature alone. 
To do this we used our empirical relation (Eq.~\ref{OIII_cor}) to correct the measured temperature and then applied the relationship between $T_e$[\ion{O}{iii}] and 12+log(O/H) given by Eq.~\ref{ZT_OIII} to obtain the oxygen abundance. The result is shown in Fig.~\ref{Z_Z[OIII]}. For abundances 12+log(O/H) $> 8.0$ the abundances are recovered to an accuracy of $\pm0.1$ dex. At lower abundances, significant scatter is introduced by variations in the ionization parameter $q$, and it becomes essential to determine this parameter independently.

\begin{table*} 
\caption{\label{empirical} Results for the oxygen abundance, in the form \abox, according to the most-commonly used empirical SEL calibrations.}
\scriptsize{
\begin{tabular}{ c@{\hspace{6pt}} c@{\hspace{6pt}}c@{\hspace{6pt}}c@{\hspace{6pt}}c@{\hspace{6pt}}c@{\hspace{6pt}} c@{\hspace{6pt}}c@{\hspace{6pt}} c@{\hspace{6pt}} c@{\hspace{4pt}}c@{\hspace{4pt}} c@{\hspace{4pt}}c@{\hspace{4pt}} c@{\hspace{4pt}} c@{\hspace{4pt}}c@{\hspace{4pt}}c@{\hspace{4pt}}  }

\hline

Model     &    Te  & BRANCH &    M91 &  KD02o &   KK04 &  KN2O2 &    P01 &   PT05 &  PVT10 &   PM11 &   D02  &  PP04a &  PP04b &  PP04c &    S23 &  S23O23  \\ 
 
 \hline
\noalign{\smallskip}

         A &   8.71 &   HIGH &   8.53 &   8.84 &   8.67 &   8.64 &   8.21 &   8.11 &   8.44 &   8.49 &   8.56 &   8.46 &   8.41 &   8.40 &   8.96 &   8.60   \\ 
         B &   7.62 &    LOW &   7.50 &   7.67 &   7.81 &  \nodata &   7.39 &   7.36 &   7.17 &   7.12 &   7.62 &   7.73 &   7.74 &   7.94 &   7.45 &   7.74   \\ 
         C &   8.88 &   HIGH &   9.13 &   9.31 &   9.35 &   9.24 &   8.86 &   8.43 &   9.11 &   8.94 &   8.72 &   8.59 &   8.58 &   9.25 &   8.53 &   9.65   \\ 
         D &   7.44 &    LOW &   7.23 &   7.57 &   7.53 &  \nodata &   7.34 &   7.01 &   6.73 &   6.31 &   7.71 &   7.80 &   7.84 &   8.28 &   7.28 &   8.31   \\ 
         E &   8.01 &    LOW &   7.96 &   7.85 &   8.10 &  \nodata &   8.77 &   6.94 &   7.18 &   6.74 &   8.21 &   8.19 &   8.18 &   8.51 &   8.17 &   8.50   \\ 
        Fl$^a$ &   8.40 &    LOW &   8.17 &   8.28 &   8.36 &   \nodata &   8.02 &   8.13 &   7.93 &   7.95 &   8.26 &   8.23 &   8.20 &   8.17 &   8.74 &   8.37   \\ 
        Fu$^a$ &   8.40 &   HIGH &   8.17 &   8.73 &   8.55 &   8.34 &   8.27 &   8.22 &   7.93 &   7.95 &   8.26 &   8.23 &   8.20 &   8.17 &   8.74 &   8.37   \\ 
         G &   9.03 &   HIGH &   9.07 &   9.22 &   9.16 &   9.09 &   8.84 &   8.70 &   8.73 &   8.76 &   8.54 &   8.45 &   8.39 &   8.69 &   8.46 &   9.30   \\ 
         H &   8.76 &   HIGH &   9.14 &   9.33 &   9.35 &   9.26 &   8.88 &   8.57 &   8.94 &   8.88 &   8.66 &   8.54 &   8.51 &   9.05 &   8.39 &   9.67   \\ 
         I &   8.89 &   HIGH &   8.95 &   9.02 &   9.06 &   8.92 &   8.76 &   8.69 &   8.65 &   8.67 &   8.44 &   8.37 &   8.31 &   8.45 &   8.55 &   8.94   \\ 
         J &   8.56 &   HIGH &   9.22 &   9.37 &   9.47 &   9.28 &   8.91 &   8.69 &   9.01 &   9.00 &   8.49 &   8.41 &   8.35 &   8.98 &   8.04 &   9.74   \\ 
         K &   7.74 &    LOW &   7.54 &   7.81 &   7.79 &   \nodata &   7.78 &   7.15 &   7.15 &   6.74 &   7.94 &   7.98 &   8.03 &   8.31 &   7.67 &   8.33   \\ 
         L &   8.24 &    LOW &   8.12 &   7.97 &   8.27 &   \nodata &   8.57 &   7.52 &   7.67 &   7.28 &   8.32 &   8.28 &   8.24 &   8.41 &   8.42 &   8.42   \\ 
         M &   7.68 &    LOW &   7.50 &   7.58 &   7.80 &  \nodata &   7.40 &   7.41 &   7.18 &   7.07 &   7.78 &   7.85 &   7.91 &   8.07 &   7.62 &   8.07   \\ 
         N &   8.65 &   HIGH &   8.61 &   8.77 &   8.74 &   8.63 &   8.41 &   8.36 &   8.52 &   8.55 &   8.38 &   8.32 &   8.27 &   8.26 &   8.74 &   8.51   \\ 
         O &   9.19 &   HIGH &   9.00 &   9.07 &   9.14 &   9.06 &   8.65 &   8.29 &   8.76 &   8.67 &   8.81 &   8.65 &   8.70 &   8.97 &   9.02 &   9.29   \\ 
         P &   7.37 &    LOW &   7.16 &   7.53 &   7.52 &   \nodata &   7.00 &   7.20 &   6.75 &   6.62 &   7.54 &   7.67 &   7.64 &   8.05 &   7.24 &   8.04   \\ 
         Q &   7.94 &    LOW &   7.91 &   7.98 &   8.15 &   \nodata &   7.73 &   7.74 &   7.53 &   7.52 &   7.86 &   7.92 &   7.98 &   7.98 &   7.90 &   7.86   \\ 
         R &   8.96 &   HIGH &   8.88 &   8.97 &   9.02 &   8.92 &   8.61 &   8.51 &   8.56 &   8.57 &   8.61 &   8.51 &   8.46 &   8.54 &   8.86 &   8.96   \\ 
         S &   8.19 &    LOW &   8.07 &   8.13 &   8.27 &   \nodata &   8.00 &   7.98 &   7.77 &   7.64 &   8.17 &   8.16 &   8.16 &   8.17 &   8.34 &   8.22   \\ 
         T &   8.14 &    LOW &   8.05 &   8.08 &   8.26 &   \nodata &   7.85 &   7.92 &   7.67 &   7.68 &   8.02 &   8.04 &   8.08 &   8.05 &   8.20 &   8.04   \\ 
        Ul$^a$ &   8.36 &    LOW &   8.13 &   8.22 &   8.32 &   \nodata &   7.90 &   8.02 &   7.88 &   7.89 &   8.13 &   8.12 &   8.13 &   8.08 &   8.40 &   8.13   \\ 
        Uu$^a$ &   8.36 &   HIGH &   8.13 &   8.58 &   8.54 &   8.33 &   8.34 &   8.29 &   7.88 &   7.89 &   8.13 &   8.12 &   8.13 &   8.08 &   8.40 &   8.13   \\ 
         V &   7.64 &    LOW &   7.50 &   7.65 &   7.81 &   \nodata &   7.38 &   7.36 &   7.16 &   7.11 &   7.65 &   7.75 &   7.78 &   7.97 &   7.51 &   7.83   \\ 
        Wl$^a$ &   8.45 &    LOW &   8.27 &   8.29 &   8.41 &   \nodata &   8.45 &   8.04 &   7.99 &   8.00 &   8.43 &   8.36 &   8.31 &   8.35 &   8.72 &   8.42   \\ 
        Wu$^a$ &   8.45 &   HIGH &   8.27 &   8.78 &   8.59 &   8.38 &   8.11 &   7.99 &   7.99 &   8.00 &   8.43 &   8.36 &   8.31 &   8.35 &   8.72 &   8.42   \\ 
         X &   8.30 &   HIGH &   9.32 &   9.40 &   \nodata &   9.28 &   8.92 &   8.62 &   9.12 &   9.18 &   8.36 &   8.31 &   8.26 &   9.05 &   7.71 &   9.76   \\ 
         Y &   7.32 &    LOW &   7.15 &   7.47 &   7.52 &   \nodata &   7.01 &   7.13 &   6.76 &   6.68 &   7.39 &   7.55 &   7.38 &   7.92 &   7.12 &   7.70   \\ 
         Z &   9.03 &   HIGH &   8.76 &   9.01 &   8.98 &   8.92 &   8.36 &   7.98 &   8.73 &   8.61 &   8.82 &   8.66 &   8.72 &   8.95 &   9.29 &   9.08   \\ 
        AA &   9.17 &   HIGH &   9.03 &   9.17 &   9.14 &   9.07 &   8.73 &   8.52 &   8.66 &   8.65 &   8.71 &   8.58 &   8.57 &   8.77 &   8.81 &   9.28   \\ 
        AB &   8.02 &    LOW &   7.90 &   7.84 &   8.12 &   \nodata &   7.96 &   7.69 &   7.61 &   7.36 &   8.11 &   8.11 &   8.13 &   8.22 &   8.12 &   8.25   \\ 
        AC &   9.02 &   HIGH &   8.79 &   9.02 &   8.97 &   8.92 &   8.42 &   8.18 &   8.53 &   8.53 &   8.77 &   8.63 &   8.65 &   8.70 &   9.12 &   9.00   \\ 
        AD &   7.33 &    LOW &   7.14 &   7.46 &   7.52 &   \nodata &   6.99 &   7.14 &   6.75 &   6.66 &   7.42 &   7.57 &   7.43 &   7.94 &   7.15 &   7.79   \\ 
        AE &   7.97 &    LOW &   7.90 &   7.89 &   8.14 &   \nodata &   7.75 &   7.76 &   7.55 &   7.51 &   7.94 &   7.98 &   8.03 &   8.05 &   8.02 &   8.02   \\ 
       AFl$^a$ &   8.48 &    LOW &   8.38 &   7.85 &   8.44 &   \nodata &   9.43 &   7.29 &   7.80 &   7.93 &   8.51 &   8.42 &   8.37 &   8.62 &   8.84 &   8.61   \\ 
       AFu$^a$ &   8.48 &   HIGH &   8.38 &   8.80 &   8.72 &   8.47 &   8.07 &   7.75 &   7.80 &   7.93 &   8.51 &   8.42 &   8.37 &   8.62 &   8.84 &   8.61   \\ 
        AG &   8.74 &   HIGH &   8.45 &   8.76 &   8.65 &   8.67 &   8.03 &   7.74 &   8.48 &   8.50 &   8.69 &   8.56 &   8.54 &   8.62 &   9.15 &   8.71   \\ 
 
\hline
\end{tabular}
}
\flushleft

$^a$ For these models, for which we derived \abox$\sim$8.4 following the \Te\ method, we list the results of the empirical calibrations considering both the low and the high metallicity branches.\\
 NOTE: The empirical calibrations and the parameters used for each of them are:   \\
 M91: \citet{McGaugh91} using $R_{23}$ and $y$; \\
 KD02: Kewley \& Dopita (2002) using $R_{23}$ and the ionization parameter defined in that paper, $q_{KD02o}$; \\
 KK04: Kobulnicky \& Kewley (2004) using $R_{23}$ and the ionization parameter defined in that paper, $q_{KK04}$; \\
 KN2O2: Kewley \& Dopita (2002) using the $N_2O_2$ parameter (calibration only valid for objects in the high metallicity branch);\\
 
 P01: Pilyugin (2001a,b) using $R_{23}$ and $P$; \\
 PT05: \citet{PT05} using $R_{23}$ and $P$; \\
 PVT10: \citet{PVT10} using $R_{23}$, $P$, $N_2$ and $S_2$; \\
 PM11: \citet{PM11} using $R_3$, $N_2$ and $S_2$; \\
 
 D02: Denicol\'o, Terlevich \& Terlevich (2002) using the $N_2$ parameter; \\
 PP04a: Pettini \& Pagel (2004), using a linear fit to the $N_2$ parameter; \\
 PP04b: Pettini \& Pagel (2004), using a cubic fit to the $N_2$ parameter; \\
 PP04c: Pettini \& Pagel (2004), using the $O_3N_2$ parameter;\\

 S23: \citet{Perez-MonteroDiaz05} using $S_{23}$;\\
 S23O23: \citet{Perez-MonteroDiaz05} using $S_{23}$ and $R_{23}$.

\end{table*}

\section{Application of the SEL Method}  \label{Section4}

The strong emission line (SEL) technique to derive nebular abundances is typically used when the spectra are of too low signal-to-noise ratio for the temperature sensitive emission lines to be detectable. This typically encompasses the majority of global determinations of metallicity in external galaxies, and all galaxies with redshifts large enough to be of cosmological interest.
An extensive review of 10 metallicity calibrations, including theoretical and empirical methods, has been presented by \citet{KE08}, using data gathered from the SDSS.  
\citet{LSE10b} also review the most common empirical calibrations and compare their results with those derived  using the \Te\ method.
The majority of the  
empirical calibrations rely on ratios between bright emission lines to estimate the oxygen abundance. The most commonly used of these are known by their shorthand contractions; $R_{23}$,  $S_{23}$,  $P$, $N_2$, $O_3N_2$,  $N_2O_2$ and $y$. Their definitions are:

\begin{eqnarray}
R_3=  \frac{I([\textsc{O\,iii}]) \lambda 4959+I([\textsc{O\,iii}]) \lambda 5007}{\rm H\beta},
\end{eqnarray}   
\begin{eqnarray}
R_2=  \frac{I([\textsc{O\,ii}]) \lambda\lambda 3726, 3729}{\rm H\beta},
\end{eqnarray}   
\begin{eqnarray}
R_{23} =  R_3 + R_2, 
\end{eqnarray}   
\begin{eqnarray}
P =  \frac{R_3}{R_{23}}, 
\end{eqnarray}   
\begin{eqnarray}
y =  \log \frac{R_3}{R_2} = \log \frac{1}{P^{-1}-1},
\end{eqnarray}   
\begin{eqnarray}
N_2 =  \frac{I([\textsc{N\,ii}]) \lambda 6584}{\rm H\alpha},
\end{eqnarray}   
\begin{eqnarray}
O_3N_2 = \frac{I([\textsc{O\,iii}]) \lambda 5007}{I([\textsc{N\,ii}]) \lambda 6584}, 
\end{eqnarray}   
\begin{eqnarray}
N_2O_2 = \frac{I([\textsc{N\,ii}]) \lambda 6584}{I([\textsc{O\,ii}]) \lambda\lambda 3726, 3729}, 
\end{eqnarray}   
\begin{eqnarray}
S_{23} =  \frac{I([\textsc{S\,ii}]) \lambda\lambda 6717, 6731+I([\textsc{S\,iii}]) \lambda\lambda 9076, 9532}{\rm H\beta},
\end{eqnarray}   
\begin{eqnarray}
S_{23}O_{23} =  \frac{S_{23}}{R_{23}}.
\end{eqnarray}   
\smallskip

Simple line ratios such as $R_{23}$ \citep{JSS76,Pagel79},   $S_{23}$ \citep{Vilchez96,Christensen97,Diaz00} --or indeed any forbidden line to recombination line ratio-- suffer from being two-valued, and hence a separate calibration for the low  
and high  
metallicity regimes is usually needed.
In the case of the $R_{23}$ index, the calibrations are given for \mbox{\abox$\lesssim$8.1} (low metallicity) and  \abox$\gtrsim$8.4 (high metallicity).
That means that a very large fraction of
the star-forming regions lie in the ill-defined turning zone around
\abox$\sim$8.20--8.30, where regions with the same $R_{23}$ value
have oxygen abundances that differ by almost an order of magnitude
\citep[see, for example, Figure~A.1 in][]{LSE10b}.
The reason of this behavior is that the intensity of the oxygen (or any other heavy element) lines do not 
monotonically increase with metallicity.
At low abundance, the lines of the heavy element being considered are weak relative to a hydrogen recombination line due to their low abundance. However, at high abundance they are weak due to the low electron temperature, which suppresses collisional excitations into the excited state responsible for the forbidden line. The turn-around in the line ratio can be extended to higher abundance by choosing lines with lower excitation potentials which make the ratio less temperature-sensitive. Hence the utility of $S_{23}$ over $R_{23}$. The use of multiple line ratios also helps to remove the abundance ambiguities.

Ratios such as  $O_3N_2$\citep{Alloin79,PP04} were introduced in an attempt to sidestep the abundance ambiguity entirely. This ratio is much more monotonic in abundance because of the strong secondary component of N enrichment alluded to above; see Eq.~\ref{calN_Z}. This helps at high abundance, but the ratio once again becomes ambiguous when 12 + log(O/H) $\lesssim 8.0$. A ratio such as $N_2O_2$ \citep{Do00,KD02} also provides a useful calibration for metallicity at high abundance. Its advantage is that both N$^{+}$ and  O$^{+}$ co-exist in the same zone of the nebula, the  [\ion{O}{iii}] lines become weaker at high metallicity due to the low electron temperature, while the  [\ion{N}{ii}]  lines become stronger due to the higher relative abundance of this element. However, its disadvantage is that the nitrogen and oxygen emission lines are widely separated in wavelength, making flux calibration and reddening corrections more critical.
 
The fundamental weakness in the use of all these strong-line ratios is that the overall spectrum of an \HII\ region depends not only upon the chemical abundances, but also upon the ionization parameter $q$ or $U$.
That this is so is abundantly obvious from Fig.~1. The importance of this was recognized by \citet*{BPT81}, who were amongst the first to use line ratio diagnostic diagrams.
However, the first empirical calibration involving an ionization parameter was presented by \citet{McGaugh91}, who developed models of \HII\ regions using the  photoionization code \textsc{Cloudy}.  \citet{McGaugh91} introduced the $y$ parameter to derive (together with the  $R_{23}$ ratio) the oxygen abundance using only the bright oxygen lines. The analytical expressions for these models were given by \citet{Kobulnicky99}.

It is clear from Fig.~1c, and as emphasized by  \citet{Do00} and \citet{KD02}, that the [\ion{O}{iii}]/[\ion{O}{ii}] ratio is probably the best one to use to determine the ionization parameter, particularly if the  [\ion{N}{ii}]/[\ion{O}{ii}]  ratio is also measured. 
To a lesser extent $R_3$ can also be used. \citet{KD02} provided a procedure to allow a simultaneous solution of both $\log q$ and 12 + log(O/H), 
which involves the $y$ and $R_{23}$ parameters, although they also give a parametrization of the oxygen abundance and $\log q$ using other parameters such as [\ion{S}{iii}]/[\ion{S}{ii}] (which is also a good estimator of the ionization parameter), $N_2O_2$, $N_2$, $S_{23}$ and $O_3N_2$. 
Later,  \citet{KK04} provided a parametrization of the \citet{KD02} method using  the $y$ and $R_{23}$ parameters with a form similar to that given by \citet{McGaugh91} calibration. Following this method, $\log q$ and the oxygen abundance are computed iteratively using only the bright oxygen lines.

\begin{table} 
\center
\caption{\label{empiricalfits}  Results of the comparison between the oxygen abundance given by several SEL methods and the oxygen abundance assumed by the theoretical models. We indicate the parameters, $a$ and $b$, of the linear fit $x_{SEL}=a+bx_{model}$, the correlation coefficient, $r$, of this linear fit, the dispersion of the data, $\sigma$, and the average value of the difference between the abundance given by the SEL method and that assumed by the model. In the first row, we also include the comparison with the oxygen abundance derived following the \Te\ method. The description of the SEL methods is the same than that used in Table~\ref{empirical}.}
\scriptsize{
\begin{tabular}{ c c c c c c c c c c c c c  c }   
\hline

Method &   $a$    &    $b$    & $r$   & $\sigma$ &  offset \\ 
 
 \hline
\noalign{\smallskip}

    Te &  -0.02 $\pm$ 0.14  &   1.001 $\pm$ 0.017    &    0.9959  &   0.05  &  -0.01 \\
 \noalign{\smallskip}   
   M91 &  -0.20 $\pm$ 0.17  &   1.005 $\pm$ 0.020    &    0.9940  &   0.07  &  -0.15 \\
  KD02 &   0.16 $\pm$ 0.21  &   0.981 $\pm$ 0.025    &    0.9897  &   0.10  &   0.00 \\
 KN2O2 &   0.94 $\pm$ 0.18  &   0.887 $\pm$ 0.020    &    0.9960  &   0.05  &  -0.07 \\
  KK04 &   0.77 $\pm$ 0.10  &   0.916 $\pm$ 0.012    &    0.9975  &   0.07  &   0.07 \\
    KD &   0.61 $\pm$ 0.16  &   0.922 $\pm$ 0.019    &    0.9932  &   0.09  &  -0.05 \\
    \noalign{\smallskip}
   P01 &   1.30 $\pm$ 0.64  &   0.814 $\pm$ 0.075    &    0.8890  &   0.31  &  -0.25 \\
  PT05 &   1.58 $\pm$ 0.51  &   0.752 $\pm$ 0.061    &    0.9113  &   0.28  &  -0.50 \\
 PVT10 &  -1.58 $\pm$ 0.24  &   1.134 $\pm$ 0.029    &    0.9901  &   0.14  &  -0.44 \\
  PM11 &  -2.46 $\pm$ 0.44  &   1.229 $\pm$ 0.052    &    0.9736  &   0.25  &  -0.52 \\
  \noalign{\smallskip}
   D02 &   2.11 $\pm$ 0.45  &   0.736 $\pm$ 0.055    &    0.9390  &   0.20  &  -0.04 \\
 PP04a &   3.43 $\pm$ 0.35  &   0.575 $\pm$ 0.043    &    0.9392  &   0.25  &  -0.04 \\
 PP04b &   3.39 $\pm$ 0.43  &   0.578 $\pm$ 0.052    &    0.9141  &   0.26  &  -0.05 \\
 PP04c &   1.81 $\pm$ 1.04  &   0.767 $\pm$ 0.116    &    0.8548  &   0.19  &  -0.25 \\
 \noalign{\smallskip}
   S23 &  -1.47 $\pm$ 0.66  &   1.186 $\pm$ 0.081    &    0.9481  &   0.24  &   0.04 \\
S23O23 &  -1.76 $\pm$ 0.67  &   1.208 $\pm$ 0.076    &    0.9641  &   0.17  &   0.07 \\
\hline
\end{tabular}
}

\end{table}

\begin{figure*}
\centering
\medskip
\begin{tabular}{l@{\hspace{4pt}}  l@{\hspace{4pt}}  l@{\hspace{4pt}}  l}
\includegraphics[angle=90,width=0.24\linewidth]{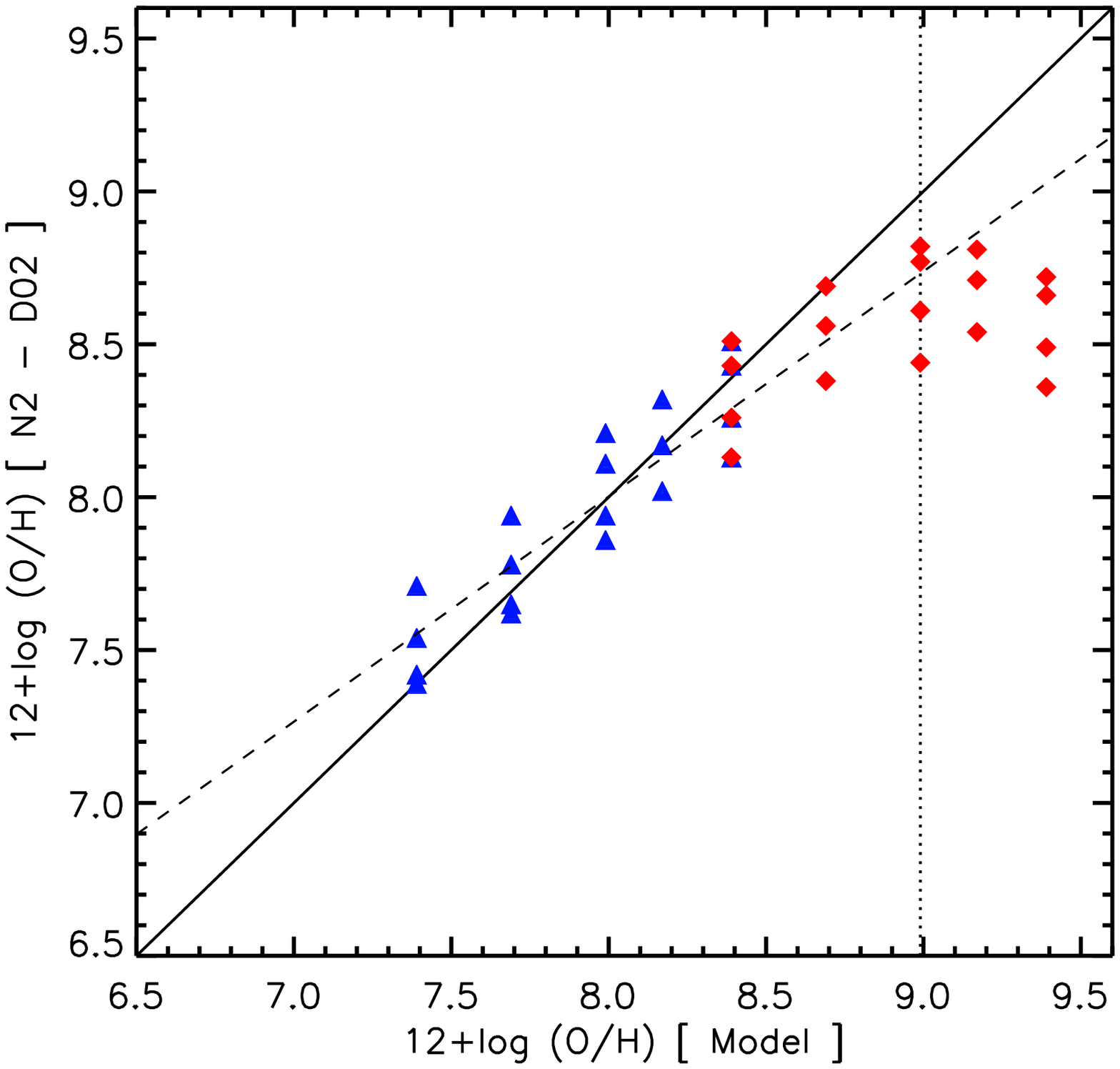} & 
\includegraphics[angle=90,width=0.24\linewidth]{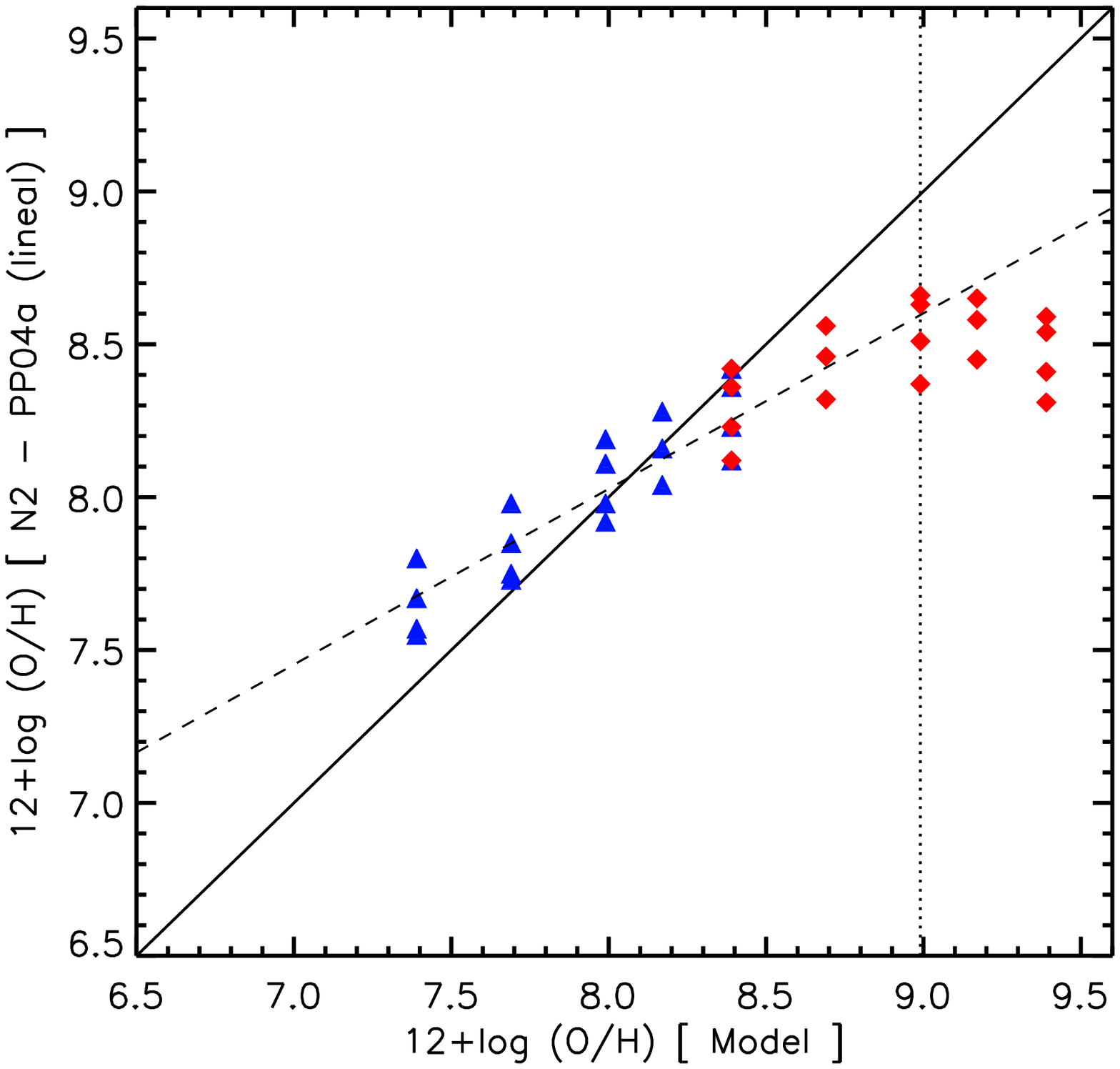} & 
\includegraphics[angle=90,width=0.24\linewidth]{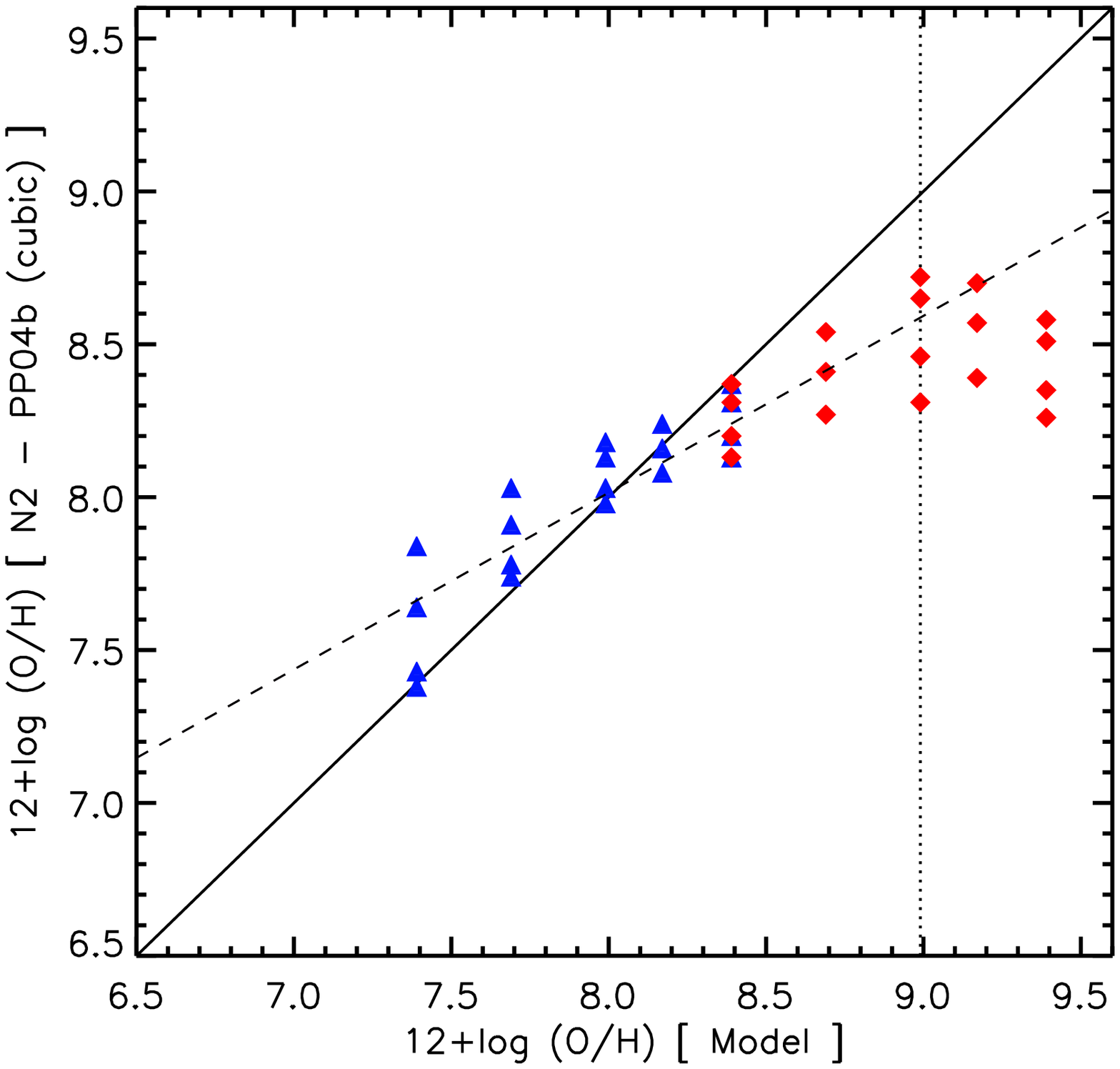} & 
\includegraphics[angle=90,width=0.24\linewidth]{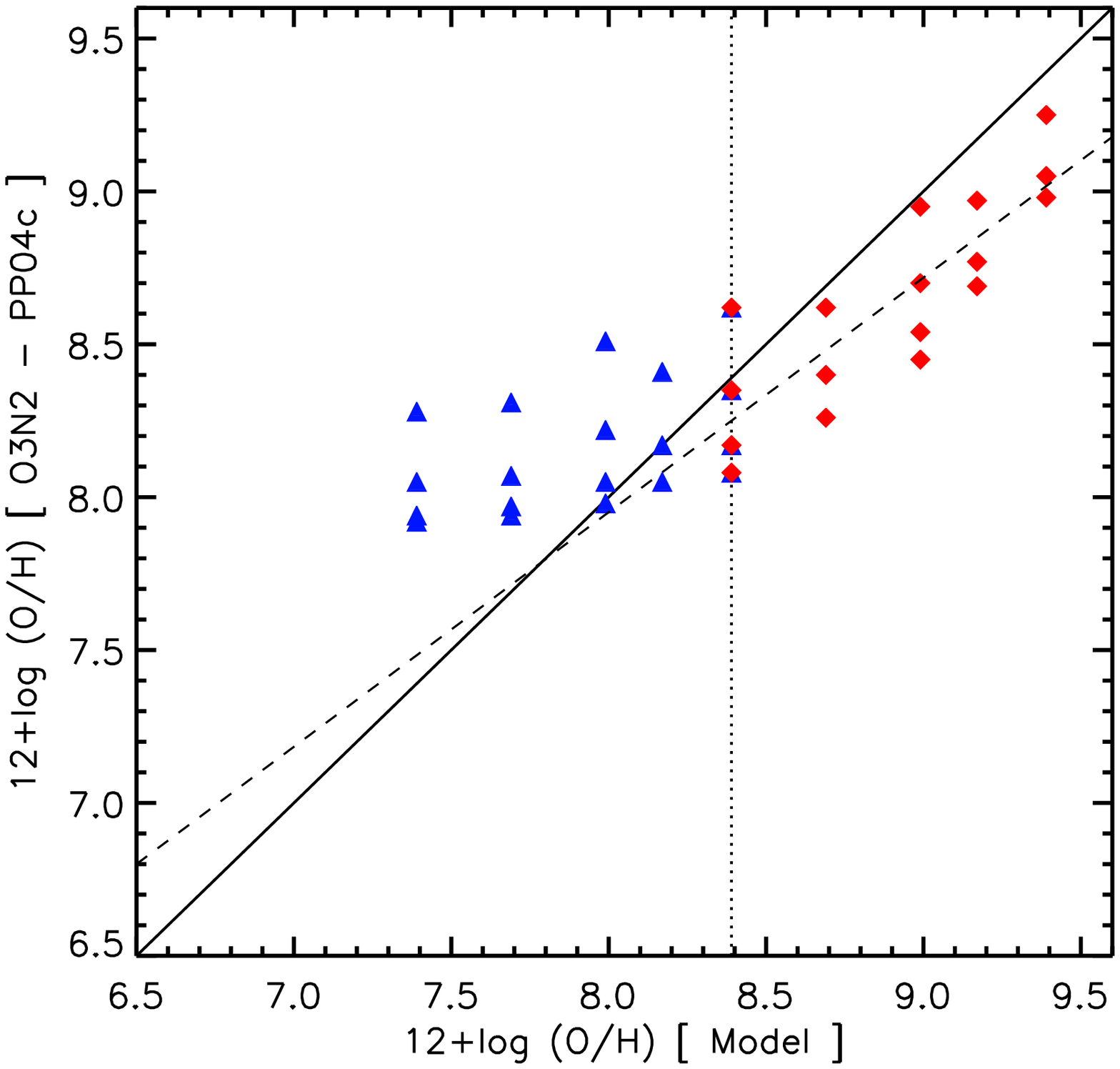} \\ 
\end{tabular}
\caption{Comparison between the total abundances used for the models (x-axis) and those computed using the empirical calibrations considering the $N_2$ and the $O_3N_2$ parameters from \citet*{D02} and \citet{PP04}. The continuous line represents $y=x$.  Red diamonds and blue triangles represent high- and low-metallicity models, respectively. 
The vertical dotted line indicates the lower (if using the $N_2$ parameter) or upper (when using the $O_3N_2$ parameter) limit of validity of the fits. The dashed line indicates the fit to the data within the appropriate metallicity range.}
\label{Z_PP04}
\end{figure*}

Pilyugin and his co workers 
have attempted an empirical calibration of the $R_{23}$  and $P$ parameters using the bright  [\ion{O}{iii}]~$\lambda\lambda$4959,5007 and [\ion{O}{ii}]~$\lambda\lambda$3726,3729 nebular lines. To get this calibration, they used a combination of photoionization models and empirical data from observed \HII\ regions where the auroral [\ion{O}{iii}]~$\lambda$4363 line was available.
Indeed, the so-called Pilyugin parameter, $P$, was introduced by \citet{P00}, after confirming that the $R_{23}$ parameter has a systematic error depending on the hardness of the ionizing radiation. Hence, the excitation parameter $P$ also considers the ionization degree of the \HII\ region.  
His first empirical calibrations involving both the $R_{23}$ and the $P$ parameter were presented in \cite{P01a} and \citet{P01b} for high- and low-metallicity \HII\ regions.  Subsequently, \citet{PT05} and \citet*{PVT10} tried to improved such calibrations including more spectroscopic measurements of \HII\ regions in spiral and irregular galaxies derived using the \Te-method. Finally, \citet{PM11} provided another empirical calibration considering only the $R_3$, $N_2$ and $S_2$ parameters (the so-called NS calibration), which was derived to estimate oxygen abundances in galaxies whose spectrum is lacking of the [\ion{O}{ii}] $\lambda\lambda$3726,3729 emission lines.

\subsection{Comparison of the SEL Techniques}

\begin{figure}
\centering
\begin{tabular}{l@{\hspace{4pt}} l}
\includegraphics[angle=90,width=0.48\linewidth]{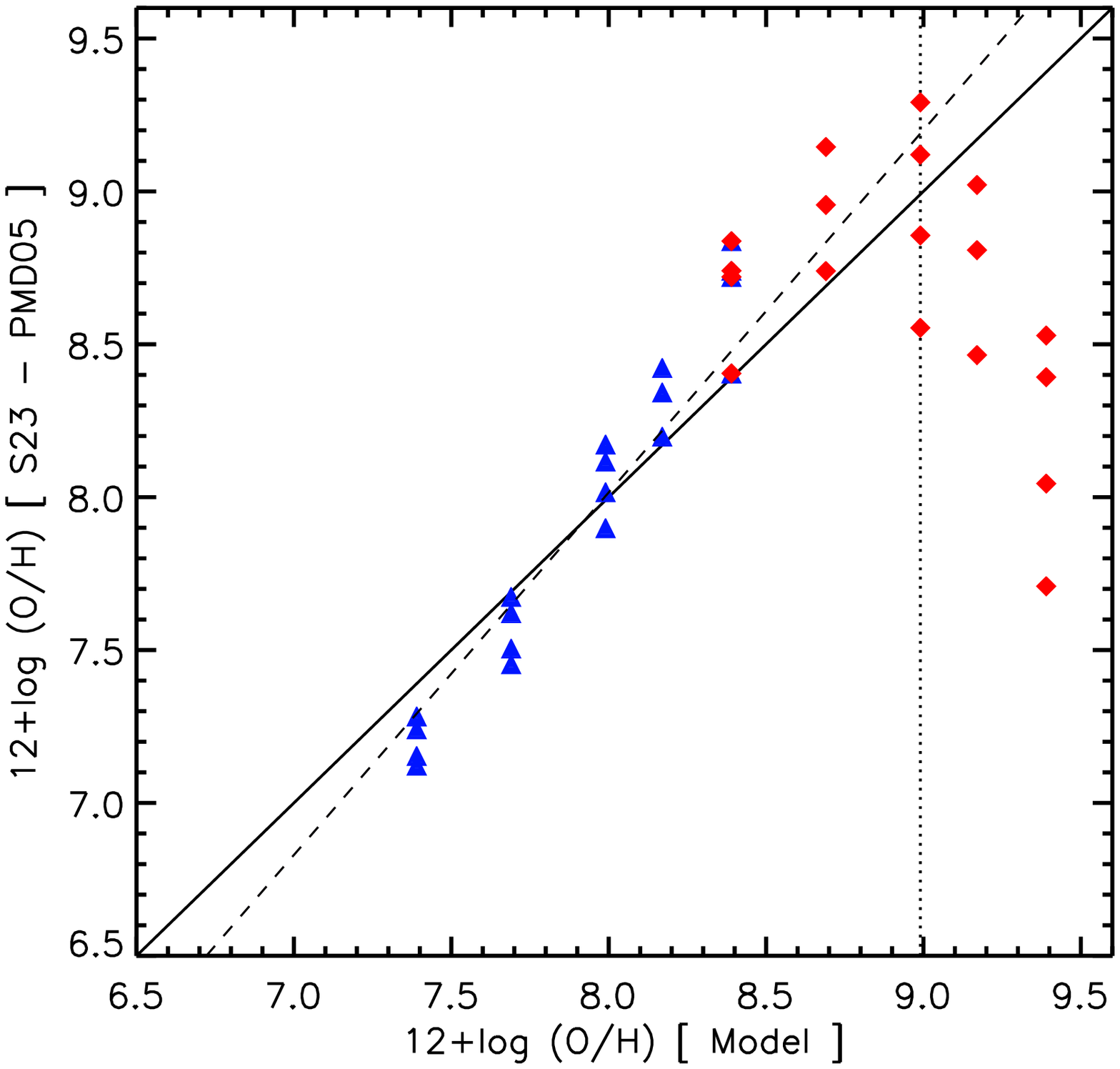} & 
\includegraphics[angle=90,width=0.48\linewidth]{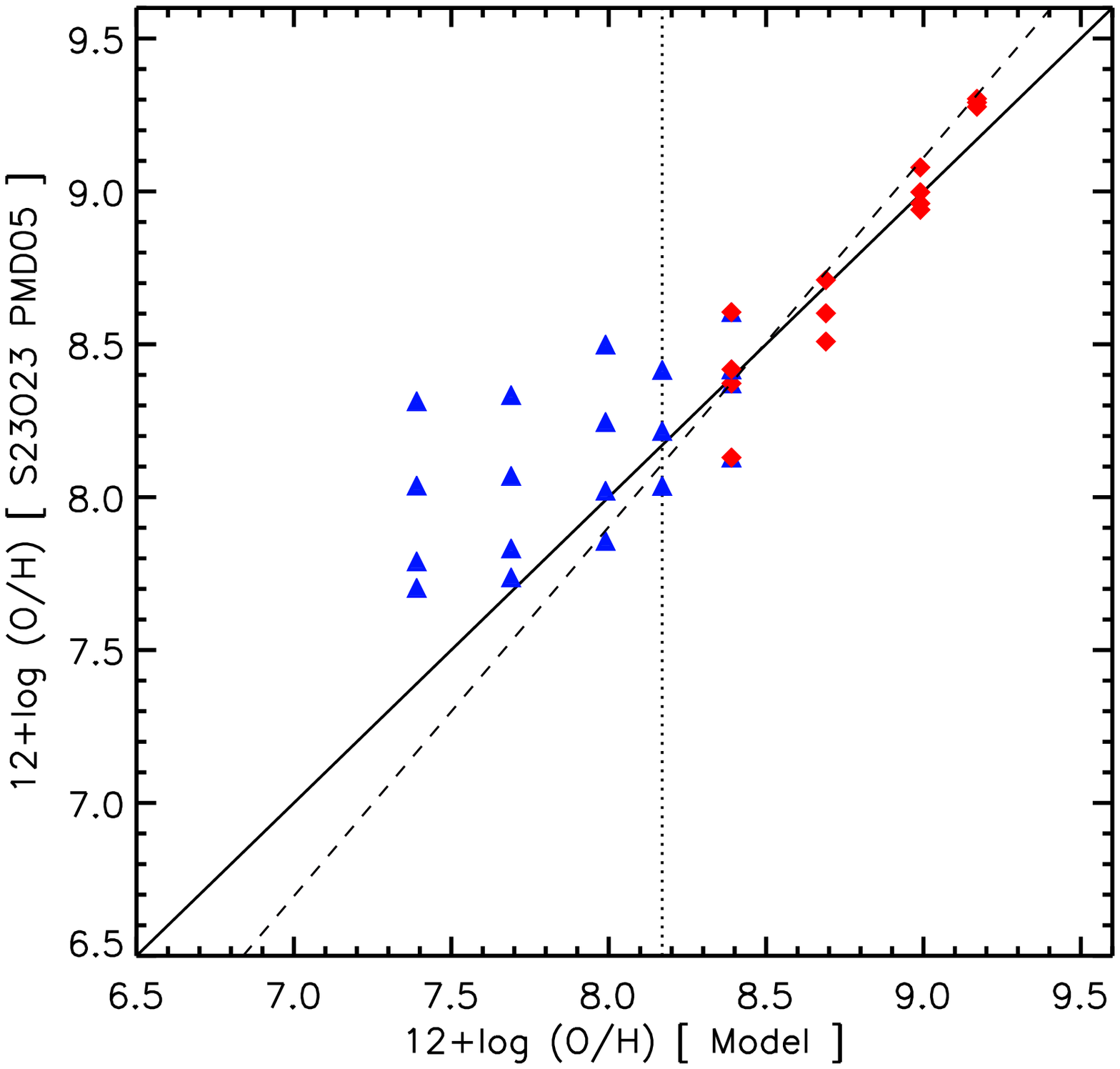} \\ 
\end{tabular}
\caption{Comparison between the total abundances used for the models (x-axis) and those computed using the empirical calibrations considering the $S_{23}$ and the $S_{23}O_{23}$ parameters from \citet{Perez-MonteroDiaz05}. The continuous line represents $y=x$. Red diamonds and blue triangles represent high- and low-metallicity models, respectively. 
The vertical dotted line indicates the upper (using the $S_{23}$ parameter) or lower (using the $S_{23}O_{23}$ parameter) limit of validity of the fits. The dashed line indicates the fit to the data within the appropriate metallicity range.}
\label{Z_S}
\end{figure}

\begin{figure*}
\centering
\medskip
\begin{tabular}{l@{\hspace{4pt}}  l@{\hspace{4pt}}  l@{\hspace{4pt}}  l}
\includegraphics[angle=90,width=0.24\linewidth]{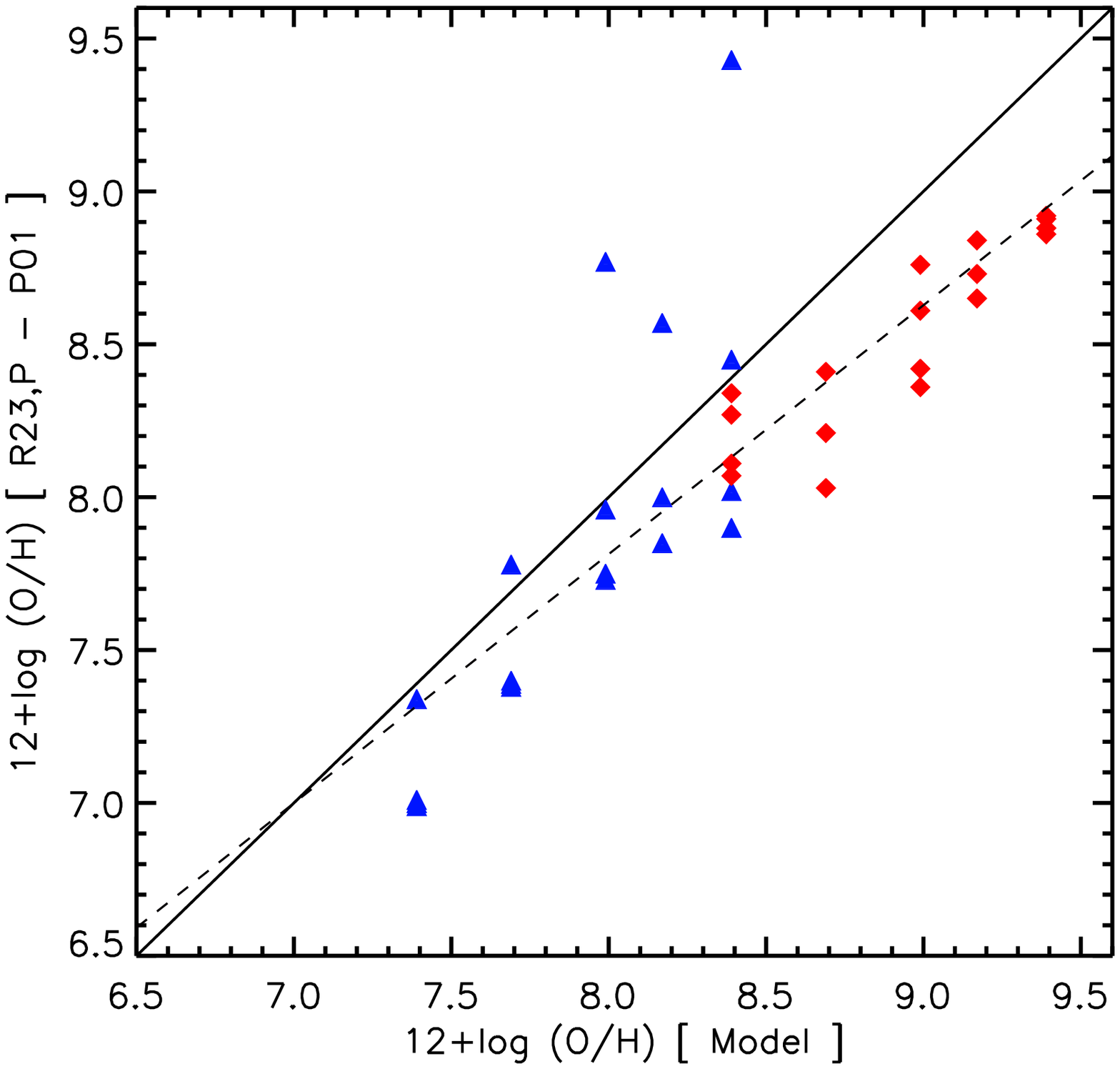} & 
\includegraphics[angle=90,width=0.24\linewidth]{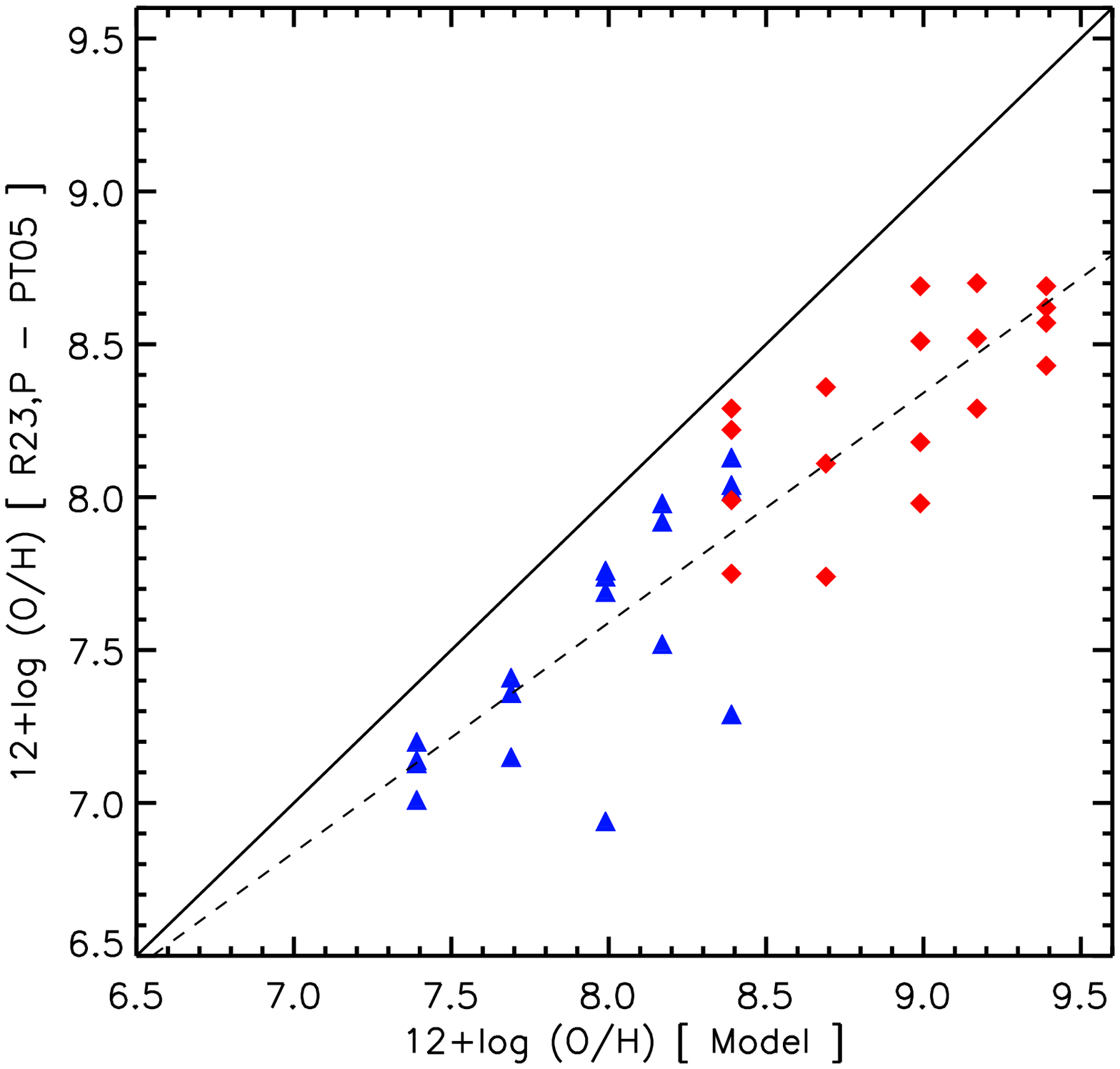} & 
\includegraphics[angle=90,width=0.24\linewidth]{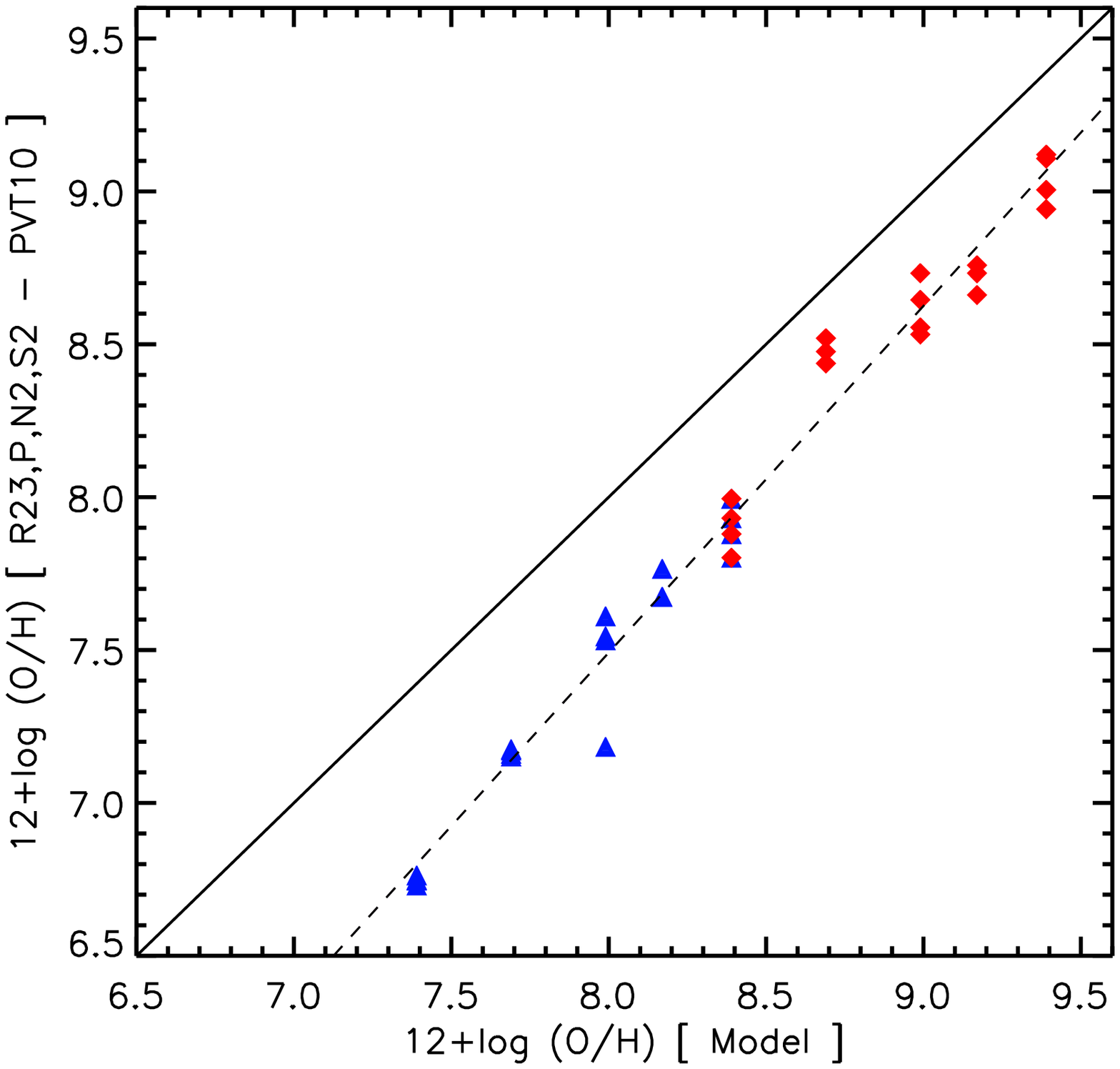} & 
\includegraphics[angle=90,width=0.24\linewidth]{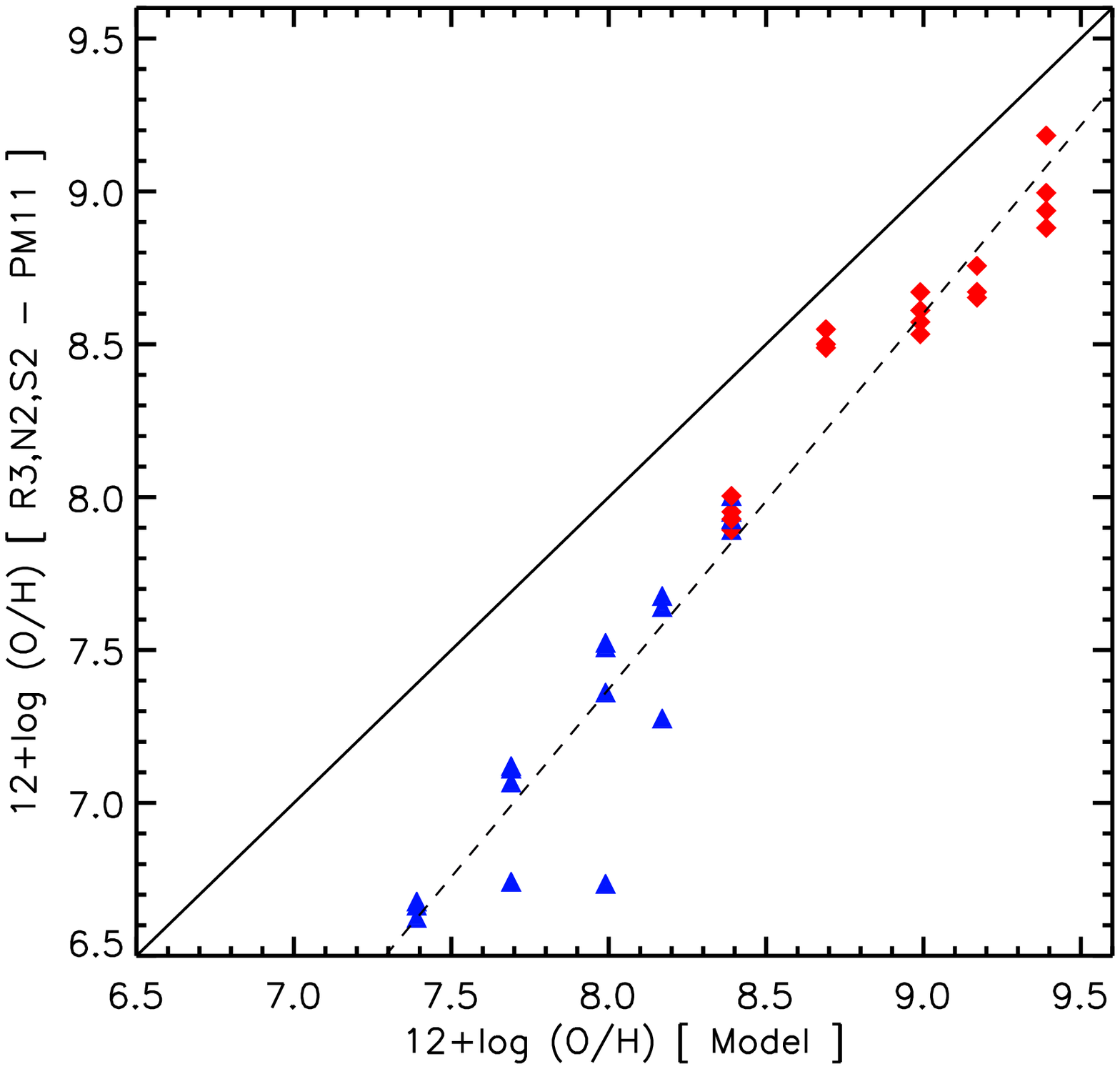} \\ 
\end{tabular}
\caption{Comparison between the abundances used for the models (x-axis) and  those computed via the Pilyugin method using the $R_{23}$ and $P$ parameters. From left to right, panels show the results for the empirical calibrations provided by  \citet{P01a,P01b}, \citet{PT05}, \citet{PVT10} and \citet{PM11}. The continuous line represents $y=x$. Red diamonds and blue triangles represent high- and low-metallicity models, respectively. The dashed line indicates the fit to the data.
Note that, in the left panel, the three data points with SEL abundances very much larger to those of the models are in the intermediate-metallicity regime and also have a relatively large $\log q$. Hence, following the \citet{P01a,P01b} method the high-metallicity calibration should be considered and averaged to the value derived using the low-metallicity calibration to get a more appropriate result of the oxygen abundance.}
\label{Z_Pil}
\end{figure*}

\begin{figure*}
\begin{tabular}{l@{\hspace{4pt}}  l@{\hspace{4pt}}  l@{\hspace{4pt}}  l}
\includegraphics[angle=90,width=0.24\linewidth]{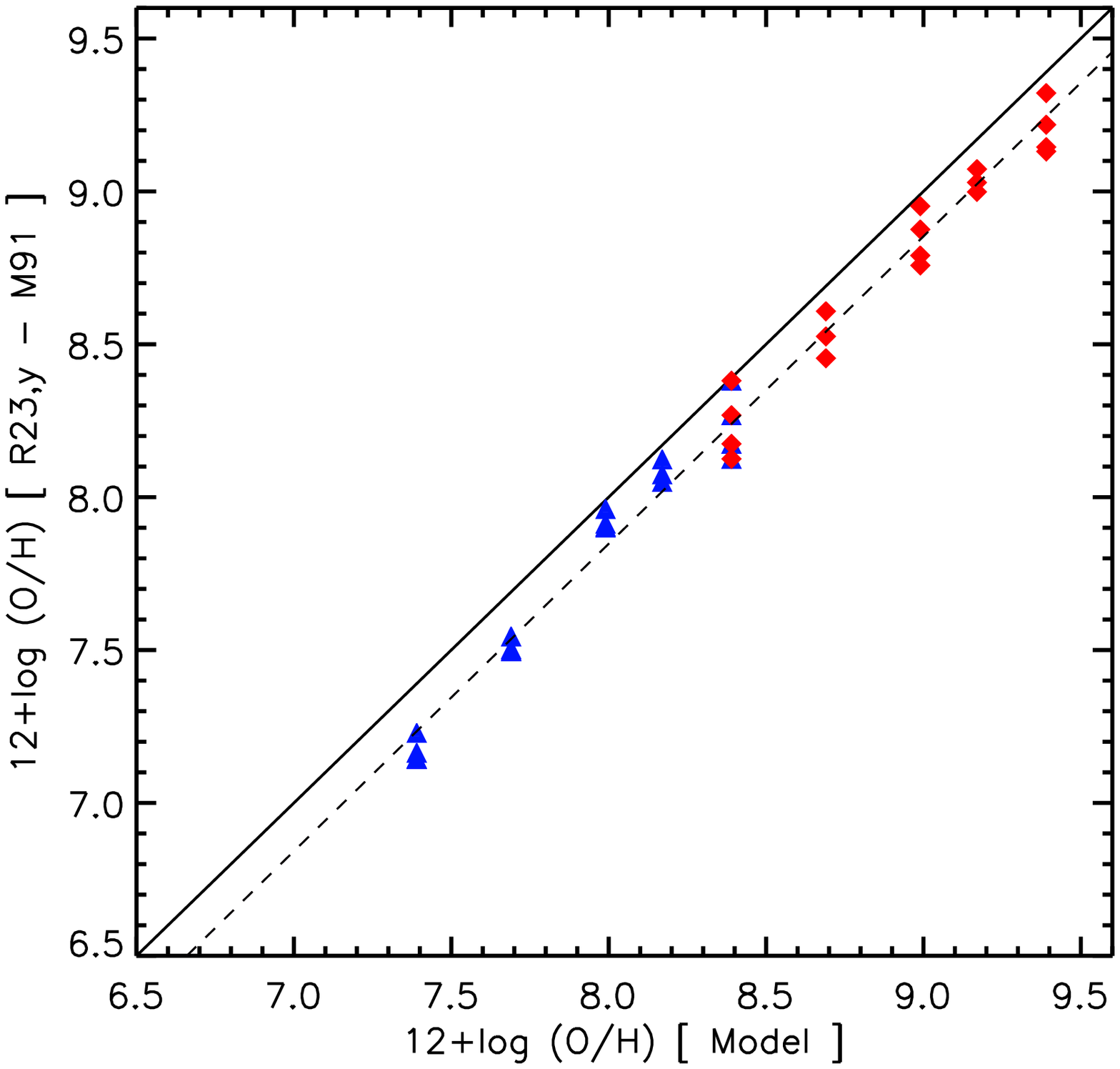} & 
\includegraphics[angle=90,width=0.24\linewidth]{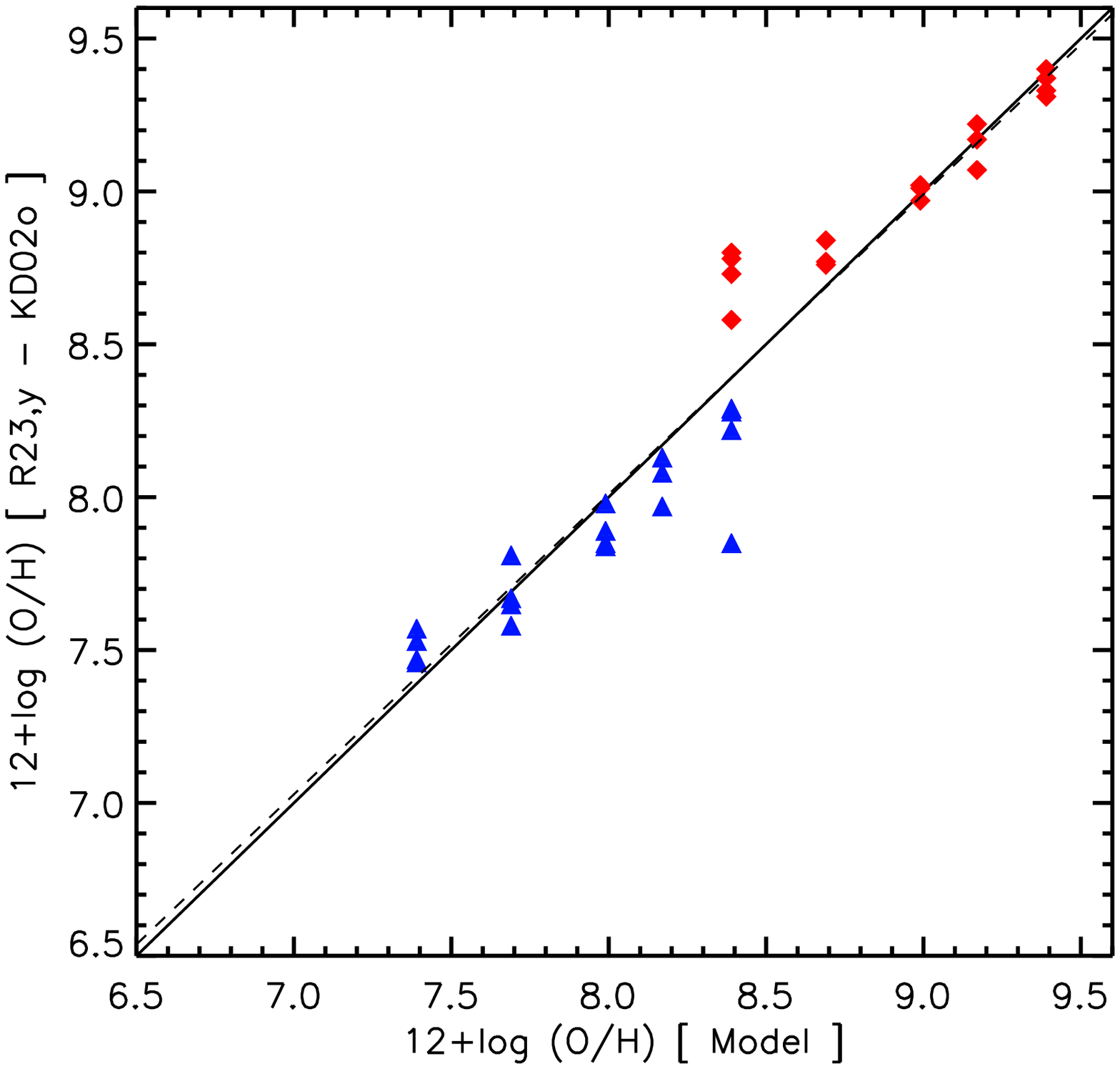} & 
\includegraphics[angle=90,width=0.24\linewidth]{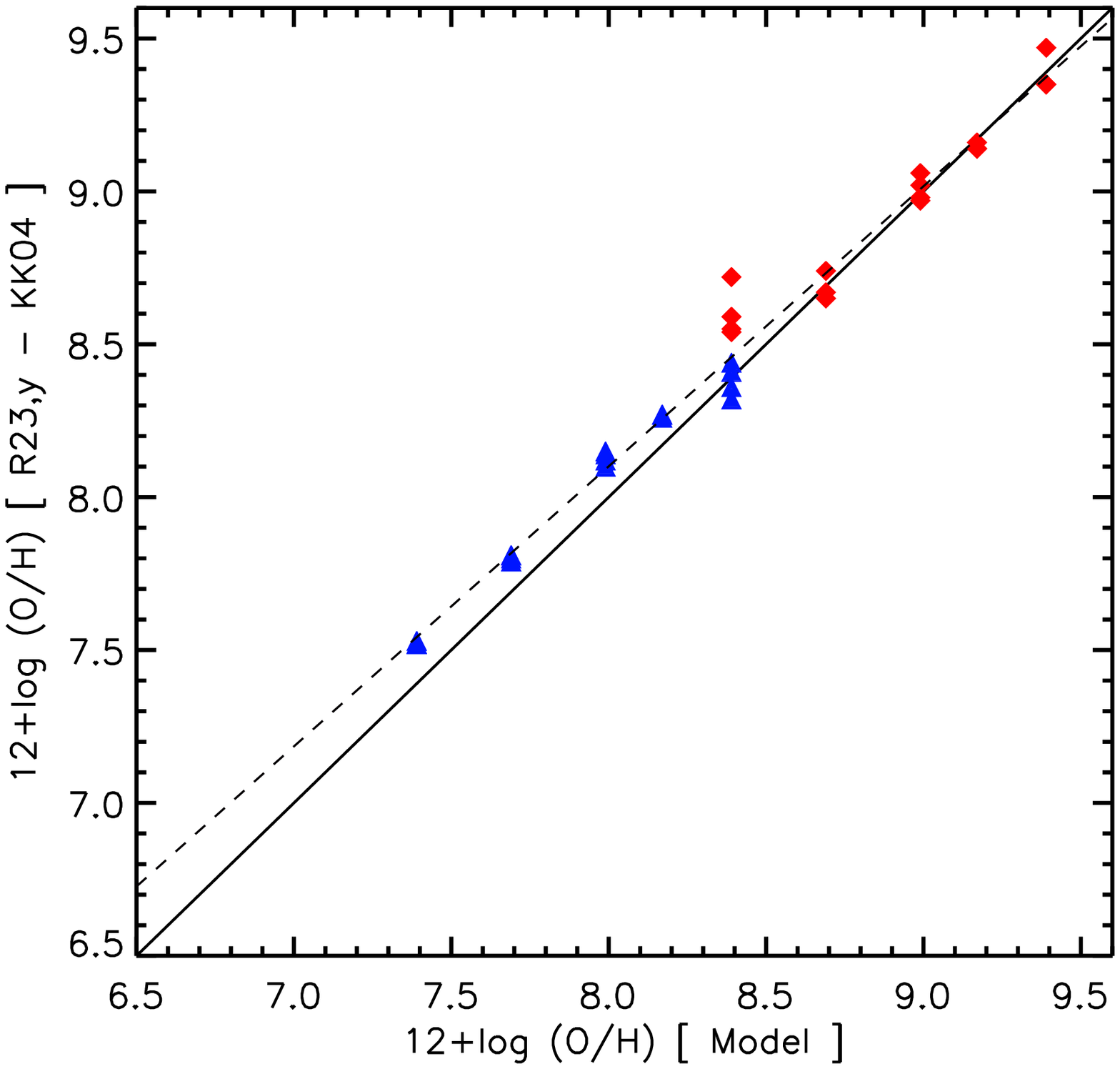} & 
\includegraphics[angle=90,width=0.24\linewidth]{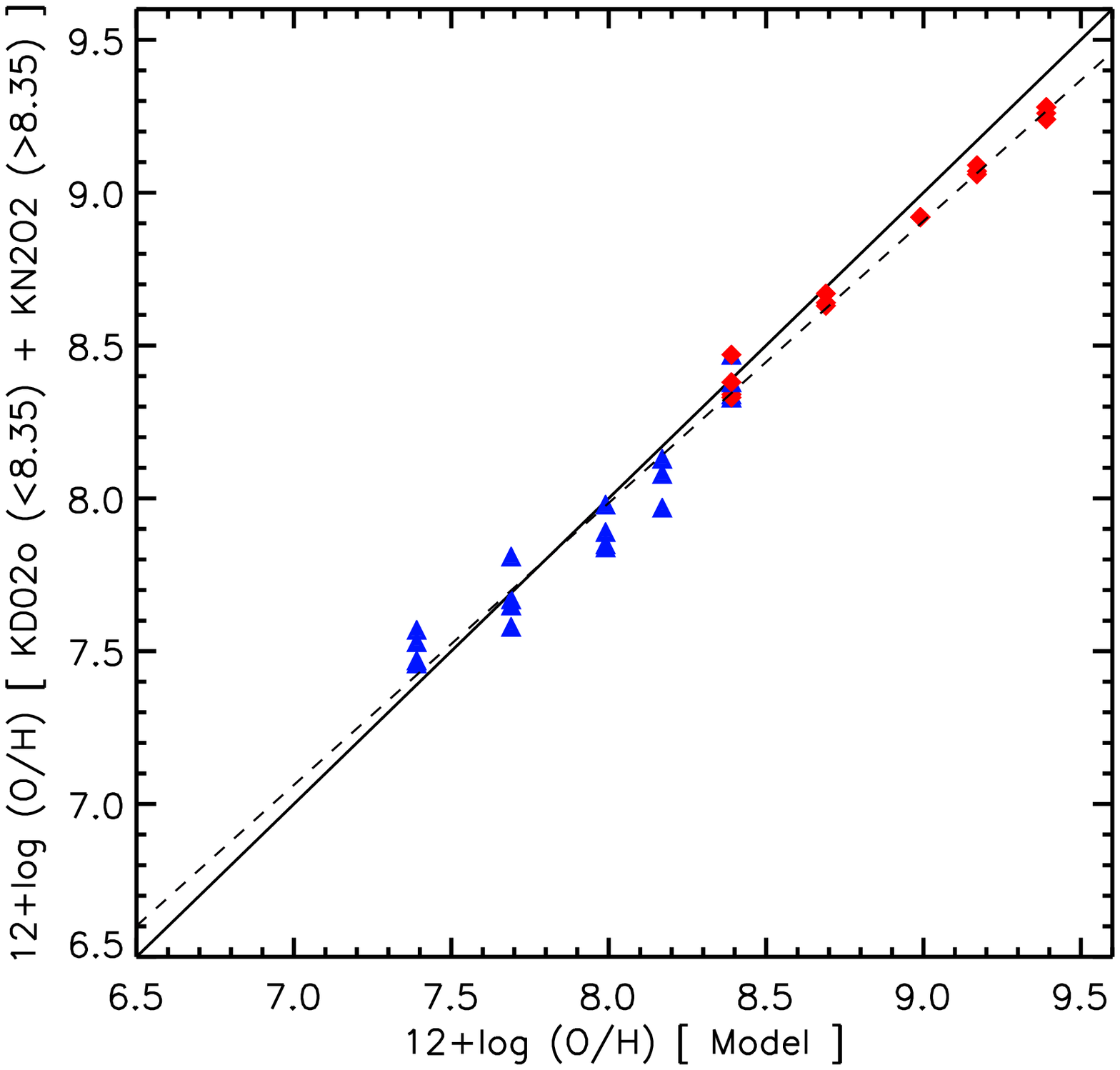} \\ 
\end{tabular}
\caption{Comparison between the total abundances used for the models (x-axis) and those computed using the \citet{McGaugh91} method (left panel) and the \citet{KD02} and  \citet{KK04} techniques, all of them based on photoionization models. The continuous line represents $y=x$. Red diamonds and blue triangles represent high- and low-metallicity models, respectively. The dashed line indicates the fit to the data. The far right panel shows the comparison with the oxygen abundance considering the  $N_2O_2$ method for the high-metallicity branch, \abox$\geq$8.35, and the $R_{23}$ and $q_{KD02o}$ for the low-metallicity branch, following \citet{KD02}.} 
\label{Z_K}
\end{figure*}

We have applied all of these commonly-used SEL techniques to our model spectra to derive oxygen abundances, again \emph{via} a double-blind procedure. 
The different SEL techniques fall into various classes, depending on which combination of strong line ratios are used. 
Table~\ref{empiricalpar} lists the values of all these parameters derived for each model. This table also includes the value derived for the $q$ parameter 
obtained from the optimal calibration provided by \citet{KD02} and \citet{KK04}. 
Table~\ref{empirical} compiles the oxygen abundances derived for each model, indicating which branch (high or low metallicity) the model spectra would fall into when using the $R_{23}$ parameter. 
We considered the values of the oxygen abundances derived from the \Te\  method 
to choose the expressions of the lower or the upper branch provided by the empirical calibrations.
For models F, U, W and AF, for which we obtained oxygen abundances of \abox$\sim$8.4 using the \Te\ method, the results provided by both the low- and high- metallicity branches are tabulated, and both sets of data are plotted in Figs.~\ref{Z_PP04} to \ref{Z_K}.
Note that the calibrations that invoke the $N_2$, $O_3N_2$, $S_{23}$ or $S_{23}O_{23}$ parameters give the same result for the low and high metallicity branches.
The \citet{KD02} calibration using the $N_2O_2$ parameter is only valid for objects in the high-metallicity branch.

To quantify the goodness of these SEL techniques, we have performed a linear fit between the oxygen abundances derived from these empirical methods and the oxygen abundances assumed by the models. In all cases, for models with \abox=8.39 we have assumed the average value between the high and low metallicity branches. 
We note that not all models have been considered in this fit, but only those within the validity range of each SEL calibration.
Table~\ref{empiricalfits} compiles the results for these linear fits, including the correlation coefficient, the dispersion of the data and the average value of the difference between the abundance given by the SEL method and that assumed by the models (offset). 
First row of Table~\ref{empiricalfits} lists the results of a linear fit between the oxygen abundances derived following the \Te\ method and those imposed by the models. This fit does not consider the models with \abox=9.39 (see Fig.~\ref{Z_Te_OIII}).

\subsubsection{$N_2$ and $O_3N_2$ methods}

Let us first consider the techniques which rely upon the  $N_2$ parameter; that of \citet{D02} and those drawn from the \citet{PP04} which involve a linear interpolation of the $N_2$ parameter, a cubic fit to this same parameter, and their fit using the $O_3N_2$ parameter. These are shown in Fig.~\ref{Z_PP04}. All these calibrations have similar scatter $\sim 0.20-0.25$~dex. However, they are well-behaved (if rather curved) in their abundance sequence. This type of curvature was also observed by \citet{KE08}, if we can interpret the mass sequence of galaxies as equivalent to an abundance sequence.


\citet{Yin+07} indicated that the $N_2$ and the $O_3N_2$ indices are only useful for calibrating metallicities of galaxies with \abox$<$8.5, while
\citet{PerezMonteroContini09} established that empirical calibrations using the $O_3N_2$ parameter are not valid for objects with \abox$\lesssim$8.0.
Following  Fig.~\ref{Z_PP04}, we suggest that the $N_2$ method should not be applied for \abox$\gtrsim$ 8.7-9.0, while the $O_3N_2$ calibration provided by \citet{PP04} is only valid for \abox$\gtrsim$ 8.7. 
Importantly, in all cases the result has a high uncertainty, $\sim$0.25~dex, which is a consequence of lacking of any parameter which considers the ionization degree of the gas.
Furthermore, in the case of galaxies showing an overabundance of nitrogen \citep[e.g.][]{Pustilnik04,LSEGRPR07,LSE10b,Monreal-Ibero+10}, the application of any $N_2$ or $O_3N_2$ calibration will provide misleading oxygen abundances.

\subsubsection{$S_{23}$ and $S_{23}O_{23}$ methods}

Figure~\ref{Z_S} compares the oxygen abundances assumed by our models with the results provided using the \citet{Perez-MonteroDiaz05} calibrations, that consider the $S_{23}$ and $S_{23}O_{23}$ parameters. In the case of using the $S_{23}$ index, the calibration seems to be valid up to \abox$\sim$9.0, having a dispersion of 0.24~dex. However, this method cannot be applied for \abox$\gtrsim$9.0. On the other hand, the $S_{23}O_{23}$ calibration works quite well at high metallicities,  \abox$\gtrsim$8.5. But it is not valid for \abox$\lesssim$8.2 as the result basically does not depend on metallicity, having a constant value which depends on $q$, and a high dispersion, $>$0.6~dex.

Much of the scatter in the  \citet{D02}, \citet{PP04} and \citet{Perez-MonteroDiaz05} calibrations is due to 
a failure to include the effects of the ionization parameter on the line ratios. 
That is specially important in the analysis of 2D spectroscopic data, as the oxygen abundance maps derived 
using these calibrations may directly reflect 
the ionization structure of the \HII\ region and not actual changes in metallicity \citep{LS+IC10+11}.
It is clear that the scatter could be reduced to $\sim$0.1~dex if the ionization parameter is taken into account, since the vertical separation of the points on the figures is almost entirely the result of the different ionization parameters used in the models.  However, this would not solve the curvature issue.

\subsubsection{Pilyugin method}

We next consider the Pilyugin et al. empirical calibrations of $R_{23}$  and $P$  \citep{P01a,P01b,PT05,PVT10,PM11}. 
As we discussed before, $P$ is an empirical parameter that traces the ionization degree of the \HII\ region.
The results we obtain are shown in Fig.~\ref{Z_Pil}. The main characteristic of these SEL calibrations is that there is a systematic offset in the sense that the Pilyugin method tends to underestimate the abundance by up to $\sim$0.25~dex \citep{P01a,P01b},  $\sim$0.4~dex \citep{PVT10} and $\sim$0.5~dex \citep{PT05,PM11}. 
It is also clear that the scatter for the  \citet{P01a,P01b} and \citet{PT05} empirical calibrations is typically 0.3~dex. 
In both calibrations, the scatter becomes considerably larger in the   
\abox$\sim$8.0--8.4\ range, as a result of the ambiguity in determining the appropriate branch on which the  $R_{23}$ is located. We note that for this intermediate-metallicity regime it is common to give an average value between the oxygen abundances derived for the high and low metallicity branches if using the \citet{P01a,P01b} and \citet{PT05} method \citep[e.g.][]{Moustakas+10,LSE10b}. 

In the case of the  \citet{PVT10} and \citet{PM11} methods (that also considers the $N_2$ and $S_2$ indices), three calibrations are provided for low, intermediate and high metallicity. In these cases, the scatter is slightly  lower, $\sim$0.14~dex and $\sim$0.25~dex, respectively, than the scatter found using the previous calibrations. 
The offset is also large, 0.4-0.5~dex, when comparing with that provided by the models.



The very interesting point here is that Pilyugin method is giving the best results when it is applied for real objects for which the oxygen abundance has been computed using the \Te\ method \citep[e.g.][]{Bresolin09,LSE10b,Moustakas+10}. However, this method is clearly failing when predicting the metallicities of the theoretical models using the bright emission lines. We will further discuss this issue in Sect.~\ref{comparaOBS}.

\begin{table*} 
\center
\caption{\label{obsdata} Results of the oxygen abundances given by the Te, RL, P and KD methods for Galactic (G) and extragalactic (E) \HII\ regions for which oxygen recombination lines have been detected.}
\scriptsize{
\begin{tabular}{ l cc cc cc cc cc }   
\hline

Object  &   $R_2$   &   $R_3$   &   $\log N_2$ &   \Te &   RL    &  &   P  & KD  & Type & Ref \\
 
 \hline
\noalign{\smallskip}

      M\,42 &  2.981 &  0.375 & -1.247 &   8.51$\pm$0.03 &   8.71$\pm$0.03 &   &  8.57 &   8.77 &  G &    E04   \\ 
NGC\,3576 &  4.169 &  2.447 & -0.767 &   8.56$\pm$0.03 &   8.74$\pm$0.06 &   &  8.51 &   8.80 &  G &   GR04   \\ 
       S\,311 &  1.318 &  5.591 & -1.720 &   8.39$\pm$0.05 &   8.54$\pm$0.10 &   &  8.28 &   8.65 &  G &   GR05   \\ 
       M\,16 &  0.668 &  3.547 & -2.055 &   8.56$\pm$0.07 &   8.81$\pm$0.07 &   &  8.45 &   8.79 &  G &   GR06   \\ 
       M\,20 &  0.350 &  0.009 & -0.549 &   8.53$\pm$0.06 &   8.71$\pm$0.07 &   &  8.45 &   8.93 &  G &   GR06   \\ 
   NGC\,3603 &  0.828 &  0.389 & -1.932 &   8.46$\pm$0.05 &   8.72$\pm$0.05 &   &  8.45 &   8.68 &  G &   GR06   \\ 
        M\,8 &  3.038 &  5.002 & -1.184 &   8.51$\pm$0.05 &   8.71$\pm$0.04 &   &  8.50 &   8.95 &  G &   GR07   \\ 
       M\,17 &  3.038 &  5.002 & -1.184 &   8.52$\pm$0.04 &   8.76$\pm$0.04 &   &  8.58 &   8.78 &  G &   GR07   \\ 
\noalign{\smallskip}

     30\,Dor &  1.823 &  1.109 & -0.692 &   8.33$\pm$0.02 &   8.54$\pm$0.05 &   &  8.39 &   8.63 &  E &   Pe03   \\
      N11\,B &  1.925 &  5.728 & -1.502 &   8.41$\pm$0.00 &   8.74$\pm$0.00 &   &  8.43 &   8.80 &  E &    T03   \\ 
       N\,66 &  2.052 &  6.349 & -1.360 &   8.11$\pm$0.00 &   8.47$\pm$0.00 &   &  8.10 &   8.40 &  E &    T03   \\ 
  NGC\,6822V &  3.000 &  3.764 & -1.296 &   8.08$\pm$0.03 &   8.37$\pm$0.09 &   &  8.09 &   8.39 &  E &  PPR05   \\ 
  NGC\,5253A &  2.429 &  3.860 & -1.020 &   8.18$\pm$0.04 &   8.42$\pm$0.13 &   &  8.12 &   8.48 &  E &   LS07   \\ 
  NGC\,5253B &  1.538 &  0.091 & -0.430 &   8.19$\pm$0.04 &   8.37$\pm$0.10 &   &  8.12 &   8.47 &  E &   LS07   \\ 
  NGC\,5253C &  0.554 &  1.229 & -2.160 &   8.28$\pm$0.04 &   8.53$\pm$0.09 &   &  8.30 &   8.59 &  E &   LS07   \\ 

    NGC\,595 &  0.596 &  0.287 & -0.792 &   8.45$\pm$0.03 &   8.69$\pm$0.05 &   &  8.51 &   8.72 &  E &    E09   \\ 
    NGC\,604 &  0.255 &  0.032 & -0.627 &   8.38$\pm$0.02 &   8.60$\pm$0.03 &   &  8.61 &   8.73 &  E &    E09   \\ 
     H\,1013 &  1.038 &  1.178 & -0.938 &   8.45$\pm$0.06 &   8.73$\pm$0.09 &   &  8.62 &   8.90 &  E &    E09   \\ 
   NGC\,5461 &  0.121 &  0.030 & -0.865 &   8.41$\pm$0.03 &   8.49$\pm$0.06 &   &  8.56 &   8.73 &  E &    E09   \\ 
      VS\,44 &  1.603 &  0.690 & -1.618 &   8.36$\pm$0.02 &   8.61$\pm$0.04 &   &  8.59 &   8.72 &  E &    E09   \\ 
   NGC\,2363 &  4.134 &  1.111 & -1.096 &   7.76$\pm$0.02 &   8.04$\pm$0.05 &   &  7.71 &   8.10 &  E &    E09   \\ 
      K\,932 &  1.085 &  2.203 & -1.842 &   8.41$\pm$0.02 &   8.62$\pm$0.03 &   &  8.52 &   8.77 &  E &    E09   \\ 

\hline
\end{tabular}
}
\flushleft

\textsc{References}:   
 E04: \citet{Esteban04}, 
  E09: \citet{Esteban09}, 
 GR04: \citet{GRE04}, 
 GR05: \citet{GRE05}, 
 GR06: \citet{GRE06}, 
 GR07: \citet{GRE07}, 
 LS07: \citet{LSEGRPR07}, 
  Pe03: \citet{Peimbert03},
  PPR05: \citet{PPR05},
  T03: \citet{Tsamis03}.

\end{table*}

\subsubsection{Kewley and Dopita method}

Although Pilyugin method considers a 
parameter which is related to ionization degree of the gas, the scatter provided by this calibration is still relatively high. 
However, the scatter is much reduced when using a SEL method which is based on photoionization models,
as those provided by the  \citet{McGaugh91}, \citet{KD02} and \citet{KK04} methods. 
All these models explicitly consider an ionization parameter, $y$.
Fig.~\ref{Z_K} shows their results. As we see, in all these cases the agreement is excellent. 
We remind the reader that we are plotting both the high and low metallicity results for the models with \abox=8.39.

Left panel of Fig.~\ref{Z_K} shows the case of the  \citet{McGaugh91} calibration using the \citet{Kobulnicky99} parametrization. This calibration systematically underestimates the abundances given by the models by $\sim$0.15~dex, but their dispersion is only 0.07~dex. 
In the case of using the \cite{KD02} or \citet{KK04} methods considering the $R_{23}$ and $y$ parameter, the results are much better. 
Their method allows to iteratively determine the ionization parameter 
and the oxygen abundance.
Both, the techniques by Kewley \& Dopita and by Kobulnicky \& Kewley, eliminate
the scatter in the SEL technique to a very large extent. For the  \citet{KD02} technique it is $\sim$0.1~dex, while for the  \citet{KK04} technique it is $\sim$0.07~dex excepting the ambiguous region around \abox=8.4.

Far right panel of Fig.~\ref{Z_K} shows a fit using the  $N_2O_2$ \mbox{calibration} provided by \citet{KD02} for the high-metallicity branch, \abox$\geq$8.35, and relied on the 
$R_{23}$ and the technique they use for obtaining the ionization parameter 
for the lower metallicities. 
This assumption provides very good results, as this method has a very low offset, 0.05~dex, and dispersion, 0.09~dex.

The slight offset in the abundances derived following the \citet{KD02} and \citet{KK04} methods is probably due to the fact that 
they used a different grid of models. 
The models we use in this paper are more closely related to the earlier \citet{KD02} grid, 
except in respect of the much more sophisticated treatment of dust physics employed here.

However, the recent analysis of the ionized gas within a sample of $\sim$40 strong star-forming galaxies performed by \citet{LSE10b} shows that the oxygen abundances provided by photoionization models \citep{McGaugh91,KD02,KK04} are, systematically, 0.2--0.3~dex higher than the oxygen abundances provided by the \Te\ method. This observational result has been also found by \citet{Yin+07}, \citet{Bresolin09}, \citet{Moustakas+10} and \citet{LS+IC10+11}; in some objects the differences reach up to 0.6~dex. 
Interesting, however, the photoionization models developed by \citet{Dors+11}  
provided 
O/H values close to those obtained using the \Te\ method. 
Detailed tailor-made photoionization models of individual galaxies \citep[e.g.][]{Perez-Montero10}
also provided results similar to those derived from the observed electron temperatures of the ionized gas.

\subsection{Comparison of the SEL techniques\\ with real data\label{comparaOBS}}

\begin{figure}
\centering
\bigskip
\includegraphics[angle=90,width=\linewidth]{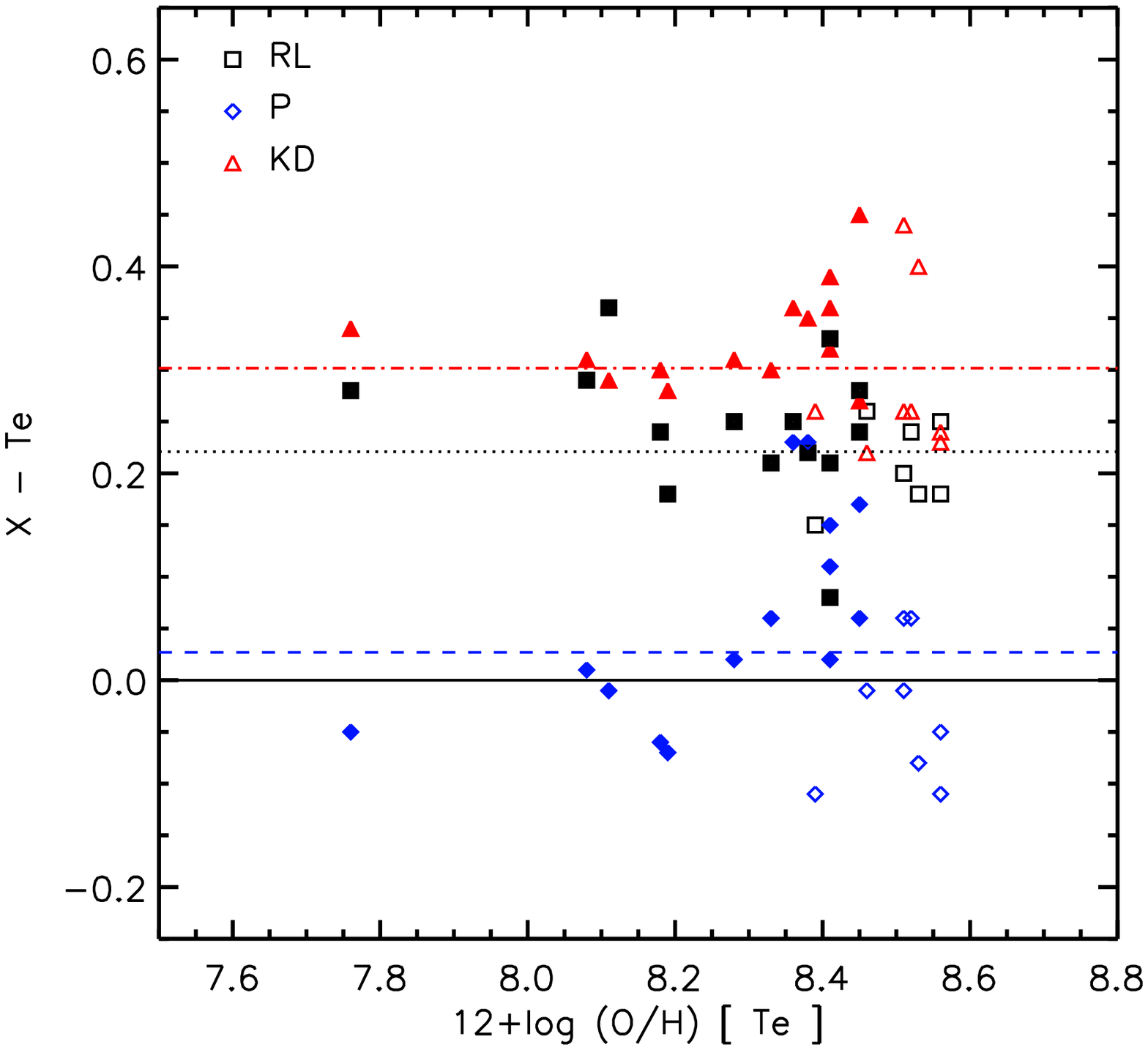} 
\caption{Comparison between the oxygen abundance derived following the \Te, RL, P and KD methods for deep, high-resolution observations of Galactic (open symbols) and extragalactic (filled symbols) \HII\ regions for which oxygen recombination lines have been detected. The   $x$-axis plots the \Te\ abundance, where the $y$-axis represents the difference between RL (black squares), P (blue diamonds) or KD (red triangles) abundances and  the \Te\ abundance. 
The black dotted line plots the average difference between the \Te\ and the RL methods. 
The blue dashed line represents the average difference between the \Te\ and the P methods. 
The red dotted-dashed line represents the average difference between the \Te\ and the KD methods.   }
\label{OBS}
\end{figure}

Hence, we have further investigated the validity of the SEL techniques using real data. For this, we aim to use the deepest, highest quality observational data of Galactic and extragalactic \HII\ regions nowadays available.
Particularly, we also want to compare the results provided by the \Te\ and the most common SEL techniques (the Pilyugin and the Kewley \& Dopita methods) with the oxygen abundances derived using oxygen recombination lines (RL).  
As it is well known \citep[e.g.][]{Peimbert03,Esteban02,Esteban04,Esteban09,GRE04,GRE05,GRE07,Peimbert07}, in all cases the oxygen abundances determined using RL are 0.2-0.3~dex higher than the oxygen abundances derived using the \Te\ method, which is based on collisionally  excited lines (CEL). The effect does not depend on metallicity, electron density, electron and Balmer temperatures or ionization degree \citep{GRE07,Mesa-Delgado09a} and it may be attributed to the presence of temperature fluctuations in \HII\ regions. 

Moreover, recent  analyses \citep{PPR05,Przybilla08,Simon-DiazStasinska11} have found that the oxygen nebular abundance based on RL 
agrees better with the abundances of the stars associated with the nebulae
than the one derived from the \Te\ method, suggesting that RL abundances are more reliable than the abundances derived from CEL.

Table~\ref{obsdata} lists all the 22 Galactic and extragalactic \HII\ regions found in the literature for which oxygen abundances are available using both the RL and the \Te\ methods. Indeed, the average offset between \Te\ and RL abundances is 0.22~dex.
Table~\ref{obsdata}  also compiles the 
derived  $R_{23}$, $P$, $y$ and $N_2$ indexes derived from the reddening-corrected data. 
We must emphasize that, in almost all the cases, the error in the [\ion{O}{ii}]~$\lambda\lambda$3726,3729 lines is less than 5\%. Using these observational parameters, we then applied the Pilyugin et al. method and the Kewley \& Dopita method to determine the oxygen abundances using these SEL techniques. Table~\ref{obsdata} compiles the average value derived for the \citet{P01a,P01b,PT05,PVT10} techniques (P) and the average value obtained for the \citet{KD02} (using the $R_{23}$, $y$ and $N_2O_2$ parameters) and the \citet{KK04} method (KD).


As we see in Table~\ref{obsdata}, we find again that the results obtained using the P technique agree quite well with the oxygen abundance determined using the \Te\ method (offset of 0.03~dex), but the results provided by the KD method are systematically $\sim$0.30~dex larger.
That is exactly the opposite behavior we discussed in the previous section. 
This result is graphically shown in Fig.~\ref{OBS}, which compares the \Te\ abundance ($x$-axis) with the oxygen abundances derived from RL (black squares), the P method (blue diamonds) and the KD technique (red triangles).

Interestingly, the results provided using the KD technique have a better agreement with the values determined from RL. These are systematically $\sim$0.08~dex lower than the oxygen abundances given by the KD method, but both agree within the errors (the dispersion of the RL data is $\sim$0.06~dex, while the dispersion of the KD data is $\sim$0.07~dex). This observational result, which was previously noted by \citet{Peimbert07} and \citet{LSE10b}, is actually suggesting that the real metallicities of the \HII\ regions may be not those given by the standard \Te\ method based on line ratios of collisionally excited lines (CELs), but the values derived using recombination lines (RL).
If this is correct, then the KD method is the only SEL technique that is reproducing those metallicities.

As we have seen along this paper, the oxygen abundances of our model spectra are well reproduced using the \Te\ method and SEL techniques based on photoionization models \citep{McGaugh91,KD02,KK04} but 0.3--0.5 dex higher than those estimated using the Pilyugin method. However, the situation is inverted in other works based on the comparison of observations with the results of the SEL methods \citep[e.g.][]{Peimbert07,Yin+07,Bresolin09,LSE10b,Moustakas+10,LS+IC10+11,RO2011}, i.e., the oxygen abundance in real nebulae is better reproduced by those SEL methods which rely in an empirical calibration derived using \HII\ regions where a direct estimation of the electron temperature exists.
 
A possible explanation of this puzzling situation is that current photoionization models are not properly reproducing the behavior of collisionally excited lines (CELs) in real nebulae. As we said before, several authors have invoked the presence of temperature fluctuations in the nebulae \citep{P67}
in order to address the differences between the abundances determined from CELs and RLs.  Due to the strong dependence on \Te\ of the intensity of CELs, such temperature fluctuations cause the metallicities derived from CELs to be underestimated. Although our models consider global temperature gradients due to the hardening of the ionizing radiation field along the nebula, they did not reproduce the values of the temperature fluctuations parameter ($t_2$) that are necessary to make the O/H ratios determined from RLs and CELs to converge 
\citep[e.g.][]{GRE07}. These hypothetical fluctuations must be of small spatial scale and may be produced by, for example, density inhomogeneities or localized deposition of mechanical energy that heat the gas. 

Summarizing, assuming the validity of the scenario outlined above, the oxygen abundances determined from RLs and using SEL techniques based on photoionization models (KD method) would be closer to the real ones, while values determined from CELs and empirical calibrations (P method) would provide abundances about 0.2--0.3~dex lower. This should have a profound impact in current galaxy metallicity estimations that have been obtained using CELs or any empirical calibration--such as those provided by Pilyugin et al. or \citet{PP04} -- that should be revised upwards.

Finally, we want to note that the small offset of $\sim$0.08~dex between the RL and the KD may be a consequence of the depletion of the oxygen atoms in dust grains. Our models are taking into account this effect, but observations suggest that oxygen depletion in dust grains may be decreasing the actual gas-phase metallicity (i.e., that derived from RL) of \HII\ regions by 0.08-0.12~dex \citep{Mesa-Delgado09b,PP10,Simon-DiazStasinska11}.


\section{Conclusions}  \label{Section5}

In this paper, we have subjected the commonly-used techniques for deriving chemical abundances to a critical ``double-blind'' analysis. We find that the originally-used \Te\ method is capable of delivering reliable abundances for O, N, S, and Cl. However, the abundances inferred for the noble gases Ne and Ar are considerably less secure, although probably for different physical reasons.  

The electron temperature itself is a reasonably good indicator of the oxygen abundance, as the one is the dominant controller of the other. Error arises because there is also a weaker dependence of electron temperature on the ionization parameter. For the [\ion{O}{iii}] lines, this renders an abundance obtained from the electron temperature alone unreliable below \abox~$\lesssim 8.0$. 

The temperature derived from the temperature-sensitive collisionally excited line ratios is always somewhat different from emission-weighted temperature defined for the principal emission lines --$\lambda4959+\lambda5007$ in the case of  [\ion{O}{iii}]--. This is due to the variability of electron temperature through the zone containing the ion of interest due to hardening of the radiation field and spherical divergence of the radiation field (which changes the photoelectric heating rate), and change in ionization state (which influences the cooling rate of the plasma). 
In this sense, the models naturally produce electron temperature gradients inside \HII\ regions, such as those discussed in \citet{S78}, \citet{Garnett92} and \citet{Sta05}.


Although our models can reproduce these ``global'' temperature fluctuations, they cannot simulate the small-scale temperature fluctuations,  such as those first discussed by \citet{P67} and \cite{PC69}, produced by turbulent or photo-evaporative flows, shocks, or radiation shielding by dense inclusions or globules of un-ionized material. Thus we can be reasonably sure that the true temperature fluctuations in real \HII\ regions should be greater than the models predict. The effect of this would be to raise the temperature as measured by the [\ion{O}{iii}] ($\lambda$4959+$\lambda$5007)/$\lambda$4363, which would then feed into a systematic underestimate in the oxygen abundance as delivered by the  \Te\ method.
Hence, accepting the existence of small-scale temperature fluctuations within the ionized gas and assuming that the RLs give the true metallicities,
those oxygen abundances determined from CELs and empirical calibrations (P method) would provide abundances which are 0.2--0.3~dex lower than the real ones. 
That is because of the strong dependence of the intensity of CELs on \Te: the existence of such small-scale temperature fluctuations will cause an underestimation of the oxygen abundances using CELs lines or strong-line methods which are based on the bright nebular lines.
However, oxygen abundances derived 
using SEL techniques based on photoionization models (such as the KD method, for which the \Te\ is fixed a priori) would be closer to the real values.

\begin{figure}
\centering
\includegraphics[angle=0,width=0.85\linewidth]{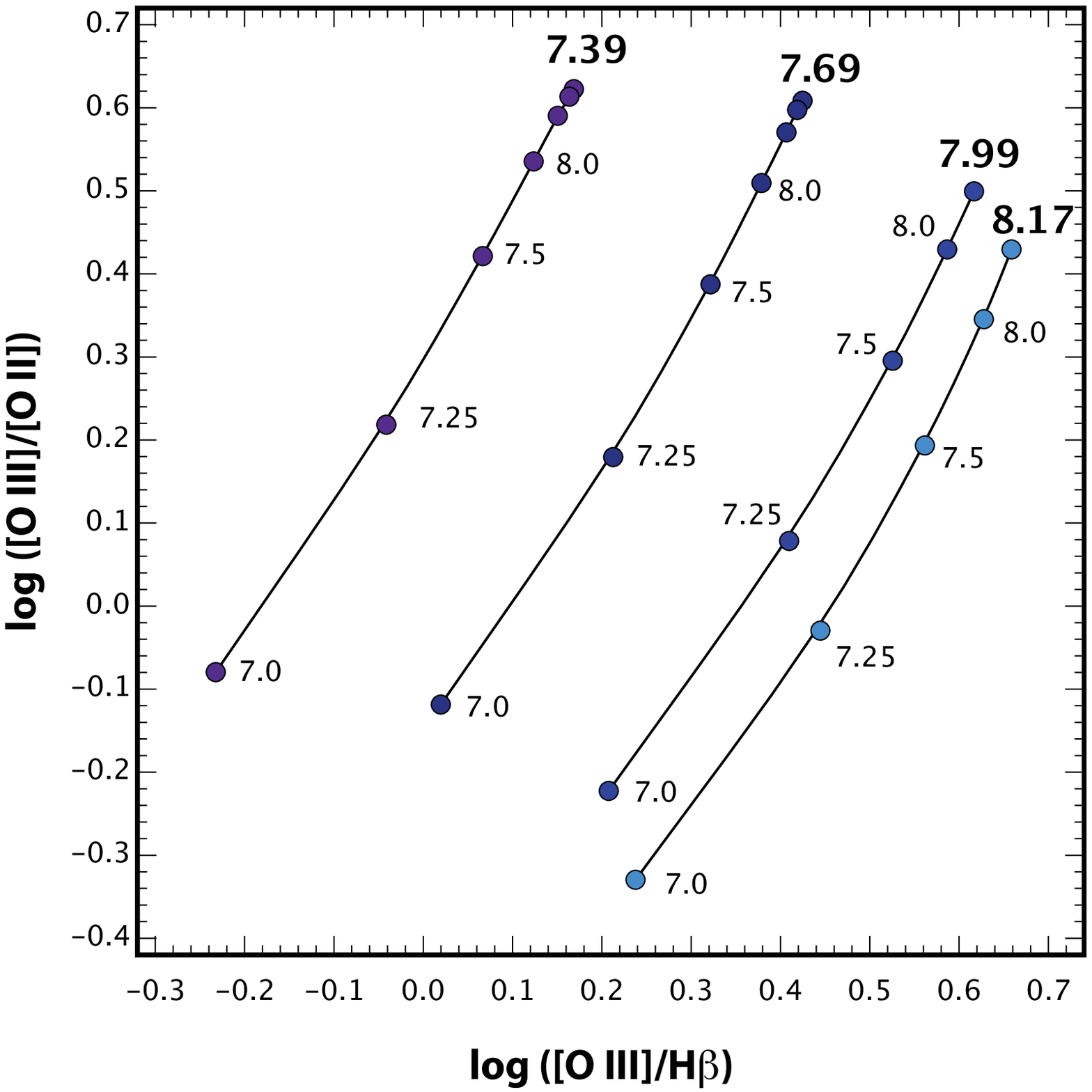}   
\caption{The model grid of  [\ion{O}{iii}] $\lambda$5007/H$\beta$ vs.  [\ion{O}{iii}] $\lambda$5007/[\ion{O}{ii}] $\lambda\lambda$3726,3729 at the low-abundance end. This confirms that both 12 + log(O/H) and $\log q$ can be derived from this diagnostic diagram for 12 + log(O/H) $\lesssim 8.0$.}
\label{Zlow-q}
\end{figure}

We have demonstrated that those SEL techniques which do not explicitly solve for the ionization parameter are quite unreliable, resulting in large scatter, systemic error, and abundance-dependent errors. 
%
%
To obtain the ionization parameter, we need a measurement of the [\ion{O}{iii}]~$\lambda$5007/[\ion{O}{ii}]~$\lambda\lambda$3726,3729 ratio. For abundances \abox$\lesssim8.0$ the oxygen abundance could then be estimated from the [\ion{O}{iii}] $\lambda$5007/H$\beta$ line ratio. This is verified in Fig~\ref{Zlow-q}, which provides a new diagnostic plot for this low abundance regime.

At high abundance, the most sensitive diagnostic plot is that of [\ion{O}{iii}]~$\lambda$5007/[\ion{O}{ii}] $\lambda$3726, 3729  \emph{vs.} {\mbox [\ion{N}{ii}]~$\lambda$6584/[\ion{O}{ii}]~$\lambda\lambda$3726,3729,} shown in panel~c of Fig.~\ref{Diagnostic_Diags}. The main error here is the degree to which the N/O abundance has been accurately calibrated against the O abundance. We believe that the relationship given in Eq.~\ref{calN_Z} is probably good, since the model grids reproduce the \citet{VO87} diagnostic plots shown in Fig.~\ref{Diagnostic_Diags}, panels (a) and (b). 

Figures  \ref{Diagnostic_Diags}, panel (c) and \ref{Zlow-q} thus provide a SEL technique that is applicable to the determination of both metallicity and ionization parameter over the full abundance range, with the caveat that the sensitivity of this technique is rather poor in the range $8.0 \lesssim $ 12 + log(O/H) $\lesssim 8.4$. In this range, common to all SEL techniques, rather accurately-measured line ratios are required.

\section*{Acknowledgements}

We thank our anonymous referee for his/her helpful comments.
We thank Maritza Lara-L\'opez, Mercedes Moll\'a, Enrique P\'erez-Montero, Jos\'e M. V\'{\i}lchez, C\'esar \mbox{Esteban}, 
\mbox{Stuart} Ryder, Guillermo H\"agele, and Fabi\'an Rosales-Ortega, 
for very fruitful discussions and comments about this study.
Dopita acknowledges support the Australian Research Council (ARC) for support under Discovery  project DP0984657.  
This research has made extensive use of the SAO/NASA Astrophysics Data System Bibliographic Services (ADS).

\appendix

\section{Electron temperatures, electron densities and ionic abundances for the model \HII\ regions.} 

Here we compile the derived parameters for the model \HII\ regions. The first table, Table \ref{temden}, gives the derived densities and temperatures in the various zones of the \HII\ region. Table \ref{ionic} gives the derived ionic abundances. Finally, Table \ref{empiricalpar} gives the line ratio values used to compute the oxygen abundance using the empirical calibrations as well as the ionization parameters derived for the models using the \cite{KD02} and \cite{KK04} techniques.

\begin{table*}
\caption{Electron density and temperature obtained from the analysis of emission line ratios of model spectra using IRAF task {\it temden}.  \label{temden}}
\label{physical} 
\begin{tabular}{c cccc cc@{\hspace{6pt}} ccc cc@{\hspace{6pt}} ccc  c }   %
\hline

      &  \multicolumn{5}{c}{ $n_e$   [ cm$^{-2}$ ]  }   &  &   \multicolumn{4}{c} {$T_{\rm e}$ High [ K ] }  &  &  \multicolumn{4}{c}{  $T_{\rm e}$ Low [ K ] } \\

\noalign{\smallskip}  \cline{2-6} \cline{8-11} \cline{13-16}

\noalign{\smallskip}

 Model              &      [\ion{S}{ii}]    &  [\ion{O}{ii}]  &  [\ion{N}{i}]   &   [\ion{Cl}{iii}]  & {\bf Adopted}         
            
             &  &      [\ion{O}{iii}]  &  [\ion{S}{iii}]   &   [\ion{Ar}{iii}]  & {\bf Adopted}  
               
             &  &    [\ion{O}{ii}]  &  [\ion{S}{ii}]   &   [\ion{N}{ii}]  & {\bf Adopted}   \\

\hline
\noalign{\smallskip}

 A    &   6  & 213   &  12   &  113   &  {\bf 86  }   & &   8465  & 8037   &  8389   &  {\bf 8297  }   &  &  8419  & 6419   &  9274  &  {\bf 8037  }   \\
 B     &   0  & 243   &  7   &  136   &  {\bf 96  } &  &   14947  & 12222   &  13307   &  {\bf 13492  }  &  &   12141  & 7621   &  13658  &  {\bf 11140  }   \\
 C    &   30  & 0   &  0   &  68   &  {\bf 49  } &  &   0  & 4425   &  4519   &  {\bf 4472  }  &  &   0  & 3843   &  4680  &  {\bf 4261$^a$  }   \\
 D     &   0  & 236   &  7   &  135   &  {\bf 126  }   &   & 12663  & 11009   &  11706   &  {\bf 11792  }   &  &   11122  & 7337   &  12820  &  {\bf 10426  }   \\
 E   &   0  & 228   &  8   &  130   &  {\bf 122  } &  &   11459  & 10134   &  10732   &  {\bf 10775  }  &  &   10311  & 7086   &  11802  &  {\bf 9733  }   \\
 F   &   0  & 226   &  9   &  130   &  {\bf 121  } &  &   10498  & 9625   &  10159   &  {\bf 10094  }  &  &   9834  & 7023   &  11254  &  {\bf 9370  }   \\
 G    &   25  & 182   &  23   &  76   &  {\bf 76  }   &  & 4855  & 4993   &  4908   &  {\bf 4918  }   &  &  5423  & 4535   &  5660  &  {\bf 5206  }   \\
 H    &   33  & 0   &  0   &  68   &  {\bf 50  }   &  & 4285  & 4210   &  4236   &  {\bf 4243  }   &  &  0  & 3737   &  4474  &  {\bf 4105$^a$ }     \\
 I   &   19  & 196   &  21   &  87   &  {\bf 80  } &  &   5830  & 5873   &  5973   &  {\bf 5892  }  &  &   6424  & 5313   &  6986  &  {\bf 6241  }   \\
 J    &   35  & 0   &  0   &  71   &  {\bf 53  }   &  & 0  & 3910   &  3768   &  {\bf 3839  }   &  &  0  & 3633   &  4269  &  {\bf 3951$^a$  }   \\
 K     &   0  & 234   &  7   &  135   &  {\bf 125  }   &  & 12413  & 10843   &  11519   &  {\bf 11591  }   &  &  10968  & 7304   &  12633  &  {\bf 10301  }   \\
 L   &   0  & 229   &  8   &  130   &  {\bf 122  }   &  & 11394  & 10139   &  10737   &  {\bf 10756  }   &  &  10297  & 7122   &  11796  &  {\bf 9738  }   \\ 
 M     &   0  & 240   &  7   &  138   &  {\bf 128  }   &  & 13637  & 11749   &  12613   &  {\bf 12666  }   &  &  11614  & 7514   &  13408  &  {\bf 10845  }   \\
 N   &   6  & 214   &  12   &  113   &  {\bf 86  }   &  & 8129  & 7811   &  8144   &  {\bf 8028  }   &  &  8409  & 6435   &  9280  &  {\bf 8041  }   \\
 O     &   21  & 182   &  21   &  77   &  {\bf 75  }   &  & 5769  & 5542   &  5665   &  {\bf 5658  }   &  &  5681  & 4660   &  5996  &  {\bf 5445  }   \\
 P   &   0  & 241   &  7   &  138   &  {\bf 128  }   &  & 14096  & 12034   &  12953   &  {\bf 13027  }   &  &  11833  & 7556   &  13658  &  {\bf 11015  }   \\
 Q  &   0  & 239   &  7   &  135   &  {\bf 127  }   &  & 13718  & 11640   &  12568   &  {\bf 12642  }   &  &  11462  & 7454   &  13202  &  {\bf 10706  }   \\
 R   &   16  & 194   &  19   &  86   &  {\bf 78  }   &  & 6225  & 6158   &  6283   &  {\bf 6222  }   &  &  6549  & 5322   &  6996  &  {\bf 6289  }   \\
 S   &   0  & 232   &  8   &  135   &  {\bf 125  }   &  & 11890  & 10586   &  11248   &  {\bf 11241  }   &  &  10661  & 7284   &  12268  &  {\bf 10071  }   \\
 T  &   0  & 234   &  8   &  136   &  {\bf 126  }   &  & 12181  & 10781   &  11506   &  {\bf 11489  }   &  &  10804  & 7314   &  12451  &  {\bf 10189  }   \\
 U  &   0  & 228   &  9   &  131   &  {\bf 122  }   &  & 10525  & 9641   &  10182   &  {\bf 10116  }   &  &  9913  & 7044   &  11364  &  {\bf 9440  }   \\
 V &   0  & 241   &  7   &  139   &  {\bf 129  }   &  & 14562  & 12172   &  13223   &  {\bf 13319  }   &  &  11840  & 7525   &  13640  &  {\bf 11001  }   \\
 W &     0  & 225   &  9   &  129   &  {\bf 121  }   &  & 10537  & 9589   &  10116   &  {\bf 10080  }   &  &  9748  & 6962   &  11119  &  {\bf 9276  }   \\
 X &   40  & 0   &  0   &  78   &  {\bf 59  }   &  & 0  & 3930   &  0   &  {\bf 3930  }   &  &  0  & 3695   &  4338  &  {\bf 4016$^a$  }   \\
Y  &   0  & 243   &  7   &  136   &  {\bf 128  }   &  & 15740  & 12547   &  13728   &  {\bf 14005  }   &  &  12101  & 7554   &  13892  &  {\bf 11182  }   \\
Z  &   13  & 194   &  17   &  90   &  {\bf 78  }   &  & 6941  & 6580   &  6790   &  {\bf 6770  }   &  &  6768  & 5356   &  7252  &  {\bf 6458  }  \\
AA  &   22  & 181   &  23   &  75   &  {\bf 75  }   &  & 5364  & 5295   &  5338   &  {\bf 5332  }   &  &  5498  & 4588   &  5788  &  {\bf 5291  }   \\
AB  &   0  & 234   &  8   &  135   &  {\bf 125  }   &  & 12452  & 10927   &  11628   &  {\bf 11669  }   &  &  11001  & 7359   &  12678  &  {\bf 10346  }   \\
AC &   15  & 194   &  17   &  89   &  {\bf 78  }   &  & 6741  & 6483   &  6654   &  {\bf 6626  }   &  &  6712  & 5378   &  7189  &  {\bf 6426  }   \\
AD  &   0  & 243   &  6   &  138   &  {\bf 129  }   &  & 15240  & 12497   &  13644   &  {\bf 13793  }   &  &  12078  & 7563   &  13888  &  {\bf 11176  }   \\
AE  &   0  & 238   &  7   &  139   &  {\bf 128  }   &  & 13248  & 11484   &  12330   &  {\bf 12354  }   &  &  11363  & 7459   &  13115  &  {\bf 10645  }   \\
AF  &   0  & 222   &  10   &  125   &  {\bf 119  }   &  & 10289  & 9308   &  9814   &  {\bf 9803  }   &  &  9470  & 6799   &  10740  &  {\bf 9003  }   \\
AG  &   5  & 212   &  12   &  113   &  {\bf 85  }   &  & 8765  & 8182   &  8545   &  {\bf 8497  }   &  &  8477  & 6389   &  9318  &  {\bf 8061  }   \\

\hline
\end{tabular}
\flushleft

$^a$ {\it ionic} task does not allow electron temperatures lower than 5,000\,K for the low ionization ions, so we adopted  \Te(low) = 5,000\,K for these cases.

\end{table*}

\begin{table*}
\scriptsize{
\caption{ Ionic abundances derived for the results of the photoionization models using IRAF task {\it ionic}
considering the electron temperature derived from all the high-ionization species, \Te(high), and the low ionization species, \Te(low), separately.
All numbers are expressed in the form of 12 + log (X/H). The label \emph{disp.} indicates the dispersion of the data when considering the values given by different emission lines.  \label{ionic}}

\begin{tabular}{cc cc c cc c cc cc }   %
\hline




 Model       &       &     O$^+$/H$^+$    
                         &     O$^{++}$/H$^+$
                         &     N$^+$/H$^+$  
                         &     S$^+$/H$^+$              
                         &     S$^{++}$/H$^+$
                         &     Ne$^{++}$/H$^+$          
                         &     Ar$^+$/H$^+$
                         &     Ar$^{++}$/H$^+$
                         &     Cl$^{++}$/H$^+$
                         &     Cl$^{3+}$/H$^+$
                           \\

\hline
\noalign{\smallskip}

A   &   &    8.717   &   8.126     &    7.333   &   6.327     &    7.074   &   7.683     &    6.458   &   4.364     &    4.272   &   3.578   \\
  &   {\it disp.}  &   {\it   0.031 }  &  {\it  0.016 }    &   {\it   0.085 }  &  {\it  0.095 }    &   {\it   0.018 }  &  {\it  0.071 }    &   {\it   0.007 }  &  {\it  0.001 }    &   {\it   0.001 }  &  {\it  0.090 } \\  
B  &  &    7.283   &   7.609     &    5.635   &   4.928     &    5.873   &   6.941     &    5.326   &   4.915     &    2.810   &   3.251   \\
    &   {\it disp.} &   {\it   0.036 }  &  {\it  0.049 }    &   {\it   0.084 }  &  {\it  0.122 }    &   {\it   0.036 }  &  {\it  0.071 }    &   {\it   0.006 }  &  {\it  0.000 }    &   {\it   0.001 }  &   {\it  0.090 }   \\    
C  &  &    8.873   &   7.101     &    8.257   &   7.224     &    7.365   &   7.347     &    6.586   &   0.000     &    4.575   &   2.487   \\
   &   {\it disp.}   &   {\it   0.115 }  &  {\it  4.100 }    &   {\it   0.070 }  &  {\it  0.189 }    &   {\it   0.011 }  &  {\it  0.071 }    &   {\it   0.013 }  &  {\it  0.000 }    &   {\it   0.001 }  &  {\it  0.090 }  \\   
 D  &  &    7.489   &   6.809     &    5.840   &   5.208     &    5.727   &   6.468     &    5.132   &   2.969     &    2.882   &   2.502   \\
    &   {\it disp.}  &   {\it   0.029 }  &  {\it  0.039 }    &   {\it   0.091 }  &  {\it  0.119 }    &   {\it   0.028 }  &  {\it  0.071 }    &   {\it   0.004 }  &  {\it  0.001 }    &   {\it   0.000 }  &  {\it  0.090 }  \\  
 E   &  &    8.173   &   6.915     &    6.606   &   6.003     &    6.282   &   6.884     &    5.703   &   2.871     &    3.517   &   2.695  \\ 
  &   {\it disp.}  &   {\it   0.029 }  &  {\it  0.037 }    &   {\it   0.092 }  &  {\it  0.113 }    &   {\it   0.028 }  &  {\it  0.071 }    &   {\it   0.003 }  &  {\it  0.001 }    &   {\it   0.000 }  &  {\it  0.090 }  \\  
 F & &    8.252   &   8.127     &    6.717   &   5.796     &    6.752   &   7.544     &    6.157   &   4.787     &    3.868   &   3.666   \\
    &   {\it disp.}  &   {\it   0.026 }  &  {\it  0.026 }    &   {\it   0.091 }  &  {\it  0.106 }    &   {\it   0.625 }  &  {\it  0.071 }    &   {\it   0.003 }  &  {\it  0.001 }    &   {\it   0.000 }  &  {\it  0.090 }  \\ 
G  &  &    9.089   &   8.315     &    8.038   &   6.764     &    7.525   &   7.803     &    6.747   &   4.544     &    4.725   &   3.535   \\
  &   {\it disp.}  &   {\it   0.038 }  &  {\it  0.018 }    &   {\it   0.079 }  &  {\it  0.088 }    &   {\it   0.015 }  &  {\it  0.071 }    &   {\it   0.003 }  &  {\it  0.000 }    &   {\it   0.000 }  &  {\it  0.090 }  \\   
H  &  &  8.710   &   7.821     &    8.152   &   7.003     &    7.564   &   7.658     &    6.802   &   3.624     &    4.771   &   3.019   \\
    &   {\it disp.} &   {\it   0.144 }  &  {\it  0.017 }    &   {\it   0.122 }  &  {\it  0.219 }    &   {\it   0.008 }  &  {\it  0.071 }    &   {\it   0.002 }  &  {\it  0.000 }    &   {\it   0.001 }  &  {\it  0.090 }  \\  
I &  &    8.963   &   8.309     &    7.709   &   6.447     &    7.240   &   7.754     &    6.559   &   4.909     &    4.416   &   3.763   \\ 
  &   {\it disp.}  &   {\it   0.061 }  &  {\it  0.076 }    &   {\it   0.137 }  &  {\it  0.058 }    &   {\it   0.048 }  &  {\it  0.071 }    &   {\it   0.041 }  &  {\it  0.000 }    &   {\it   0.000 }  &  {\it  0.090 }  \\
J  &  &    8.372   &   8.105     &    7.892   &   6.667     &    7.676   &   7.710     &    6.781   &   4.095     &    4.855   &   3.231   \\   
   &   {\it disp.}  &   {\it   0.170 }  &  {\it  4.680 }    &   {\it   0.178 }  &  {\it  0.238 }    &   {\it   0.022 }  &  {\it  0.071 }    &   {\it   0.028 }  &  {\it  0.002 }    &   {\it   0.001 }  &  {\it  0.090 }  \\ 
K & &    7.778   &   7.081     &    6.155   &   5.502     &    6.028   &   6.754     &    5.433   &   3.223     &    3.194   &   2.765   \\   
   &   {\it disp.}  &   {\it   0.026 }  &  {\it  0.039 }    &   {\it   0.087 }  &  {\it  0.120 }    &   {\it   0.028 }  &  {\it  0.071 }    &   {\it   0.004 }  &  {\it  0.001 }    &   {\it   0.000 }  &  {\it  0.090 }  \\ 
 L  & &    8.313   &   7.388     &    6.758   &   6.015     &    6.505   &   7.129     &    5.905   &   3.456     &    3.711   &   3.024   \\
 &   {\it disp.} &   {\it   0.028 }  &  {\it  0.035 }    &   {\it   0.091 }  &  {\it  0.112 }    &   {\it   0.027 }  &  {\it  0.071 }    &   {\it   0.002 }  &  {\it  0.001 }    &   {\it   0.000 }  &  {\it  0.090 } \\ 
M  &   &    7.536   &   7.470     &    5.884   &   5.167     &    6.022   &   6.876     &    5.431   &   4.191     &    3.036   &   3.096   \\ 
     &   {\it disp.}  &   {\it   0.029 }  &  {\it  0.038 }    &   {\it   0.089 }  &  {\it  0.120 }    &   {\it   0.029 }  &  {\it  0.071 }    &   {\it   0.003 }  &  {\it  0.001 }    &   {\it   0.000 }  &  {\it  0.090 }  \\  
N &   &    8.481   &   8.378     &    7.079   &   6.029     &    7.065   &   7.804     &    6.458   &   5.041     &    4.216   &   3.858   \\ 
  &   {\it disp.}   &   {\it   0.031 }  &  {\it  0.010 }    &   {\it   0.085 }  &  {\it  0.094 }    &   {\it   0.016 }  &  {\it  0.071 }    &   {\it   0.010 }  &  {\it  0.001 }    &   {\it   0.001 }  &  {\it  0.090 }  \\  
O  &   &    9.185   &   7.426     &    8.201   &   7.180     &    7.307   &   7.405     &    6.597   &   3.204     &    4.517   &   2.846   \\   
   &   {\it disp.}  &   {\it   0.025 }  &  {\it  0.000 }    &   {\it   0.047 }  &  {\it  0.119 }    &   {\it   0.034 }  &  {\it  0.071 }    &   {\it   0.018 }  &  {\it  0.000 }    &   {\it   0.000 }  &  {\it  0.090 }  \\   
P  &   &    7.218   &   7.183     &    5.548   &   4.862     &    5.712   &   6.581     &    5.124   &   3.946     &    2.705   &   2.818   \\
    &   {\it disp.}  &   {\it   0.030 }  &  {\it  0.039 }    &   {\it   0.089 }  &  {\it  0.122 }    &   {\it   0.030 }  &  {\it  0.071 }    &   {\it   0.003 }  &  {\it  0.001 }    &   {\it   0.000 }  &  {\it  0.090 } \\
Q &  &    7.644   &   7.880     &    6.022   &   5.232     &    6.224   &   7.223     &    5.667   &   5.094     &    3.210   &   3.505   \\   
 &   {\it disp.}  &   {\it   0.029 }  &  {\it  0.042 }    &   {\it   0.090 }  &  {\it  0.120 }    &   {\it   0.032 }  &  {\it  0.071 }    &   {\it   0.003 }  &  {\it  0.001 }    &   {\it   0.000 }  &  {\it  0.090 }  \\
 R  &  &    8.984   &   8.338     &    7.788   &   6.615     &    7.368   &   7.862     &    6.711   &   4.579     &    4.579   &   3.652   \\   
&   {\it disp.}  &   {\it   0.034 }  &  {\it  0.001 }    &   {\it   0.082 }  &  {\it  0.089 }    &   {\it   0.008 }  &  {\it  0.071 }    &   {\it   0.009 }  &  {\it  0.000 }    &   {\it   0.000 }  &  {\it  0.090 }  \\   
 S &   &    8.110   &   7.855     &    6.516   &   5.674     &    6.535   &   7.300     &    5.927   &   4.379     &    3.643   &   3.414   \\  
 &   {\it disp.}  &   {\it   0.027 }  &  {\it  0.033 }    &   {\it   0.090 }  &  {\it  0.112 }    &   {\it   0.026 }  &  {\it  0.071 }    &   {\it   0.001 }  &  {\it  0.001 }    &   {\it   0.000 }  &  {\it  0.090 }  \\   
 T &   &    7.896   &   8.008     &    6.297   &   5.444     &    6.472   &   7.373     &    5.896   &   5.017     &    3.510   &   3.603   \\
&   {\it disp.}  &   {\it   0.027 }  &  {\it  0.033 }    &   {\it   0.090 }  &  {\it  0.114 }    &   {\it   0.027 }  &  {\it  0.071 }    &   {\it   0.001 }  &  {\it  0.001 }    &   {\it   0.000 }  &  {\it  0.090 }  \\   
 U &   &    8.067   &   8.227     &    6.531   &   5.607     &    6.692   &   7.596     &    6.117   &   5.290     &    3.755   &   3.812   \\  
 &   {\it disp.} &   {\it   0.026 }  &  {\it  0.026 }    &   {\it   0.091 }  &  {\it  0.107 }    &   {\it   0.023 }  &  {\it  0.071 }    &   {\it   0.003 }  &  {\it  0.002 }    &   {\it   0.000 }  &  {\it  0.090 }  \\  
 V &  &  7.349   &   7.579     &    5.698   &   4.983     &    5.921   &   6.924     &    5.369   &   4.758     &    2.868   &   3.221   \\ 
   &   {\it disp.}  &   {\it   0.030 }  &  {\it  0.043 }    &   {\it   0.089 }  &  {\it  0.123 }    &   {\it   0.033 }  &  {\it  0.071 }    &   {\it   0.004 }  &  {\it  0.001 }    &   {\it   0.000 }  &  {\it  0.090 }  \\
W &  &      8.481   &   7.804     &    6.972   &   6.106     &    6.759   &   7.406     &    6.150   &   3.992     &    3.946   &   3.351   \\  
  &   {\it disp.}  &   {\it   0.026 }  &  {\it  0.029 }    &   {\it   0.091 }  &  {\it  0.107 }    &   {\it   0.024 }  &  {\it  0.071 }    &   {\it   0.002 }  &  {\it  0.001 }    &   {\it   0.000 }  &  {\it  0.090 }  \\
 X  & &    8.199   &   7.632     &    7.698   &   6.368     &    7.457   &   7.095     &    6.138   &   3.674     &    4.599   &   2.834  \\  
&   {\it disp.} &   {\it   0.152 }  &  {\it  4.407 }    &   {\it   0.159 }  &  {\it  0.227 }    &   {\it   0.000 }  &  {\it  0.071 }    &   {\it   0.607 }  &  {\it  0.003 }    &   {\it   0.000 }  &  {\it  0.090 } \\   
 Y  & &    7.004   &   7.312     &    5.337   &   4.657     &    5.561   &   6.641     &    5.015   &   4.628     &    2.484   &   2.956   \\ 
 &   {\it disp.}  &   {\it   0.031 }  &  {\it  0.053 }    &   {\it   0.088 }  &  {\it  0.126 }    &   {\it   0.039 }  &  {\it  0.071 }    &   {\it   0.009 }  &  {\it  0.001 }    &   {\it   0.000 }  &  {\it  0.090 }  \\
Z  &  &    9.179   &   7.161     &    8.024   &   7.151     &    7.158   &   7.350     &    6.502   &   2.898     &    4.387   &   2.758   \\
&   {\it disp.}  &   {\it   0.036 }  &  {\it  0.025 }    &   {\it   0.087 }  &  {\it  0.098 }    &   {\it   0.020 }  &  {\it  0.071 }    &   {\it   0.002 }  &  {\it  0.001 }    &   {\it   0.001 }  &  {\it  0.090 }  \\
 AA  &  &    9.292   &   8.103     &    8.240   &   7.056     &    7.484   &   7.746     &    6.769   &   4.105     &    4.698   &   3.336   \\
  &   {\it disp.} &   {\it   0.037 }  &  {\it  0.008 }    &   {\it   0.083 }  &  {\it  0.090 }    &   {\it   0.006 }  &  {\it  0.071 }    &   {\it   0.001 }  &  {\it  0.000 }    &   {\it   0.000 }  &  {\it  0.090 }  \\
 AB   &  &    8.016   &   7.570     &    6.405   &   5.639     &    6.349   &   7.093     &    5.742   &   3.905     &    3.478   &   3.168   \\
 &   {\it disp.}  &   {\it   0.028 }  &  {\it  0.036 }    &   {\it   0.090 }  &  {\it  0.116 }    &   {\it   0.027 }  &  {\it  0.071 }    &   {\it   0.002 }  &  {\it  0.001 }    &   {\it   0.000 }  &  {\it  0.090 }  \\
  AC &  &    9.138   &   7.906     &    7.964   &   6.903     &    7.303   &   7.666     &    6.654   &   3.869     &    4.529   &   3.294  \\ 
 &   {\it disp.}  &   {\it   0.035 }  &  {\it  0.017 }    &   {\it   0.084 }  &  {\it  0.093 }    &   {\it   0.016 }  &  {\it  0.071 }    &   {\it   0.003 }  &  {\it  0.001 }    &   {\it   0.001 }  &  {\it  0.090 }  \\  
  AD &   &     7.042   &   7.285     &    5.374   &   4.690     &    5.604   &   6.625     &    5.055   &   4.492     &    2.535   &   2.931  \\   
  &   {\it disp.} &   {\it   0.031 }  &  {\it  0.047 }    &   {\it   0.089 }  &  {\it  0.125 }    &   {\it   0.035 }  &  {\it  0.071 }    &   {\it   0.005 }  &  {\it  0.001 }    &   {\it   0.000 }  &  {\it  0.090 }  \\   
 AE   &  &    7.766   &   7.814     &    6.141   &   5.348     &    6.304   &   7.189     &    5.722   &   4.717     &    3.322   &   3.422   \\
 &   {\it disp.}  &   {\it   0.028 }  &  {\it  0.037 }    &   {\it   0.090 }  &  {\it  0.118 }    &   {\it   0.029 }  &  {\it  0.071 }    &   {\it   0.002 }  &  {\it  0.001 }    &   {\it   0.000 }  &  {\it  0.090 }  \\
 AF  &  &    8.585   &   7.112     &    7.118   &   6.439     &    6.669   &   7.157     &    6.084   &   3.024     &    3.914   &   2.857   \\
   &   {\it disp.} &   {\it   0.027 }  &  {\it  0.033 }    &   {\it   0.091 }  &  {\it  0.107 }    &   {\it   0.026 }  &  {\it  0.071 }    &   {\it   0.001 }  &  {\it  0.001 }    &   {\it   0.000 }  &  {\it  0.090 }  \\
 AG  &  &    8.851   &   7.570     &    7.503   &   6.639     &    7.005   &   7.456     &    6.400   &   3.515     &    4.239   &   3.153   \\ 
  &   {\it disp.}  &   {\it   0.033 }  &  {\it  0.024 }    &   {\it   0.085 }  &  {\it  0.098 }    &   {\it   0.021 }  &  {\it  0.071 }    &   {\it   0.004 }  &  {\it  0.001 }    &   {\it   0.001 }  &  {\it  0.090 }  \\

 \hline
\end{tabular}

}
\end{table*}

\begin{table*} 
\caption{ Parameters used to compute the oxygen abundance using empirical calibrations. \label{empiricalpar}}
\scriptsize{
\begin{tabular}{ c c c c c c c c c c c c c  }   %
\hline

Model &   $R_2$   &   $R_3$   &   $R_{23}$  &   $P$   &  $y$    & $\log N_2$ & $\log O_3N_2$ & $\log O_2N_2$ &   $S_2$   &  $S_{23}$   & $\log q_{KD02}$$^a$  & $\log q_{KK04}$$^b$ \\ 
 
 \hline
\noalign{\smallskip}

           A &  4.169 &  2.447 &  6.616 &  0.370 & -0.231 & -0.640 &  1.026 &  1.387 &  0.623 &  2.445 &   7.72 &   7.32  \\
            B &  0.668 &  3.547 &  4.215 &  0.842 &  0.725 & -1.928 &  2.481 &  1.880 &  0.060 &  0.367 &   8.40 &   7.87  \\
            C &  0.350 &  0.009 &  0.359 &  0.025 & -1.583 & -0.422 & -1.619 &  0.093 &  1.001 &  1.560 &   6.69 &   6.47  \\
            D &  0.828 &  0.389 &  1.216 &  0.319 & -0.329 & -1.805 &  1.397 &  1.850 &  0.098 &  0.269 &   7.16 &   7.13  \\
            E &  2.981 &  0.375 &  3.356 &  0.112 & -0.901 & -1.120 &  0.691 &  1.721 &  0.510 &  1.028 &   7.30 &   6.87  \\
           F$^c$ &  3.038 &  5.002 &  8.040 &  0.622 &  0.217 & -1.057 &  1.754 &  1.667 &  0.284 &  1.960 &   7.81 &   7.65 , 7.74  \\
            G &  0.596 &  0.287 &  0.883 &  0.325 & -0.318 & -0.665 &  0.120 &  0.567 &  0.352 &  1.452 &   7.47 &   7.83  \\
            H &  0.255 &  0.032 &  0.288 &  0.112 & -0.898 & -0.500 & -0.993 &  0.034 &  0.620 &  1.337 &   6.78 &   7.52  \\
            I &  1.038 &  1.178 &  2.215 &  0.532 &  0.055 & -0.811 &  0.879 &  0.954 &  0.266 &  1.604 &   7.78 &   8.09  \\
            J &  0.121 &  0.030 &  0.152 &  0.200 & -0.603 & -0.738 & -0.783 & -0.051 &  0.294 &  0.875 &   6.94 &   8.58  \\
            K &  1.603 &  0.690 &  2.293 &  0.301 & -0.366 & -1.491 &  1.328 &  1.823 &  0.191 &  0.522 &   7.08 &   7.14  \\
            L &  4.134 &  1.111 &  5.245 &  0.212 & -0.571 & -0.969 &  1.013 &  1.712 &  0.524 &  1.384 &   7.61 &   7.09  \\
            M &  1.085 &  2.203 &  3.288 &  0.670 &  0.308 & -1.715 &  2.056 &  1.877 &  0.098 &  0.481 &   8.00 &   7.56  \\
            N &  2.429 &  3.860 &  6.289 &  0.614 &  0.201 & -0.893 &  1.477 &  1.405 &  0.313 &  1.959 &   7.90 &   7.85  \\
            O &  1.538 &  0.091 &  1.629 &  0.056 & -1.227 & -0.303 & -0.739 &  0.617 &  1.397 &  2.610 &   6.87 &   6.99  \\
            P &  0.554 &  1.229 &  1.782 &  0.689 &  0.346 & -2.033 &  2.120 &  1.903 &  0.050 &  0.247 &   7.54 &   7.54  \\
            Q &  1.318 &  5.591 &  6.909 &  0.809 &  0.627 & -1.593 &  2.338 &  1.840 &  0.110 &  0.721 &   8.31 &   7.90  \\
            R &  1.823 &  1.109 &  2.932 &  0.378 & -0.216 & -0.565 &  0.608 &  0.953 &  0.532 &  2.210 &   7.55 &   7.73  \\
            S &  3.000 &  3.764 &  6.764 &  0.556 &  0.099 & -1.169 &  1.742 &  1.773 &  0.260 &  1.262 &   7.94 &   7.53  \\
            T &  1.925 &  5.728 &  7.653 &  0.748 &  0.473 & -1.375 &  2.131 &  1.787 &  0.158 &  1.061 &   8.34 &   7.81  \\
           U$^c$ &  2.052 &  6.349 &  8.401 &  0.756 &  0.490 & -1.233 &  2.034 &  1.672 &  0.188 &  1.356 &   8.37 &   7.85 , 7.97  \\
            V &  0.744 &  3.232 &  3.976 &  0.813 &  0.638 & -1.883 &  2.390 &  1.882 &  0.066 &  0.400 &   8.32 &   7.80  \\
           W$^c$ &  4.906 &  2.354 &  7.260 &  0.324 & -0.319 & -0.814 &  1.183 &  1.632 &  0.565 &  1.919 &   7.62 &   7.28 , 7.34  \\
            X &  0.079 &  0.012 &  0.092 &  0.134 & -0.812 & -0.961 & -1.002 & -0.012 &  0.145 &  0.550 &   7.12 &  \nodata  \\
            Y &  0.357 &  1.968 &  2.325 &  0.846 &  0.741 & -2.242 &  2.533 &  1.922 &  0.033 &  0.192 &   8.42 &   7.82  \\
            Z &  3.360 &  0.107 &  3.467 &  0.031 & -1.495 & -0.288 & -0.683 &  0.941 &  2.049 &  3.382 &   6.90 &   6.76  \\
           AA &  1.085 &  0.275 &  1.361 &  0.202 & -0.595 & -0.434 & -0.128 &  0.597 &  0.742 &  2.104 &   7.01 &   7.49  \\
           AB &  2.707 &  2.180 &  4.887 &  0.446 & -0.094 & -1.257 &  1.593 &  1.816 &  0.257 &  0.959 &   7.91 &   7.36  \\
           AC &  2.975 &  0.545 &  3.520 &  0.155 & -0.737 & -0.354 &  0.088 &  0.954 &  1.127 &  2.874 &   7.44 &   7.22  \\
           AD &  0.389 &  1.796 &  2.185 &  0.822 &  0.664 & -2.206 &  2.458 &  1.923 &  0.035 &  0.206 &   8.34 &   7.76  \\
           AE &  1.710 &  4.535 &  6.245 &  0.726 &  0.424 & -1.486 &  2.140 &  1.846 &  0.141 &  0.844 &   8.11 &   7.73  \\
          AF$^c$ &  5.403 &  0.431 &  5.834 &  0.074 & -1.098 & -0.709 &  0.342 &  1.569 &  1.125 &  2.169 &   8.18 &   6.82 , 6.88 \\
           AG &  5.738 &  0.737 &  6.475 &  0.114 & -0.891 & -0.465 &  0.330 &  1.351 &  1.296 &  2.945 &   7.15 &   6.98  \\
                      
\hline
\end{tabular}
}
\flushleft

$^a$ Value derived for the $q$ parameter (in units of cm s$^{-1}$) obtained using the optimal calibration given by \citet{KD02}. \\
$^b$ Value derived for the $q$ parameter (in units of cm s$^{-1}$) obtained using the iterative procedure described in \citet{KK04}. \\
$^c$ For these models, for which we derived \abox$\sim$8.4 following the \Te\ method, we list the results of the empirical calibrations considering both the low and the high metallicity branches. The $q_{KK04}$ parameters listed here for these models are for the low and high metallicity branches, respectively.\\

\end{table*}

\end{document}